\documentclass[11pt]{article}
\usepackage{fullpage}

\usepackage{graphicx}

\usepackage{latexsym}
\usepackage{epsfig}
 \usepackage{times}
\usepackage{amsmath,amsthm, amssymb,color,graphicx,epsf,verbatim}

\newcommand{\ignore}[1]{}
\newcommand{\notinproc}[1]{#1}
\newcommand{\onlyinproc}[1]{}

\newtheorem{thm}{Theorem}[section]
\newtheorem{theorem}{Theorem}[section]

\newtheorem{lemma}[thm]{Lemma}
\newtheorem{conjecture}[thm]{Conjecture}

\newtheorem{definition}[thm]{Definition}

\def\pri{\mbox{\sc pri}}
\def\ws{\mbox{\sc ws}}
\def\rc{\mbox{\sc RC}}

\def\cR{{\cal R}}
\def\cW{{\cal W}}
\def\cS{{\cal S}}

\newcommand{\SV}{{\Sigma}V}
\def\sigv{\SV}
\newcommand{\nSV}{n{\Sigma}V}

\def\pr{\mbox{\sc pr}}
\def\var{\mbox{\sc var}}
\def\cov{\mbox{\sc cov}}
\def\HT{\mbox{\sc HT}}
\def\HTp{\mbox{\sc HTp}}
\def\E{{\textsf E}}

\def\largest{{\texttt largest}}
\def\smallest{{\texttt smallest}}
\def\top{{\texttt{top}\mbox{-}}}

\def\AW{\mbox{\sc AW}-summary}
\def\AWs{\mbox{\sc AW}-summaries}

\def\varopt{\mbox{\sc VarOpt}}

\def\EXP{\mbox{\sc exp}}
\def\IPPS{\mbox{\sc ipps}}

\def\pr{\mbox{\sc pr}}
\def\var{\mbox{\sc var}}
\def\cov{\mbox{\sc cov}}

\begin{document}


\title{Coordinated Weighted Sampling: \\ Estimation of Multiple-Assignment Aggregates}


\ignore{
\author{
\alignauthor Edith Cohen\\
       \affaddr{AT\&T Labs--Research}\\
       \affaddr{180 Park Avenue}\\
       \affaddr{Florham Park, NJ 07932, USA}\\
       \email{edith@research.att.com}
\alignauthor  Haim Kaplan \\
       \affaddr{School of Computer Science}\\
       \affaddr{Tel Aviv University}\\
       \affaddr{Tel Aviv, Israel}\\
       \email{haimk@cs.tau.ac.il}
\alignauthor Subhabrata Sen\\
       \affaddr{AT\&T Labs--Research}\\
       \affaddr{180 Park Avenue}\\
       \affaddr{Florham Park, NJ 07932, USA}\\
       \email{sen@research.att.com}
} }

\author{
Edith Cohen\thanks{
    AT$\&$T Labs-Research, 180 Park Avenue, Florham Park, NJ.
} \and Haim Kaplan \thanks{
    The Blavatnik School of Computer Science, Tel Aviv University.}
\and Subhabrata Sen\thanks{
    AT$\&$T Labs-Research, 180 Park Avenue, Florham Park, NJ.
}   }
\date{}


 \maketitle
\begin{abstract}
Many data sources are naturally modeled by multiple weight assignments
over a set of keys: snapshots of an evolving database at multiple
points in time, measurements collected over multiple time periods,
requests for resources served at multiple locations, and records
with multiple numeric attributes. Over such vector-weighted
data we are interested
in aggregates with respect to one set of weights, such as weighted sums,
and  aggregates over multiple sets of weights such as the $L_1$
difference.

Sample-based summarization is highly effective for data sets that
are too large to be stored or manipulated.  The summary facilitates
approximate processing of queries that may be specified
after the summary was generated.
 Current designs, however,  are geared for data sets
where a single {\em scalar} weight is associated with each key.

 We develop a sampling framework based on
{\em coordinated weighted samples}
that is suited for multiple weight assignments and obtain estimators
that are {\em orders of magnitude tighter} than previously possible.
  We demonstrate the power of our methods through an extensive empirical
evaluation on diverse data sets ranging from IP network  to   stock
quotes data.

\end{abstract}


\section{Introduction} \label{intro:sec}

Many business-critical applications  today are based on extensive  use of computing and  communication network resources. These systems are  instrumented to collect a  wide range  of different types of data. Examples include performance  or environmental measurements, traffic traces, routing  updates,  or SNMP traps in an IP network, and
transaction logs,  system resource (CPU, memory) usage statistics, service level end-end performance statistics in
an end- service infrastructure.   Retrieval of useful information
from this vast amount of data  is critical to  a wide range of  compelling
applications including  network and service management, troubleshooting and root cause analysis,  capacity provisioning,  security,  and  sales and marketing.


Many of these  data sources produce data sets consisting of numeric
vectors ({\em weight vectors}) associated with a set of identifiers
({\em keys}) or equivalently as a set of {\em weight assignments}
over {\em keys}. Aggregates over the data are specified using this
abstraction.

We distinguish between data sources with
{\em co-located} or {\em dispersed} weights.
A data source has
{\bf dispersed weights} if
entries of the weight vector of each key occur in different times or
locations:
(i) Snapshots of a database that is modified over time (each snapshot is a
weight assignment, where the weight of a key is the value of
a numeric attribute in a record with this key.)
(ii) measurements of a set of parameters (keys)
 in different time periods (weight assignments). (iii) number of
 requests for different objects (keys) processed at multiple servers
 (weight assignments).
A data source has {\bf co-located weights} when
a complete weight vector is ``attached'' to each key:
(i)~Records with multiple numeric attributes such as
IP flow records generated by a statistics module at
an IP router, where the attributes are
the number of bytes, number of packets, and unit.
(ii) Document-term datasets,
where keys are documents and weight attributes
are terms or features (The weight value of a term in a document can be the
respective number of occurrences).  (iii) Market-basket datasets, where
keys are baskets and weight attributes are goods (The
weight value of a good in a basket can be its multiplicity).  (iv)
Multiple numeric functions over one
(or more) numeric measurement of a parameter.
For example, for measurement $x$ we
might be interested in both first and second moments, in which case we
can use the weight assignments $x$ and $x^2$.


  A very useful common type of query involves properties of a {\em
sub-population} of the monitored data that are {\em additive} over
keys.  These aggregates can be broadly categorized as : (a) {\em
Single-assignment}  aggregates, defined with respect to a
single attribute, such as the weighted sum or selectivity of a
subpopulation of the keys.  An example over IP flow records
is the total bytes of all IP
traffic with a certain destination Autonomous
System~\cite{DLT:sigcomm03,dlt:pods05,KSXZ:sigmetrics05,bottomk07:ds,bottomk:VLDB2008}.
(b) {\em Multiple-assignment} aggregates include similarity or divergence
metrics such as the $L_1$ difference between two weight assignments
or maximum/minimum weight over a subset of
assignments~\cite{KSZC:IMC03,CM:ton05,Charikar:stoc02,CM:esa03}.
Figure~\ref{example1:fig} (A) shows an example of three weight assignments over
a set of keys and key-wise values for multiple-assignment aggregates
including the minimum or maximum value of a key over subset of assignments
and the $L_1$ distance.  The aggregate value over selected keys
is the sum of key-wise values.

Multiple-assignment aggregates are used for clustering, change detection, and mining emerging patterns.
Similarity over  corpus of documents,
according to a selected subset of features, can be used
to detect near-duplicates and reduce redundancy~\cite{Manber:usenix94,CFGM:ACMtis02,swa:sigmod03,CS:sigir04,KCA:sigkdd04,MJS:WWW07}.
A retail merchant may want to cluster locations according to
sales data for a certain type of merchandise.
In IP networks, these
aggregates are used for monitoring, security, and
planning~\cite{fcb_summary:ton00,CM:ton05,CCHKS:KDD08,MSDFPS:sigcomm08}:
An increase in the amount of distinct flows on a certain port might
indicate a worm activity, increase in traffic to a certain set of
destinations might indicate a flash crowd or a DDoS attack, and an
increased number of flows from a certain source may indicate scanner
activity.
A network security application might track the increase in traffic to a customer site  that originates from a certain suspicious  network or geographic area.


Exact computation of such aggregates can be prohibitively resource-intensive:
Data sets are often too large to be either stored for long time periods or to be collated across many locations.
Computing multiple-assignment aggregates may
require  gleaning information across data sets
from different times and locations.
 We therefore aim at
concise summaries of the data sets, that can be computed in
a scalable way and facilitate approximate query processing.

  Sample-based summaries
\cite{Knu69,Vit85,Broder:CPM00,BRODER:sequences97,ECohen6f,GT:spaa2001,Gibbons:vldb2001,BHRSG:sigmod07,DasuJMS:sigmod02,HYKS:VLDB2008,bottomk:VLDB2008,DLT:jacm07,varopt:CDKLT08,CK:sigmetrics09}
are more flexible than other formats: they naturally facilitate
subpopulation queries by focusing on sampled keys that are members of
the subpopulation and are suitable when the exact query of interest is
not known beforehand or when there are multiple attributes of
interest.  When keys are weighted, weighted sampling, where heavier keys are more likely to be sampled, is essential for performance. 
Existing methods, however, are designed for one set of weights
and are either not applicable or perform
poorly on multiple-assignment aggregates.

\subsection*{Contributions}

We develop sample-based
summarization framework for vector-weighted data that
supports efficient approximate aggregations.
The challenges differ between  the
dispersed and co-located models due to the particular constraints
imposed on scalable summarization.

\medskip
\noindent
\textsl{{\bf Dispersed weights model:}}
Any scalable algorithm must decouple the processing of
different assignments --
collating dispersed-weights data to obtain explicit
key/vector-weight representation is too expensive.
Hence, processing of one assignment can not depend
on other assignments.

 We propose summaries based on {\em coordinated weighted samples}.
The summary contains a ``classic'' weighted sample taken with respect
to each assignment.
Coordination loosely means that a key that is sampled under one
assignment is more likely to be sampled under other assignments.
 We can tailor the sampling to be Poisson,
$k$-mins, or bottom-$k$ (order) sampling. In all three cases,
sampling is efficient on data streams, distributed data, and metric
data~\cite{ECohen6f,CoKa:jcss07,ES:IPL2006,bottomk07:ds} and there are
unbiased subpopulation weight estimators that have variance that
decreases linearly or faster with the sample
size~\cite{ECohen6f,DLT:jacm07,Szegedy:stoc06,bottomk:VLDB2008}.
Bottom-$k$
samples~\cite{Rosen1972:successive,SSW92,Rosen1997a,ECohen6f,bottomk:VLDB2008,Ohlsson_SPS:1998,DLT:jacm07},
with the advantage of a fixed sample size, emerge as a better choice.
 Our design has the following important properties:
\begin{description}
\item[$\bullet$] {\bf Scalability:}
The processing of each assignment  is a
simple adaptation of single-assignment weighted sampling
algorithm.  Coordination is achieved by
using the same hash function across assignments.

\ignore{ 
}

\item[$\bullet$]{\bf Weighted sample for each assignment:}
Our design is especially appealing for
 applications where sample-based summaries
are already used, such as periodic (hourly) summaries of IP flow records.
The use of our framework versus independent sampling in different periods
facilitates support for queries on the
relation of the data across time periods.

\item[$\bullet$]
{\bf Tight estimators:}
 We provide a principled ``template'' derivation of estimators,
tailor it to obtain tight unbiased
estimators
for the min, max, and  range ($L_1$), and bound the variance.

\end{description}

\smallskip
\noindent
\textsl{{\bf Colocated weights model:}}
For colocated data, the full weight vector of each key is readily
available to the summarization algorithm and can be easily incorporated
in the summary.  We discuss the shortcomings of applying
previous methods to summarize this data.
One approach is to sample records according to one
particular weight assignment. Such a
sample can be used to estimate aggregates that involve other
assignments,\notinproc{\footnote{This is standard, by multiplying the
per-key estimate with an appropriate ratio~\cite{SSW92}}} but
estimates may have large variance and be biased.
Another approach is to concurrently compute
multiple weighted samples, one for each
assignment. In this case, single-assignment
aggregates can be computed over the respective sample but
no unbiased estimators for
multiple-assignment aggregates were known.
Moreover, such a summary is wasteful in terms of storage as
different assignments are often correlated (such as number of bytes
and number of IP packets of an IP flow).

\ignore{
  The challenges are:
(i) Exploit correlation between weight assignments to obtain
a summary that
contains a size-$k$ scalar weighted sample
of each assignment but has a smaller combined sample size (in terms
of distinct keys).
(ii) derive tighter statistical estimators that are able to utilize
samples collected with respect to ``other''
weight assignments to
obtain tighter estimates for scalar and multiple-assignment
aggregates.
}

We consider summaries where the set of included keys embeds
a weighted sample with respect to
each assignment.  The set of embedded samples
can be independent or coordinated.
Such a summary can be computed in a scalable way
by a stream algorithm or distributively.

\begin{description}

\item[$\bullet$]

 We derive estimators, which we refer
to as {\em inclusive estimators},
that utilize all keys included in the summary.
An inclusive estimator of a single-assignment aggregate
applied to a summary that embeds a certain weighted sample
from that assignment is at least as tight, and typically significantly tighter, than an
estimator directly applied to the embedded sample.
 Moreover, inclusive estimators are applicable to multiple-assignment
aggregates, such as the $\min$, $\max$, and $L_1$.
\item[$\bullet$]
 We show that when the embedded samples are
coordinated, the
number of distinct keys in the summary (in the union of the embedded samples) is minimized.
\ignore{For example, if we want
size-$k$ weighted samples that can be used to estimate each of
the number of bytes, number of
packets, and number of distinct flows, coordinated samples
have the smallest number of distinct keys in these samples.}
\end{description}

\noindent \textsl{{\bf Empirical evaluation.}} We performed a comprehensive
empirical evaluation using IP packet traces, movies' ratings data
set (The Netflix Challenge~\cite{netflix}), and stock quotes data
set.
These data sets and queries also demonstrate potential applications.
For dispersed data  we achieve {\em orders of magnitude}
reduction in variance over previously-known estimators and
estimators applied to independent weighted samples. The variance of
these estimators is comparable to the variance of a weighted sum
estimator of a single weight assignment.

  For co-located data, we demonstrate that the size of our combined
  sample is significantly smaller than the sum of the sizes of
  independent samples one for each weight assignment.
  We also demonstrate
that even for single assignment aggregates, our estimators which use
the combined sample are much tighter than the estimators that use
only a sample for the particular  assignment.

\ignore{
To our best knowledge, our framework is the first to support
bounded-variance multiple-assignment aggregates,
such as the $L_1$ difference, over
sample-based summaries.
}


\smallskip
\noindent
{\bf Organization.}
The remainder of the paper is arranged as follows. Section ~\ref{related:sec} reviews related work, Section~\ref{prelim:sec}  presents key background concepts and Section~\ref{model:sec} presents our sampling approach.\notinproc{
Sections~\ref{genest:sec}-\ref{dispersed:sec} present our estimators:
Section~\ref{genest:sec} presents a template estimator which we apply to colocated summaries in Section~\ref{coloc:sec}
and to dispersed summaries in Section~\ref{dispersed:sec}.
 Section~\ref{varbounds:sec} provides bounds on the variance.
This is an extended version of~\cite{multiw:VLDB2009}.
 }

 \section{Related work}\label{related:sec}

\noindent
{\bf Sample coordination.} Sample coordination was used in survey
sampling for almost four decades.  {\em Negative coordination} in
repeated surveys was used to decrease the likelihood that the same
subject is surveyed (and burdened) multiple times.  {\em Positive
coordination} was used to make samples as similar as possible when
parameters change in order to reduce overhead.  Coordination is
obtained using the PRN (Permanent Random Numbers) method  for
Poisson samples~\cite{BrEaJo:1972} and order (bottom-$k$)
samples~\cite{Saavedra:1995,Ohlsson:2000,Rosen1997a}.  PRN resembles
our ``shared-seed'' coordination method.
The challenges of massive data sets, however, are different from those
of survey sampling and in particular, we are not aware of previously
existing unbiased estimators for multiple-assignment aggregates over
coordinated weighted samples.

 Coordination (of Poisson, $k$-mins, and bottom-$k$ samples)
was (re-)introduced in computer science as a
method to support aggregations that involve multiple sets
\cite{Broder:CPM00,BRODER:sequences97,ECohen6f,GT:spaa2001,Gibbons:vldb2001,BHRSG:sigmod07,bottomk:VLDB2008,HYKS:VLDB2008,CK:sigmetrics09}
and as a form of locality sensitive hashing \cite{Charikar:stoc02}.
Coordination addressed the issue that 
independent samples of different sets over the same universe
provide weak estimators for multiple-set aggregates such as
intersection size or similarity.
Intuitively, two large but almost identical sets are likely
to have disjoint independent samples -- the sampling does not retain any
information on the relations between the sets.

This previous work, however, considered restricted weight models:
{\em uniform}, where
all weights are $0/1$, and {\em global weights}, where a key has the
same weight value across all assignments where its weight is strictly positive
(but
the weight can vary between keys).  Allowing the
same key to assume different positive weights in
different assignments is clearly essential for our applications.

While these methods can be applied with general weights, by ignoring
weight values and performing coordinated uniform sampling, resulting
estimators are weak.  Intuiti vely, uniform sampling performs poorly on
weighted data because it is likely to leave out keys with dominant
weights.  Weighted sampling, where keys with larger weights are more
likely to be represented in the sample, is essential for boundable
variance of weighted aggregates.

  The structure of coordinated samples for general weights
turns out to be more involved than with global weights, where
essentially the samples of different assignments (sets) are derived from a single
``global'' (random) ranking of keys.  The derivation of unbiased
estimators was also more challenging: global weights allow us to make
simple inferences on inclusion of keys in a set when the key is not
represented in the sample.  These inferences facilitate the derivation
of estimators but do not hold under general weights.

\ignore{ General weights pose a significant challenge. Firstly,
coordinated sketches for global weights are derived using {\em
global rank assignments over keys} (essentially, a random order of
keys where heavier keys are more likely to have smaller rank
values).  Coordinated samples are determined by this global ranking,
where a sample of a set includes the least ranked key(s) that is a
member of the set. With general weights, it is not possible to use a
global ranking of keys and still obtain a valid weighted sample for
each set.  This is because an key can have large weight with respect
to one assignment and a small weight on another.

 The global ranking used with uniform/global weights determines
the full joint distribution of coordinated samples and also
facilitates efficient sampling algorithms. Part of our challenge in
handling general weights was to determine appropriate ``joint
distributions'' that are both easy to generate in a scalable way and
have desirable estimation qualities.

  Finally, estimators derived for global weights rely on the property that
the full set membership (weight assignment) information can be
deduced for all keys included in the union sketch.  This is because
once we see one weight assignment where a key assumes positive
weight we obtain an immediate ``lower bound'' on the smallest
positive weight a key can assume. This is not the case with general
weights, which required a different design of estimators.

}

\notinproc{
\noindent
{\bf Unaggregated data.}  Sample-based
sketches~\cite{GM:sigmod98,CDKLT:pods07,CDKLT:IMC07} and sketches that
are not samples were also proposed for {\em unaggregated} data streams
(the scalar weight of each key appears in multiple data
points)~\cite{ams99}. This is a more general model with weaker
estimators than ``aggregated'' keys.  We leave for future work
summarization of unaggregated data set with vector-weights.

\noindent
 \varopt\ is a
weighted sampling design~\cite{Cha82,varopt:CDKLT08} that
realizes all the advantages of other schemes but it is not clear
if it can be applied with coordinated samples (even with global weights).
}

\noindent {\bf Sketches that are not samples.}  Sketches that are
not {\em sample based}
\cite{Manber:usenix94, CFGM:ACMtis02,swa:sigmod03,CS:sigir04,KCA:sigkdd04,MJS:WWW07,CM:esa03,FKSV:99}
are effective point solutions for particular metrics such as
max-dominance~\cite{CM:esa03} or $L_1$~\cite{FKSV:99} difference.
Their disadvantage is less flexibility in terms of supported
aggregates and in particular, no support for aggregates over
selected subpopulations of keys: we can estimate the overall $L_1$
difference between two time periods but we can not estimate the
difference restricted to a subpopulation such as flows to particular
destination or certain application. There is also no mechanism to
obtain ``representatives'' keys\cite{Schweller:IMC04}. 
Lastly, even a practical implementation of~\cite{CM:esa03,FKSV:99}
involves constructions of stable distributions or range summable
random variables (whereas for our sample-based summaries all is
needed is ``random-looking'' hash functions).  \ignore{When compared with
these methods, for example, to estimate the max-dominance norm
between weight assignments, our methods feature the same asymptotic
dependence between approximation and sample size
($k=O(\epsilon^{-2})$) in order to support the queries supported by
these methods. }

Bloom filters~\cite{bloomfilters:cacm70,fcb_summary:ton00}
also support estimation of similarity
metrics but summary size is not tunable and
grows linearly with the number of keys.
\ignore{
The specific application of sketches that support change detection for IP
traffic were studied in~\cite{KSZC:IMC03,CM:ton05}.  The
sketch format, based on~\cite{Charikar:stoc02,CM:esa03}, is
not sample based. \cite{Schweller:IMC04}  offers some partial remedy to the
sketches not providing ``representative samples.''
In contrast,  we offer a principled general approach based on the
classic notion of weighted samples.
}

\section{Preliminaries} \label{prelim:sec}
\notinproc{
\begin{figure}
{\small
\begin{minipage}{6in}
{\bf weighted set} $(I,w)$ with {\bf keys} $I=\{i_1,\ldots,i_6\}$ and a {\bf rank assignment} r\\
\begin{tabular}{|c|r|r|r|r|r|r|}
\hline
key $i$: & $i_1$ & $i_2$ & $i_3$ & $i_4$ & $i_5$ & $i_6$ \\
\hline
weight $w(i)$ & $20$ & $10$ & $12$ & $20$ &  $10$ & $10$ \\
 $u(i)\in U[0,1]$ & $0.22$ & $0.75$ & $0.07$ & $0.92$ & $0.55$ & $0.37$ \\
 $r(i)=u(i)/w(i)$ & $0.011$ & $0.075$   & $0.0583$ & $0.046$ &  $0.055$ & $0.037$ \\
\hline
\end{tabular}

\smallskip
{\bf Poisson samples with expected size $k=1,2,3$ and \AWs:}\\
$p(i)=\min\{1,w(i)\tau\}$, $k=\sum_i p(i)$, $a(i)=w(i)/p(i)$\\
\begin{tabular}{|l|l|l|l|r|r|r|r|r|r|}
\hline
$k$ & sample & $\tau$ & $i:$ & $i_1$ & $i_2$ & $i_3$ & $i_4$ & $i_5$ & $i_6$ \\
\hline
$1$ & $i_1$ & $\frac{1}{82}$ & $p(i)$: & $0.24$ & $0.12$ & $0.15$ &$0.24$ & $0.12$ & $0.12$  \\
 & & & $a(i)$: & $82$ & $0$ & $0$  & $0$ & $0$ & $0$ \\
\hline
$2$ & $i_1$ & $\frac{2}{82}$ & $p(i)$: & $0.49$ & $0.24$ & $0.29$ &$0.49$ & $0.24$ & $0.24$  \\
 & & & $a(i)$: & $41$ & $0$ & $0$  & $0$ & $0$ & $0$ \\
\hline
$3$ & $i_1$ & $\frac{3}{82}$ & $p(i)$: & $0.73$ & $0.37$ & $0.44$ & $0.73$ & $0.37$ & $0.37$  \\
 & & & $a(i)$: & $27.40$ & $0$ & $0$  & $0$ & $0$ & $0$ \\
\hline
\end{tabular}

\smallskip
{\bf Bottom-$k$ samples of size $k=1,2,3$ and \AWs:}\\
$p(i)=\min\{1,w(i)r_{k+1}\}$, $a(i)=w(i)/p(i)$\\
\begin{tabular}{|l|l|l|l|r|r|r|r|r|r|}
\hline
$k$ & sample & $r_{k+1}$ & $i:$ & $i_1$ & $i_2$ & $i_3$ & $i_4$ & $i_5$ & $i_6$ \\
\hline
$1$ & $i_1$ & $0.037$ & $p(i)$: & $0.74$ & & & & &  \\
 & & & $a(i)$: & $27.02$ & $0$ & $0$  & $0$ & $0$ & $0$ \\
 \hline
 $2$ & $i_1,i_6$ & $0.046$ & $p(i)$: & $0.92$ & & & & & $0.46$ \\
 & & & $a(i)$: & $21.74$ & $0$ & $0$  & $0$ & $0$ & $21.74$ \\
\hline
 $3$ & $i_1,i_6,i_4$ & $0.055$ & $p(i)$: & $1$ & & & $1$ & & $0.55$ \\
 & & & $a(i)$: & $20.00$ & $0$ & $0$  & $20.00$ & $0$ & $18.18$ \\
\hline
\end{tabular}
\end{minipage}
}
\caption{Example of a weighted set, a random rank assignment with \IPPS\ ranks, Poisson and bottom-$k$ samples, and respective \AWs.\label{ex0:fig}}
\end{figure}
}
A {\em weighted set\/} $(I,w)$ consists of a set of keys $I$ and a
function $w$ assigning a scalar weight value
 $w(i)\geq 0$ to each key $i\in I$.
We review components of sample-based summarizations of a weighted set: sample distributions,
respective {\em sketches}, that in our context are samples with some auxiliary information, and estimators for weight queries in the form of
{\em adjusted weights} associated with sampled keys.

Sample distributions are defined through
{\em random rank assignments}~\cite{ECohen6f,Rosen1997a,bottomk07:ds,DLT:jacm07,bottomk:VLDB2008,CK:sigmetrics09} that map each key $i$ to a rank value $r(i)$.  The rank assignment
is defined with respect to a family of probability density functions ${\bf f}_w$ ($w\geq 0$), where
each $r(i)$ is drawn independently according to  ${\bf f}_{w(i)}$.
 We say that ${\bf f}_w$ ($w\geq 0$) are {\em monotone}
if for all $w_1 \geq w_2$, for all $x$, ${\bf F}_{w_1}(x)\geq {\bf
F}_{w_2}(x)$ (where $\bf{ F}_w$ are the respective cumulative
distributions). For a set $J$ and a rank assignment $r$ we denote by
$r_i(J)$ the $i$th smallest rank of a key in $J$, we also abbreviate
and write $r(J) = r_1(J)$.
\begin{description}
\item[$\bullet$] A {\em Poisson}-$\tau$ sample of $J$ is
defined with respect to a rank assignment $r$.  The sample is the set of keys
with $r(i)<\tau$.  The sample
has {\em expected} size $k=\sum_i {\bf F}_{w(i)}(\tau)$.
Keys have independent inclusion probabilities.
The sketch includes the pairs $(r(i),w(i))$ and may include
key identifiers with attribute values.
\item[$\bullet$] A {\em bottom-$k$ (order-$k$)} sample
of $J$  contains the $k$ keys $i_1,\ldots,i_k$ of smallest ranks in $J$.
The sketch $s_k(J,r)$  consists of the $k$ pairs
 $(r(i_j),w(i_j))$, $j=1,\ldots,k$, and
$r_{k+1}(J)$. (If $|J| \le k$ we store only $|J|$ pairs.), and may include
the key identifiers $i_j$ and additional attributes.
\item[$\bullet$] A {\em $k$-mins sample\/} of $J\subset I$ is produced
from $k$ independent rank assignments, $r^{(1)},\ldots,r^{(k)}$. The
sample is the set of (at most $k$) keys) with
minimum rank values $r^{(1)}(J)$, $r^{(2)}(J)$,
$\ldots$, $r^{(k)}(J)$.  The sketch
includes the minimum rank values and,
depending on the application, may include
corresponding key identifiers and attribute values.
\end{description}

When weights of keys are uniform,  a $k$-mins
sample is the result of $k$ uniform draws with replacement, bottom-$k$ samples are
$k$ uniform draws without replacements, and Poisson-$\tau$ samples
are independent Bernoulli trials.
  The particular family ${\bf f}_w$  matters
when weights are not uniform.
Two families with
special properties are:
\begin{description}
\item[$\bullet$]
 \EXP\ ranks:
${\bf f}_w(x) = w e^{-wx}$ (${\bf F}_w(x) = 1- e^{-wx}$) are
exponentially-distributed with parameter $w$ (denoted by
 $\EXP[w]$).
Equivalently, if $u\in U[0,1]$ then $-\ln(u)/w$ is an exponential
random variable with parameter $w$. $\EXP[w]$ ranks have the
property that the minimum rank $r(J)$  has distribution
$\EXP[w(J)]$, where $w(J)=\sum_{i\in J} w(i)$\notinproc{. (The
minimum of independent exponentially distributed random variables is
exponentially distributed with parameter equal to the sum of the
parameters of these distributions).}  This property is useful for
designing estimators and efficiently computing
sketches~\cite{ECohen6f,CoKa:jcss07,ES:IPL2006,bottomk07:ds,bottomk:VLDB2008}.
The $k$-mins sample~\cite{ECohen6f} of a set is a sample drawn {\em
with replacement} in $k$ draws where a key is selected with
probability equal to the ratio of its weight and the total weight.
A bottom-$k$ sample is the result of $k$ such draws performed {\em
without replacement}, where keys are selected according to the ratio
of their weight and the weight of remaining keys
\cite{Rosen1972:successive,Hajekbook1981,Rosen1997a}. \ignore{We use
the term \ws\ sketch for a bottom-$k$ sketch with $\EXP$ ranks.}
\item[$\bullet$] \IPPS\ ranks:
\looseness=-1 ${\bf f}_w$ is the uniform distribution $U[0,1/w]$
(${\bf F}_w(x) = \min\{1,wx\}$). This is the equivalent to choosing
rank value $u/w$, where
 $u\in U[0,1]$.
The Poisson-$\tau$ sample is an \IPPS\ sample~\cite{Hajekbook1981} (Inclusion Probability Proportional to Size). The bottom-$k$ sample
is a priority sample~\cite{Ohlsson_SPS:1998,DLT:jacm07} (\pri).
\end{description}

\notinproc{Figure~\ref{ex0:fig} shows an example of a weighted set with 6 keys
and a respective rank assignment with \IPPS\ ranks.  The figure
shows the corresponding
Poisson samples of expected size $k=1,2,3$.  The value $\tau$ is
calculated according to the desired expected sample size.  The sample
includes all keys with rank value that is below $\tau$.
This particular rank assignment yielded a
Poisson sample of size $1$ when the expected size was $1,2,3$.
The figure also shows bottom-$k$ samples of sizes $k=1,2,3$, containing
the $k$ keys with smallest rank values.}

\noindent {\bf Adjusted weights.}
A technique to obtain estimators for the weights of keys is by
assigning an adjusted weight $a(i)\geq 0$ to each key $i$ in the sample
(adjusted weight $a(i) = 0$ is implicitly assigned to keys not in the sample).
 The adjusted
weights are assigned such that $\E[a(i)] = w(i)$, where the
expectation is over the randomized algorithm choosing the sample.
 We refer to the (random variable) that
combines a weighted sample of $(I,w)$
together with adjusted weights as an {\em adjusted-weights summary}
(\AW) of $(I,w)$.  An \AW\ allows us
to obtain an unbiased estimate on the weight of {\em any}
subpopulation $J\subset I$.
The estimate $\sum_{j\in J} a(j) = \sum_{j\in J |
a(j)>0} a(j)$ is easily computed from the summary provided that
we have sufficient auxiliary
information to tell for each key in the summary whether it
belongs to $J$ or not.
Figure~\ref{ex0:fig} shows example \AWs\ for the Poisson and bottom-$k$
samples.  The set $J=\{i_2,i_4,i_6\}$
with weight $w(J)=w(i_2)+w(i_4)+w(i_6)=10+20+10=40$ has estimate
of $0$ using the three Poisson \AWs\ and estimates $0,21.74,38.18$
respectively by the three bottom-$k$ \AWs.
Moreover,
for any secondary numeric function $h()$ over keys' attributes such that
$h(i)>0\implies w(i)>0$ and any
subpopulation $J$,
$\sum_{j\in J | a(j)>0} a(j)h(j)/w(j)$ is  an unbiased estimate of
$\sum_{j\in J} h(j)$.

\smallskip
\noindent {\bf  Horvitz-Thompson (\HT).}
\looseness=-1 Let $\Omega$ be the
distribution over samples such that if $w(i)>0$ then
$p^{(\Omega)}(i)=\Pr\{i\in s | s\in\Omega\}$ is positive.
If we know $p^{(\Omega)}(i)$ for every $i\in s$,
we can assign to $i\in s$ the
adjusted weight
$a(i)=\frac{w(i)}{p^{(\Omega)}(i)}\ .$
Since $a(i)$ is $0$ when $i\not\in s$,
$\E[a(i)] = w(i)$ ($a(i)$ is an unbiased estimator of $w(i)$).
 These adjusted weights are
called the Horvitz-Thompson (\HT) estimator \cite{HT52}.
For a particular $\Omega$,
the \HT\ adjusted weights minimize $\var[a(i)]$ for all $i\in I$.
The \HT\ adjusted weights for Poisson $\tau$-sampling are $a(i)=w(i)/{\bf
F}_{w(i)}(\tau)$.
\notinproc{Figure~\ref{ex0:fig} shows the inclusion probability ${\bf
F}_{w(i)}(\tau)$ and a corresponding \AW\ for the Poisson samples.}
Poisson sampling with \IPPS\ ranks and \HT\ adjusted weights
are known to
minimize the sum $\sum_{i\in I} \var(a(i))$ of per-key variances over all
\AWs\ with the same expected size.

\smallskip
\noindent {\bf \HT\ on a partitioned sample space (\HTp)~\cite{bottomk:VLDB2008}.} This is a
method to derive adjusted weights when we cannot determine
$\Pr\{i\in s | s\in\Omega\}$ from the information contained in the
sketch $s$ alone. 

\looseness=-1  For each key $i$ we
consider a partition of $\Omega$ into equivalence classes.
For a sketch $s$, let $P^i(s)\subset \Omega$
be the equivalence class of $s$.
This partition must
satisfy the following requirement:
Given $s$ such that $i\in s$, we can compute the
conditional probability $p^i(s)=\Pr\{i\in s' \mid s'\in P^i(s)\}$
from the information included in $s$.

 We can therefore compute for all $i\in s$
the assignment $a(i)=w(i)/p^i(s)$
(implicitly, $a(i)=0$ for $i\not\in s$.)
It is easy to see that within each equivalence class,
$\E[a(i)]=w(i)$.  Therefore,
also over $\Omega$ we have $\E[a(i)]=w(i)$.

\ignore{
The variance
of the adjusted weight $a(i)$ obtained using \HTp\
depends on the particular partition in the following way. (This follows
from the convexity of the variance.)
\begin{lemma} \label{coarser:lemma} \label{lem:coarse}\cite{bottomk:VLDB2008}
Consider two partitions of the sample space, such that one partition
is a refinement of the other.
Then the variance of $a(i)$ using \HTp{} with the coarser partition
is at most that of the \HTp{} with the  finer partition.
\end{lemma}
}

\noindent {\bf Rank Conditioning (\rc)} [\cite{bottomk:VLDB2008}].
When $s$ is a bottom-$k$ sketch
of $(I,w)$, then $\Pr\{i\in s | s\in\Omega\}$
generally depends on all the
weights $w(i)$ for $i\in I$ and therefore cannot be determined from $s$.
Therefore, we can not directly apply the \HT\ estimator.  
\rc\ is an \HTp{}  method designed for bottom-$k$ sketches.
 For each $i$ and
possible rank value $\tau$ we
have an equivalence class $P^i_{\tau}$ containing all
sketches in which the $k$th
smallest rank value assigned to a key other than $i$ is $\tau$.
Note that if $i\in s$ then $\tau$
is the $(k+1)$st smallest rank which is included in the sketch.
The inclusion probability of $i$ in a sketch
in $P^i_{\tau}$ is $p^i_{\tau}={\bf F}_{w(i)}(\tau)$ and it can be
computed from the sample.

Applying \rc, consider
$s$ containing keys
 $i_1,\ldots,i_k$ and the
$(k+1)$st smallest rank value $r_{k+1}$.
Then for key $i_j$, we have $s\in P^{i_j}_{r_{k+1}}$ and
$a(i_j)=\frac{w(i_j)}{\mbox{\bf F}_{w(i_j)}(r_{k+1})}$.
 The \rc\ estimator for bottom-$k$ samples with \IPPS\ ranks \cite{DLT:jacm07}
has a sum of per-key variances that is at most
that of an \HT\ estimator applied to a Poisson sample  with \IPPS\ ranks
and expected size $k+1$ \cite{Szegedy:stoc06}.

This \rc{} method was extended in~\cite{CK:sigmetrics09} to obtain
estimators over coordinated bottom-$k$ sketches with
global weights~\cite{CK:sigmetrics09}.
The \rc{} method facilitated the derivation and analysis of unbiased
estimators over bottom-$k$ samples -- it provided a way to ``get around''  the  dependence between inclusions of different keys, typically, 
without sacrificing accuracy with respect to estimators over Poisson samples.

Figure~\ref{ex0:fig} shows the $(k+1)$st smallest rank value, the
conditional inclusion probability ${\bf F}_{w(i)}(r_{k+1})$ and
the corresponding \AW\ for each bottom-$k$ sample in the example.

We subsequently use the notation $\Omega(i,r^{-i})$ for the probability
subspace of rank assignments that contains all rank assignments $r'$
that agree on $r$ for
all keys in $I\setminus\{i\}$.

\ignore{
 The \pri\ \rc\ adjusted weight for an key $i_j$
(obtained by a tailored derivation
in~\cite{dlt:pods05}), is $\max\{w(i_j),1/r_{k+1}\}.$
}

 \noindent
{\bf Sum of per-key variances}
\looseness=-1  Different \AWs\ are compared based
on their {\em estimation quality}.
 Variance is the standard metric for the quality of an estimator for
a single quantity.
 For a subpopulation $J$
and \AWs\ $a()$, the variance is $\var[a(J)]=\E[a(J)]^2-w(J)^2$.
Since our application is for subpopulations that may not be
specified a priori, the notion of a good metric is more subtle.
Clearly there is no single \AW\ that dominates all others of the same size
(minimizes the variance) for all subpopulations $J$.


The metric we use in our performance evaluation is
the {\em sum of per-key variances} $\sigv[a]\equiv
\sum_{i\in I} \var[a(i)]$~\cite{DLT:jacm07,bottomk:VLDB2008}.
For \AWs\ with
{\em zero covariances} (for any two keys $i,j$,
$\cov[a(i),a(j)]=\E[a(i)a(j)]-w(i)w(j)=0$),
$\sigv[a]$
also measures {\em average variance} over
subpopulations of any given weight or size~\cite{SzTh:esa07}.
\rc\ adjusted weights on single-assignment~\cite{bottomk:VLDB2008} and
their extension to coordinated sketches with
global weights~\cite{CK:sigmetrics09} have zero covariances.
\HT\ adjusted weights for Poisson sketches have
zero covariances (this is immediate from independence).
When covariances are zero,
the variance of $a(J)$ for a particular subpopulation $J$
is equal to $\sum_{i,j\in J}\cov[a(i),a(j)]= \sum_{i\in J}\var[a(i)]$.

\looseness=-1 Estimators for
Poisson, $k$-mins, and bottom-$k$ sketches with \EXP\ or \IPPS\ ranks
have $\sigv[a]\leq \frac{w(I)^2}{k-2}$
(where $k$ is the
(expected) sample size)~\cite{ECohen6f,bottomk07:ds,DLT:jacm07,Szegedy:stoc06}.
For a subpopulation $J$
with expected $k'$ samples in the sketch, the variance on estimating
$w(J)$ is bounded by $w(J)^2/(k'-2)$.
For bottom-$k$ and Poisson sketches, the variance is smaller  
when the weight distribution is more skewed
~\cite{bottomk07:ds,DLT:jacm07}.

\section{Model and summary formats} \label{model:sec}

  We model the data using
a set of keys $I$ and a set $\cW$ of {\em weight assignments} over
$I$.  For each $b\in \cW$, $w^{(b)}:I \rightarrow {\cal{R}}_{\ge 0}$
maps keys to nonnegative reals. Figure~\ref{example1:fig} shows a
 data set with $I=\{i_1,\ldots,i_6\}$ and $\cW=\{1,2,3\}$.
For $i\in I$ and $\cR\subset \cW$, we use the notation
$w^{(\cR)}(i)$ for the {\em weight vector} with entries $w^{(b)}(i)$
ordered by $b\in\cR$.

\begin{figure*}
{\tiny
\begin{minipage}{2.4in}
{\bf keys:} $I=\{i_1,\ldots,i_6\}$\\
{\bf weight assignments:} $w^{(1)},w^{(2)},w^{(3)}$\\
\begin{tabular}{|c|r|r|r|r|r|r|}
\hline
assignment/key & $i_1$ & $i_2$ & $i_3$ & $i_4$ & $i_5$ & $i_6$ \\
\hline
$w^{(1)}$ & 15 & 0  & 10 &  5 & 10 & 10 \\
$w^{(2)}$ & 20 & 10 & 12 & 20 &  0 & 10 \\
$w^{(3)}$ & 10 & 15 & 15 &  0 & 15 & 10 \\
\hline
\multicolumn{7}{l}{{\bf Example functions} $f(i_j)$}\\
\hline
$w^{(\max \{1,2\})}$   & 20 & 10 & 12 & 20 & 10 & 10 \\
$w^{(\max \{1,2,3\})}$ & 20 & 15 & 15 & 20 & 15 & 10 \\
$w^{(\min \{1,2\})}$   & 15 &  0 & 10 &  0 & 0  & 10 \\
$w^{(\min \{1,2,3\})}$ & 10 &  0 & 10 &  0 & 0  & 10 \\
$w^{(L_1 \{1,2\})}$ & 5 & 10 & 2 & 15 & 10 & 0 \\
$w^{(L_1 \{2,3\})}$ & 10 & 5 & 3 & 20 & 15 & 0 \\
\hline
\end{tabular}
\centerline{(A)}
\end{minipage}
\begin{minipage}{3.0in}
{\bf Consistent shared-seed \IPPS\ ranks:}\\
\begin{tabular}{|c|l|l|l|l|l|l|}
\hline
key: & $i_1$ & $i_2$ & $i_3$ & $i_4$ & $i_5$ & $i_6$ \\
\hline
$u$ & $0.22$ & $0.75$       & $0.07$ & $0.92$ & $0.55$ & $0.37$ \\
\hline
$r^{(1)}$ & $0.0147$ & $+\infty$ & $0.007$ &  $0.184$  & $0.055$ & $0.037$ \\
$r^{(2)}$ & $0.011$ & $0.075$   & $0.0583$ & $0.046$ &  $+\infty$ & $0.037$ \\
$r^{(3)}$ & $0.022$ & $0.05$    & $0.0047$ &  $+\infty$ & $0.0367$ & $0.037$ \\
\hline
\end{tabular}

{\bf Independent \IPPS\ ranks:}\\
\begin{tabular}{|c|l|l|l|l|l|l|}
\hline
key: & $i_1$ & $i_2$ & $i_3$ & $i_4$ & $i_5$ & $i_6$ \\
\hline
$u^{(1)}$ & $0.22$ & $0.75$ & $0.07$ & $0.92$ & $0.55$ & $0.37$ \\
$r^{(1)}$ & $0.0147$ & $+\infty$ & $0.007$ &  $0.184$  & $0.055$ & $0.037$ \\
\hline
$u^{(2)}$ & $0.47$ & $0.58$ & $0.71$ & $0.84$ & $0.25$ & $0.32$ \\
$r^{(2)}$ & $0.0235$ & $0.058$   & $0.0592$ & $0.042$ &  $+\infty$ & $0.032$ \\
\hline
$u^{(3)}$ & $0.63$ & $0.92$ & $0.08$ & $0.59$ & $0.32$ & $0.80$ \\
$r^{(3)}$ & $0.063$  & $0.0613$  & $0.0053$ &  $+\infty$ & $0.0213$ & $0.08$ \\
\hline
\end{tabular}
\\\centerline{(B)}
\end{minipage}
\begin{minipage}{1in}
bottom-$3$ samples:\\
\begin{tabular}{ll}
$w^{(1)}$ & $i_3$, $i_1$, $i_6$ \\
$w^{(2)}$ & $i_1$, $i_6$, $i_4$ \\
$w^{(3)}$ & $i_3$, $i_1$, $i_5$
\end{tabular}

\medskip\medskip\medskip
bottom-$3$ samples:\\
\begin{tabular}{ll}
$w^{(1)}$ & $i_3$, $i_1$, $i_6$ \\
$w^{(2)}$ & $i_1$, $i_6$, $i_4$ \\
$w^{(3)}$ & $i_3$, $i_5$, $i_2$
\end{tabular}
\end{minipage}
}
\caption{(A): Example data set with keys $I=\{i_1,\ldots,i_6\}$ and weight assignments $w^{(1)},w^{(2)},w^{(3)}$ and per-key values for example aggregates. (B): random rank assignments and corresponding bottom-$3$ samples.\label{example1:fig}}
\end{figure*}
\noindent

We are interested in aggregates  of the form $\sum_{i|d(i)=1} f(i)$
where $d$ is a selection predicate and $f$ is a numeric function,
both defined over the set of keys $I$.  $f(i)$ and $d(i)$ may depend
on the attribute values associated with key $i$ and on the weight
vector $w^{(\cW)}(i)$.

We say that
the function $f$/predicate $d$ is  {\em single-assignment}
if it depends on $w^{(b)}(i)$ for a single $b\in \cW$.
Otherwise we say that it is {\em multiple-assignment}.
The {\em  relevant assignments} of $f$ and $d$ are those necessary for
determining all keys $i$ such that $d(i)=1$ and
evaluating $f(i)$ for these keys.

\noindent 
The {\em maximum} and {\em minimum} with respect to
a set of assignments $\cR\subset\cW$,
are defined by $f(i)$ as follows:
\begin{equation}
w^{(\max \cR)}(i) \equiv  \max_{b\in\cR} w^{(b)}(i) \;\;\;\;
w^{(\min \cR)}(i) \equiv   \min_{b\in\cR} w^{(b)}(i) \label{maxdef}\
.
\end{equation}
The relevant assignments for $f$ in this case are $\cR$. Sums over
these $f$'s are also known as the {\em max-dominance} and {\em
min-dominance} norms \cite{CM:esa03,CM:ton05} of the selected
subset. \notinproc{The maximum reduces to the size of set union and
the minimum to the size of set intersection for the special case of
global weights.}

\noindent 
The ratio ${\sum_{i\in J} w^{(\min \cR)}(i)}/{\sum_{i\in J} w^{(\max \cR)}(i)}$
when $|\cR|=2$ is the {\em weighted Jaccard similarity} of the assignments
$\cR$ on $J$.  The {\em range} ($L_1$ difference when $|\cR|=2$)
can be expressed as a sum aggregate by choosing $f(i)$ to be
\begin{equation}\label{L1def}
w^{(L_1 \cR)}(i) \equiv w^{(\max \cR)}(i) - w^{(\min \cR)}(i)\ .
\end{equation}
For the example in Figure~\ref{example1:fig}, the max dominance norm
over even keys (specified by a predicate $d$ that is true for $i_2,i_4,i_6$)
and assignments $\cR=\{1,2,3\}$ is
$w^{(\max \{1,2,3\})}(i_2)+w^{(\max \{1,2,3\})}(i_4)+w^{(\max \{1,2,3\})}(i_6)=15+20+10=45$, the $L_1$ distance between assignments $\cR=\{2,3\}$
over keys $i_1,i_2,i_3$ is $w^{(L_1 \{2,3\})}(i_1)+w^{(L_1 \{2,3\})}(i_2)+w^{(L_1 \{2,3\})}(i_3)=10+5+3=18$.

\ignore{
(\cite{CM:esa03} proposed sketches of data streams that support estimates
on the max-dominance norms.  These sketches do not support subpopulations and also estimates are not unbiased.  For the min they just show impossibility results. \cite{CM:ton05} discusses our same framework of detecting ``Deltoids'' and what is new between times and locations. They only evaluate ``independent'' sampling and show that ofcourse it is bad.  Their approach does not support a posteriori specification of groups.  Also designed for ``cash register'' unaggregated data.)  }


\ignore{
In our discussion of applications, we distinguished between data sources with
(i)~colocated weights, where the full weight vector is contained in
the ``record'' of each key, or (ii)~dispersed weights, where each key
occurs in multiple locations or time periods $b\in \cW$ and only the
weight value $w^{(b)}(i)$ is available at the ``occurrence'' that
corresponds to $b\in \cW$.}

  This classification of dispersed and colocated models differentiates
the summary formats that can
be computed in a scalable way:
With colocated weights, each key is processed
once, and samples for different assignments $b\in \cW$ are generated
together and can be coupled.  Moreover, the (full) weight vector can
be easily incorporated with each key included in the final summary.
With dispersed weights, any scalable summarization algorithm must
decouple the sampling for different $b\in\cW$. The process and result
for $b\in \cW$ can only depend on the values $w^{(b)}(i)$ for $i\in I$.
The final summary is generated from the results of these disjoint processes.

\noindent {\bf Random rank assignments for $(I,\cW)$.} A {\em random
rank assignment} for $(I,\cW)$ associates a rank value $r^{(b)}(i)$
for each $i\in I$ and $b\in \cW$. If $w^{(b)}(i)=0$,
$r^{(b)}(i)=+\infty$. The {\em rank vector} of $i\in I$,
$r^{(\cW)}(i)$, has entries $r^{(b)}(i)$ ordered by $b\in\cW$. The
distribution $\Omega$ is defined with respect to a monotone family
of density functions $\bf{ f}_w$ ($w\geq 0$) and has the following
properties: (i) For all $b$ and $i$ such that $w^{(b)}(i)>0$, the
distribution of $r^{(b)}(i)$ is $\bf{ f}_{w^{(b)}(i)}$. (ii) The
{\em rank vectors} $r^{(\cW)}(i)$ for $i\in I$ are independent.
(iii) For all $i\in I$, the distribution of the rank vector
$r^{(\cW)}(i)$ depends only on the weight vector $w^{(\cW)}(i)$.

  It follows from (i) and (ii) that for each $b\in\cW$,
$\{r^{(b)}(i) | i\in I\}$ is a random rank assignment for the
weighted set $(I,w^{(b)})$ with respect to the family $\bf{ f}_w$
($w\geq 0$).
 The distribution $\Omega$ is specified
by the mapping (iii) from weight vectors to distributions of rank vectors
specifies $\Omega$.

\noindent
{\bf Independent or consistent ranks.}
  If for each key $i$, the entries $r^{(b)}(i)$ ($b\in \cW$) of the
rank vector of $i$ are independent we say that the rank assignment
 has {\em independent ranks}. In this case $\Omega$ is
the product distribution of independent rank assignments $r^{(b)}$ for
$(I,w^{(b)})$ ($b\in\cW$).

A rank assignment has {\em consistent ranks} if for each key $i\in I$
and any two weight assignments $b_1,b_2\in \cW$,
$$w^{(b_1)}(i) \geq w^{(b_2)}(i)\Rightarrow r^{(b_1)}(i) \leq
r^{(b_2)}(i)\ .$$ (in particular, if entries of the weight vector are
equal then corresponding rank values
are equal, that is, $w^{(b_1)}(i) = w^{(b_2)}(i)\Rightarrow
r^{(b_1)}(i) = r^{(b_2)}(i)$.)

\notinproc{
 In the special case of global (or uniform) weights,  consistency
means that the entries of each rank vector are equal and distributed
according to ${\bf f}_{w(i)}$ for all $b\in \cW$ such that
$w^{(b)}(i)>0$. Therefore, the distribution of the rank vectors is
determined uniquely by the family ${\bf f}_w$ ($w>0$). This is not
true for general weights. We explore the following two
distributions of consistent ranks, specified by a mapping of weight
vectors to probability distributions of rank vectors.}

\smallskip
\noindent
{\bf $\bullet$ Shared-seed:}
Independently, for each key $i\in I$:\\
\framebox{
\parbox[t]{4.2in}{
\noindent
$\bullet$  $u(i)\leftarrow U[0,1]$ (where $U[0,1]$ is the uniform distribution on $[0,1]$.)\\
\noindent $\bullet$ For $b\in \cW$, $r^{(b)}(i)\leftarrow {\bf
F}^{-1}_{w^{(b)}(i)}(u(i))$. }}

That is, for $i\in I$, $r^{(b)}(i)$ ($b\in\cW$) are determined using
the same ``placement'' ($u(i)$) in ${\bf F}_{w^{(b)}(i)}$.

Consistency of this construction is an immediate
consequence of
the monotonicity property of ${\bf f}_w$.
\ignore{
That is, for each $i$, the relation $\{(r^{(b)}(i),w^{(b)}(i))\vert b\in\cW\}$
is monotone non increasing.}

 Shared-seed assignment for
\IPPS\ ranks is $r^{(b)}(i)=u(i)/w^{(b)}(i)$
and for \EXP\ ranks, is $r^{(b)}(i)=-\ln (1-u(i))/w^{(b)}(i)$.

\noindent {\bf $\bullet$ Independent-differences} is specific to
\EXP\ ranks. Recall that $\EXP[w]$
denotes the exponential distribution with parameter $w$.
Independently, for each key $i$: \\
\smallskip
\framebox{
\parbox[t]{6.2in}{
\noindent
Let $w^{(b_1)}(i)\leq \cdots \leq w^{(b_h)}(i)$ be the entries of the
weight vector of $i$.\\
$~\quad\bullet$ For $j\in 1\ldots h$, $d_j\leftarrow \EXP[w^{(b_j)}(i)-w^{(b_{j-1})}(i)]$, where $w^{(0)}(i)\equiv 0$ and $d_j$ are
independent.\\
$~\quad\bullet$ For $j\in 1\ldots h$,
$r^{(b_j)}(i)\leftarrow \min_{a=1}^{j} d_j$.
}}
\smallskip

For these ranks consistency is immediate from the construction.
Since the distribution of the minimum of independent exponential
random variables is exponential with parameter that is equal to the
sum of the parameters, we have that for all $b\in\cW$, $i\in I$,
$r^{(b)}(i)$ is exponentially distributed with parameter
$w^{(b)}(i)$.


\smallskip

\noindent
{\bf Coordinated and independent sketches.}
Coordinated sketches
are derived from assignments with
consistent ranks and independent sketches
from assignments with independent ranks.
$k$-mins sketches:
 An ordered set of $k$
rank assignments for $(I,\cW)$
defines a set of $|\cW|$ $k$-mins sketches, one for each
assignment $b\in \cW$.
Bottom-$k$ and Poisson sketches:
   A single rank assignment $r$ on $(I,\cW)$ defines a bottom-$k$ sketch
(and a Poisson $\tau^{(b)}$-sketch)
for each $b\in \cW$,
(using the rank values $\{r^{(b)}(i) | i\in I\}$).
Figure~\ref{example1:fig} shows examples of independent and shared-seed consistent rank assignments for the example data set and the corresponding
bottom-$3$ samples.

Independent differences ranks allow for a
generalization of
the estimator for unweighted Jaccard similarity~\cite{BRODER:sequences97}:
\begin{theorem}
 $k$-mins sketches derived from rank
assignments with independent-differences
consistent ranks have the following property:
For any $b_1,b_2\in \cW$,
the probability that both assignments have the same minimum-rank
key is equal to
the weighted Jaccard similarity of the two weight assignments.
\end{theorem}
\begin{proof}
 Let $\cR=\{b_1,b_2\}$.
 Independent differences rank distribution is equivalent to drawing (independently) two exponentially distributed random variables for each key $i$, $r'(i)$ and $r''(i)$ according to weights $w^{(\min \cR)}(i)$ and $w^{(L_1\ \cR)}(i)$.  If $w^{(b_h)}(i)=\min\{w^{(b_1)}(i),w^{(b_2)}(i)\}$ we set 
$r^{(b_h)}(i)\leftarrow r'(i)$ and otherwise 
$r^{(b_h)}(i)\leftarrow \min\{r'(i),r''(i)\}$.  This follows from 
the property that the minimum 
of independent and exponentially distributed variables 
is exponentially distributed with parameter equal to 
the sum of the parameters.

 To establish the lemma, 
we need to show that
$$\pr[\min_i r^{(b_1)}(i)= \min_i r^{(b_2)}(i)] = 
\pr[\min_i r'(i) < \min_i r''(i)]\ .$$
We again use
the property that the minimum 
of independent and exponentially distributed variables 
is exponentially distributed with the sum of the parameters.
Hence, $\min_i r'(i)$ and $\min_i r''(i)$ are independent and exponentially
distributed with parameters $\sum_i w^{(\min \cR)}(i)$ and 
$\sum_i w^{(L_1\ \cR)}(i)$, respectively.
Therefore, 
$$\pr[\min_i r'(i) < \min_i r''(i)]=
\frac{\sum_i w^{(\min \cR)}(i)}{\sum_i w^{(\min \cR)}(i)+\sum_i w^{(L_1\ \cR)}(i)} = \frac{\sum_i w^{(\min \cR)}(i)}{\sum_i w^{(\max \cR)}(i)}\ .$$
The last equality is correct because
for two independent exponentially distributed random variables with parameters
$y_1,y_2$, the probability that the $i$th is smaller is $y_i/(y_1+y_2)$.
\end{proof}
It follows that the fraction of common keys in the two $k$-mins sketches
is an unbiased estimator of the weighted Jaccard similarity of the two assignments.

 For Poisson sketches, 
shared-seed consistent ranks 
{\em maximize the sharing of keys between sketches}.  In fact,
they are the only joint distribution over assignments that does so.
We conjecture that this holds also for bottom-$k$ and $k$-mins sketches.
\begin{theorem} \label{sharing:lemma}
Consider  all distributions of
rank assignments on $(I,\cW)$ obtained using a family ${\bf F}_w$.
Shared-seed consistent ranks (is the unique rank distribution that) minimize
the expected number of distinct keys in the
union of the  sketches for $(I,w^{(b)})$, $b\in\cW$.
\end{theorem}
\begin{proof}
We first consider Poisson sketches.
 Consider Poisson-$\tau^{(b)}$ sketches ($b\in\cR$).
Since the inclusion of different keys are independent, it suffices to
show the claim for a single key $i$.
Let $p^{(b)}={\bf F}_{w^{(b)}(i)}(\tau^{(b)})$.
With any distribution of rank assignments,
the probability that $i$ is included in at least
one sketch for $b\in \cR$ is at least $\max_{b\in\cR} p^{(b)}$.  
With
shared-seed ranks, this probability  equals
$\max_{b\in\cR} p^{(b)}$, and hence, it is minimized.

  For bottom-$k$ sketches, 
there is no joint distribution with the
property that the inclusion probability of a key in at least one sketch is
equal to its maximum inclusion probability over sketches ($b\in\cR$).
What we can show is that
shared-seed is optimal for a key $i$ {\em when fixing the distribution of
$r(I\setminus\{i\})$}. The argument uses
the independence of the rank vectors of different keys.  
Specifically, for each key $i$, the joint distribution of 
$\tau^{(b)}=r^{(b)}_{k}(I\setminus\{i\})$ ($b\in\cR$)
is independent of the joint distribution of $r^{(b)}(i)$ ($b\in\cR$).
Thus, conditioned on the outcome of 
$r^{(b)}_{k}(I\setminus\{i\})$ ($b\in\cR$), the inclusion probability
of $i$ in the sketch of $b$ is $p^{(b)}={\bf F}_{w^{(b)}(i)}(\tau^{(b)})$.
We can now reuse the argument for Poisson sketches to show that
with shared-seed ranks, $i$ is included in at least one sketch from $\cR$
with probability exactly $\max_{b\in\cR} p^{(b)}$.  Hence, this choice is optimal under this conditioning.

\ignore{  ## partial proof to be checked later
    A sufficient condition for optimality is that for any key $i$ and set
of assignments $\cR$, the sample distribution maximizes the
probability of the event $\forall b\in\cR, i\in s_k(I,r^{(b)})$.
We show that this holds for $k=1$.  A general sample distribution is
determined by the seeds $u^{(b)}(i)$.  These seeds are from $U[0,1]$
and are independent
for different keys but can be dependent for different assignment of
the same key.  For shared-seed ranks $u^{(b)}(i)$ are identical for all $b$.

  We consider $k=1$.  
$r^{(b)}(i)<r^{(b)}(j)$ if and only if
$u^{(b)}(j) \geq f_{j}^{(b)}(u^{(b)}(i))$.  
We refer to this as event
$V^{(b)}_j$.
For any two keys $i,j$, $\pr[\forall b\in \cR, V^{(b)}_j]$ is maximized
for shared-seed ranks.  It is equal to $\min_{b\in \cR} \pr[V^{(b)}_j]$.

Key $i$ is included in all the sketches if and only if this condition holds 
for all $b\in\cR$ and 
all $j\in I\setminus\{i\}$, that is $\forall b\forall j V^{(b)}_j$, which
can be expressed as 
the intersection of the events 
$\forall b\in \cR, V^{(b)}_j$.

The probability is maximized if it is maximized for each $V_j$

(For $k$-mins sketches, since
coordinates are independent, we can
consider each one separately.  The argument for each coordinate
then follows from the proof for bottom-$1$ sketches.)}
\end{proof}

\noindent
{\bf Sketches for the maximum weight.}
For $\cR\subset\cW$, let $r^{(\min \cR)}(i)=\min_{b\in \cR} r^{(b)}(i)$.
 The following
holds for all consistent rank assignments:
\begin{lemma} \label{ranksformax}
Let $r$ be a consistent rank assignment for $(I,\cW)$ with respect
to ${\bf f}_{w}$ ($w>0$). Let $\cR\subset \cW$.
Then $r^{(\min \cR)}(i)$
 is a rank
assignment for the weighted set $(I,w^{(\max \cR)})$ with respect to
${\bf f}_{w}$ ($w>0$).
\end{lemma}
\notinproc{
\begin{proof}
From the definition of consistency, $r^{(\min \cR)}(i)\equiv
r^{(b)}(i)$ where $b=\arg\max_{b\in \cR} w^{(b)}(i)$. Therefore, the
distribution of $r^{(\min \cR)}(i)$ is ${\bf
f}_{w^{(\max \cR)}(i)}$. It remains to show that
$\{r^{(\min \cR)}(i)|i\in I\}$ are independent. This immediately follows
from the definition of a rank assignment: if sets of random
variables are independent (rank vectors of different keys), so are
the respective maxima.
\end{proof}
}

A consequence of Lemma \ref{ranksformax} is the following:
\begin{lemma} \label{maxsketch:coro}
From coordinated Poisson $\tau^{(b)}$-/bottom-$k$-/$k$-mins sketches for $\cR\subset\cW$,
we can obtain a  Poisson $\min_{b\in\cR}\tau^{(b)}$-/bottom-$k$-/$k$-mins
sketch for $(I,w^{(\max \cR)})$.
\end{lemma}
\begin{proof}
{\bf $k$-mins sketches:} we take the coordinate-wise minima (and
respective keys) of the $k$-mins sketch vectors of $(I,w^{(b)})$, $b
\in \cR$.

Given a rank assignment $r$ for $(I,\cW)$  then by Lemma
\ref{ranksformax} $r^{(\min \cR)}(i)$ is a rank assignment for
 $(I,w^{(\max \cR)})$. So by the definition of
 a $k$-mins sketch we should take the key achieving
 $\min_{i\in I} r^{(\min \cR)}(i)$ to the $k$-mins sketch of $(I,w^{(\max \cR)})$, and repeat this for $k$ different rank assignments.

Let $j_b$ be the key such that $r^{(b)}(j_b)$ is minimum among all
$r^{(b)}(i)$. The lemma follows since $$\min_{b\in \cR} r^{(b)}(j_b)
= \min_{b\in \cR} \min_{i\in I} r^{(b)}(i) =
 \min_{i\in I} \min_{b\in \cR} r^{(b)}(i) = \min_{i\in I}
 r^{(\min \cR)}(i) \ .$$

{\bf Poisson $\tau^{(b)}$-sketches}: we include all keys with
rank value at most $\min_{b\in\cR}\tau^{(b)}$ in the union of the sketches.

{\bf Bottom-$k$ sketches}: we take the $k$ distinct keys
with smallest rank values in the union of the sketches.
The proof is deferred and is a consequence of Lemma~\ref{scschar:lemma}.
\end{proof}
 This property of coordinated sketches
generalizes the union-sketch property of coordinated sketches for
global and uniform weights, which
facilitates multiple-set aggregates~\cite{ECohen6f,Broder:CPM00,BRODER:sequences97,bottomk:VLDB2008,CK:sigmetrics09}.

\ignore{ 
\noindent {\bf Independent sketches with known or unknown seeds.}  A
stronger model of independent sketches is such that an underlying
``seed'' vector $u^{(\cR)}(i)$ is used in the randomization and can be
retrieved from the key value $i$.  This model applies when
pseudorandom seeds are obtained using indepdendent hash functions (one
for each assignment) that are applied to the key valie.  We show that
for dispersed weights, we can obtain better estimators when seeds are
known.  Intuitively, because with unknown seeds, we can not bound
$w^{(b)}(i)$ when $i\not\in s_k(I,r^{(b)})$ but with known seeds, we
can obtain precise upper bound on $w^{(b)}(i)$ by considering the
largest weight such that with seed $u^{(b)}(i)$, $i$ would not be
included in $s_k(I,r^{(b)})$.  Independent sketches with known seeds
are a middle ground between independent sketches with unknown seeds
and coordinated sketches with shared seeds.  This intermediate model
helps us demonstrate that
most of the value of coordinated sketches can be attributed to
coordination and is not simply a result of knowing the seed values.
}

\noindent {\bf Fixed number of distinct keys for colocated data.} The
number of distinct keys in a set of bottom-$k$ sketches of different assignments
is between $k$ and $|\cW| k$.  It is closer to $k$ with coordination of sketches and similarity of assignments. 
Therefore, even though $k$, which is the size of each embedded sample,
is fixed the total size varies.
A natural goal under storage constraints 
is to fix the number of distinct keys in the combined sample.

 For a rank assignment $r$, we choose $\ell$ ($\ell\geq k$) to be the
largest such that there are at most $|\cW| k$ distinct keys in the
union of the bottom-$\ell$ samples taken with respect to $r^{(b)}$
($b\in\cW$).  The total number of distinct keys is at least
$|\cW| (k-1)+1$.  Such a sample can be computed by a
simple adaptation of the stream sampling algorithm for the fixed-$k$
variant.
For Poisson sketches, we can similarly fix the {\em expected size} of the combined samples, a stream algorithm can similarly select $\tau^{(b)}$ adaptively so that the expected size of each embedded sample is at least $\ell\geq k$ and
the expected total size is in $[|\cW| (k-1)+1,|\cW| k]$.

In the sequel we develop estimators over bottom-$k$ sketches.
The treatment of Poisson sketches is similar and simpler.
Derivations extend easily to bottom-$k^{(b)}$ sketches (with different sizes for different assignments) and to colocated data sketches
with fixed number of distinct keys.
We shall
denote by $S(r)$ the summary consisting of $|\cW|$ bottom-$k$ sketches
obtained using a rank assignment $r$.

\smallskip
\noindent {\bf Computing coordinated sketches.} Coordinated
bottom-$k$ sketches can be computed by a small modification of
existing bottom-$k$ sampling algorithms. If weights are colocated
the computation is simple (for both shared-seed and
independent-\-differences), as each key is processed once.
  For dispersed weights and shared-seed, random hash functions
 must be used to ensure that the same seed $u(i)$ is used for the key $i$
in different assignments.
 We apply the common practice of
assuming perfect randomness of the rank assignment in the analysis.
This practice is justified by a general
phenomenon~\cite{RusuDobra:sigmod06,MV:soda08},  that simple
heuristic hash functions and
 pseudo-random number generators~\cite{BHRSG:sigmod07}
perform in practice as predicted by this simplified analysis. This
phenomenon  is also supported by our evaluation.

\looseness=-1 Independent-differences are not suited
for dispersed weights as they require range summable
universal hash functions~\cite{FKSV:99,RusuDobra:sigmod06}.


\section{Template Estimator} \label{genest:sec}

Consider $(I,\cW)$ a sample space of rank assignments $\Omega$, and a
summary $S$. 

 We present an unbiased template estimator for sum aggregates
$\sum_{i\in I|d(i)=1}f(i)$
 (where $f$ is a numeric function $f$ $d$ is a predicate $d$).
  The template
estimator
is an adaptation of the \HT\ estimator to our multiple-assignments setting.

 The template estimator assigns adjusted $f$-weights $a^{(f)}(i)\geq 0$ 
to the keys (we have implicit zero adjusted weights
$a^{(f)}(i)\equiv 0$ for $i\not\in S$). 
The estimate on  $\sum_{i|d(i)=1} f(i)$ is the sum 
$\sum_{i\in S \mid d(i)=1} a^{(f)}(i)$ of 
the adjusted $f$-weights of keys in $S$ that satisfy
the predicate $d$.\footnote{ With sum aggregates defined this way, the same adjusted
$f$-weights can be used for different selection predicates
$d()$. 
This is natural when
multiple queries
share the same $f$ but has different attribute-based selections.
For example, the $L_1$ distance of bandwidth (bytes) for IP destination
between two time periods,  for
different subpopulations of flows (applications, destination AS, etc.)
We note that nonetheless, the selection predicate $d(i)$ is technically redundant, as $\sum_{i|d(i)=1} f(i)=\sum_{i\in I} d(i)f(i)$, and we
can replace $f$ and $d$ with  the weight function $d(i)f(i)$ without
a predicate.}

We subsequently apply this template to different summary
types (colocated or dispersed weights), independent or coordinated
distributions of rank assignments, and $f$ and $d$ with different
dependence on the weight vector.

 We start with the template estimator for Poisson sketches.
Independence of key inclusions implies that
there is no gain in considering estimates of $f(i)$ that
depend on keys other than $i$.  We limit our attention
to a single key $i$, with weight
vector $w^{(\cW)}(i)$ and ranks
$r^{\cW}(i)$.
Let $\Omega\equiv \Omega(i)$ denote the sample space of rank
 assignments for $i$, $S(r,w)$ the sample obtained when
 the rank assignment is $r\in\Omega$ and the weight vector of $i$ is $w\equiv
 w^{(\cW)}(i)$, and by $\cS$ the universe of all possible outcomes (in terms of key $i$) over different $r$ and $w$.
 
\noindent
{\bf Template estimator (Poisson)}

\begin{center}
\noindent
\framebox{
\parbox[t]{5in}{
Identify $\cS^* \subset \cS$  and functions
$f(S)>0$ and $0 \leq p(S) \leq 1$ for all $S\in \cS^*$ 
such that the following holds:
\begin{enumerate}
\item
for  any weight vector $w$ such that $d(i)=${\em true}  and $f(i)>0$,
$\pr[\cS^*|w] > 0$.
\item
for each $S\in \cS^*$, 
for all $w$ and $r\in \Omega$ such that $S(r,w)\equiv S$.
\begin{itemize}
\item
$d(i)=\mbox{{\em true}}$, $f(i) \equiv f(S)$
\item
$p(S) \equiv \pr[\cS^*|w] > 0$
\end{itemize}
\end{enumerate}
}}
\end{center}

Estimate:
\noindent
\framebox{
\parbox[t]{3in}{
\begin{itemize}
\item
if $S \not\in \cS^*$,  $a^{(f)}(i) \equiv 0$.  
\item
if $S\in \cS^*$, $a^{(f)}(i)=f(S)/p(S)$.
\end{itemize}
}}

 Where $\pr[\cS^*|w]$ is the probability that the sample $S$ is a member of $\cS^*$ when the weight vector of key $i$ is $w$.  The first requirement is
clearly necessary.  It means that any key that has a positive contribution to the aggregate must have a positive probability of being accounted for.
  It is not hard to see that
the template estimator is unbiased.
 The expectation of $a^{(f)}(i)$ on weights $w$ is 
$\pr[\cS^*|w] * f(S)/p(S) \equiv f(i)$.  This is well defined when the first requirement is satisfied, namely, $\pr[\cS^*|w] > 0$.

 We next apply the \rc\ method to obtain a template estimator for
bottom-$k$ sketches.  The template can be twicked to handle
summaries with sketches with different sizes for each $b\in\cW$ or for 
colocated samples with fixed number of distinct keys.

 For a key $i$  and rank assignment $r^{-i}: I\setminus \{ i\}$, 
we consider  the probability space $\Omega(i,r^{-i})$  
containing all rank assignments $r':I$ that are identical to $r^{-i}$ on
$I\setminus \{ i\}$ (that is, $\forall b\in\cW, \forall
j\in I\setminus\{i\}$, $r'^{(b)}(j)=r^{(b)}(j)$).

  We denote by $S(r,w)$ the sample obtained when
 the rank assignment is $r$ and the weights are $w\equiv
 w^{(\cW)}$ and by $\cS$ the universe of all possible outcomes over 
different $r$ and $w$.  

\noindent
{\bf Template \rc\ estimator (Bottom-$k$ sketches):}

\begin{center}
\noindent
\framebox{
\parbox[t]{5in}{
Identify, for each key $i$, a subset $\cS^*(i)
  \subset \cS$ and functions
$f(S,i)>0$ and $0 \leq p(S,i) \leq 1$ for
$S\in \cS^*(i)$ such that the following holds:
\begin{enumerate}
\item
For all $i$, for all $w$ such that $d(i)=${\em true} and $f(i)>0$, and each $r^{-i}$, $\pr[\cS^*(i)|w,r^{-i}] > 0$.
\item
for all $w$ and $r$ such that $S(r,w)\equiv S$:
\begin{itemize}
\item
$d(i)=\mbox{{\em true}}$, $f(i)\equiv f(S,i)$
\item
$p(S,i) \equiv \pr[\cS^*(i)|w,r^{-i}]$.
\end{itemize}
\end{enumerate}
}}
\end{center}

Estimate:
\noindent
\framebox{
\parbox[t]{3in}{
\begin{itemize}
\item
if $S \not\in \cS^*(i)$,  $a^{(f)}(i) \equiv 0$.  
\item
if $S\in \cS^*(i)$, $a^{(f)}(i)=f(S,i)/p(S,i)$.
\end{itemize}
}}

Where for weights $w$ and ranks $r^{-i}$ for $I\setminus\{i\}$, 
 $\pr[\cS^*(i)|w,r^{-i}]$ is
the probability that a sample is in $\cS^*(i)$ over $\Omega(i,r^{-i})$.

This formulation is an instance of \HTp{} that builds on the Rank
Conditioning (\rc) method~\cite{bottomk:VLDB2008} (see
Section~\ref{prelim:sec}). Correctness, which is equivalent to
saying that for every $i\in I$, $\E[a^{(f)}(i)]=f(i)$, is immediate
from \HTp{}.  The estimate is well defined when $p(S,i)>0$, which 
follows from the first requirement.

 The template is fully specified by the selection of $\cS^*$.
A necessary condition for inclusion of $S$ in $\cS^*$ is that
$d(i)$ and $f(i)$ can be determined from $S$.  Ideally, 
$\cS^*$ would include all such samples, but it is not always possible, as
$\pr[\cS^*|w]$ may not be unique across all
 applicable $w$ that are consistent with outcome $S$.

We show that the more inclusive $\cS^*(i)$ is, the lower the
variance of $a^{(f)}(i)$.

\ignore{
Step (1) of the derivation identifies for each key $i$, a subset $\cS^*(i)$ of $\cS$
such that for each sample $S$ we can determine membership in
$\cS^{*}(i)$, and if a member, we can determine  $f(i)$, $d(i)$, and
$\pr[\cS^{*}(i)]$ in $\Omega(i,r^{-i})$ can be computed for all $w$ and $r$
such that $S(r,w)\equiv S$

$p(i,r^{-i})$ for all $i\in S^{*}(r)$ can be computed from the summary
$S(r)$. To get an unbiased estimator (i.e.\ $\E[a^{(f)}(i)]=f(i)$)
we also require that for any  key $i$ such that  $d(i)>0$ and  $f(i)
>0$, we have that $p(i,r^{-i}) > 0$.

In our tailored derivations, the inclusion event of $i$ in
$S^{*}(r')$ is typically a union or intersection of events of the
form $r'^{(b)}(i)< r_{k+1}^{(b)}(I)$ for some $b\in \cW$. We refer
to $S^{*}(r)$ as {\em the set of applicable samples}. There are
typically multiple ways to select a mapping $S^{*}$ that obeys the
requirements.

 Step (2) computes
positive adjusted $f$-weights $a^{(f)}$.  Keys $i$ where $S\not\in
\cS^{*}(i)$ obtain an implicit zero
adjusted $f$-weight and otherwise if $S\in \cS^*(i)$ they obtain a positive
adjusted $f$-weights. The requirements of Step (1) guarantee that
these adjusted weights can be computed from the summary.  For
bottom-$k$ summaries, our generic

Step (3) outputs our estimate for $\sum_{i\in I|d(i)=1} f(i)$. The
requirements of Step (1) ensure that we can evaluate the predicate
$d(i)$ for all keys $i\in S^{*}(r)$, using information in $S(r)$,
and hence, we can evaluate the estimate.
}

 To obtain tight estimators, we need to apply the template with
the {\em most inclusive suitable selection
$\cS^{*}$}.
\begin{lemma} \label{incl:lemma}
Consider two selections $\cS_1^{*}(i)$ and $\cS_2^{*}(i)$ such that
$\cS_1^{*}(i) \subseteq \cS_2^{*}(i)$.
Let $a_1^{(f)}$ and $a_2^{(f)}$ be the corresponding adjusted weights of $i$
Then for all $w$,
$\var[a_1^{(f)}(i)]\geq \var[a_2^{(f)}(i)]$.
\end{lemma}
\begin{proof}
 Fix $w$ and $r^{-i}$.
Let $p_h(i,r^{-i})$ be the
corresponding probabilities, in $\Omega(i,r^{-i})$,
that $S\in \cS_h^{*}(i)$ ($h=1,2$).

From definition,
for all $i\in I$ and $r\in \Omega$, $p_1(i,r^{-i})\leq p_2(i,r^{-i})$.

 Observe now that
it suffices to establish the relation of the variance 
for a particular $\Omega(i,r^{-i})$
(since the projection of $r$ on $I\setminus\{i\}$ is a partition of $\Omega$ and the adjusted weights are unbiased in each partition.)
We have
$\var_{\Omega(i,r^{-i})}[a_h^{(f)}(i)]=f(i)^2 (1/p_h(i,r^{-i})-1)$ for $h=1,2$
(variance of the \HT\ estimator on $\Omega(i,r^{-i})$).

\end{proof}

 When adjusted weights $a^{(f)}$ of different keys have zero covariances,
 $\var[a^{(f)}(J)] = \sum_{j\in J}\var[a^{(f)}(j)]$. In particular, this means that a most inclusive $\cS^*$ implies at most the variance for any subset $J$.


\section{Colocated weights} \label{coloc:sec}

We apply the template (Section~\ref{genest:sec}) to
summaries of data sets with colocated weights.

 We apply the template with $\cS^*(i) \equiv \{ S\in \cS | i\in S\}$, that is,
all samples where $i$ is included in the union of the single-assignment 
``embedded'' samples.  This is the most inclusive possible selection
and we refer to it as the {\em inclusive} estimator.

  A requirement for existence of
a template estimator is that
for all $f$, $d$, $w$, and $i$, 
$f(i)d(i)>0$ implies that $i$ is sampled with positive probability.
This is the first requirement of the template with the inclusive selection substituted for $\cS^*(i)$.  If the requirement holds for any other selection (must be a proper subset of the inclusive one), it must also hold for the inclusive selection.

 With all considered forms of (single-assignment) 
weighted sampling, a key has a positive 
probability to be sampled if and only if it has a positive weight.
Therefore, equivalent requirement is
\begin{equation}\label{condinclusive}
f(i)d(i)>0 \implies w^{(\max \cW)}(i)>0\ .
\end{equation}

   We show that (\ref{condinclusive}) is also sufficient, that is, the inclusive estimator is defined whenever (\ref{condinclusive}) holds.
  Recall that in the colocated model, once a key is sampled its full weight
vector $w^{(\cW)}(i)$ is available with $S$.  Hence, $f(i)$ and $d(i)$ can
be determined from $S$ for all $i\in S$.

  It remains to show how to
compute $p(S,i)$ for each $i\in S$.
  For Poisson sketches, $i\in S$ if and only if 
for at least one $b\in\cW$, $r^{(b)}(i)\leq \tau^{(b)}$.  For bottom-$k$ sketches,
$i\in S$ if and only if
for at least one $b\in\cW$, $r^{(b)}(i)\leq
 r^{(b)}_{k}(I\setminus\{i\})$.

 We provide explicit expressions for \EXP\ and \IPPS\ rank distributions 
and bottom-$k$ sketches.  The derivation for Poisson is omitted, but the
expressions can be obtained by substituting $\tau^{(b)}$ for $r^{(b)}_{k}(I\setminus\{i\})$.

 The probability that $i$ is included in $S$ over
$\Omega(i,r^{-i})$ is:
\begin{equation} \label{coloc_p}
p(i,r^{-i})=\pr[\exists b\in\cW ,
r'^{(b)}(i)<r^{(b)}_{k}(I\setminus\{i\}) \mid r'\in \Omega(i,r^{-i})]\ .
\end{equation}

To compute~(\ref{coloc_p}), 
the summary should
include, for each $b\in\cW$, the rank values $r_k^{(b)}(I)$  and
$r_{k+1}^{(b)}(I)$ and for each $i\in S(r)$ and $b\in\cW$, whether
$i$ is included in the bottom-$k$ sketch of $b$ (that is, whether
$r^{(b)}(i)<r^{(b)}_{k+1}(I)$). This information allows us to
determine the values $r^{(b)}_k(I\setminus\{i\})$ for all $i\in I$
and $b\in\cW$:  

The values $r^{(b)}_k(I\setminus\{i\})$ can be determined from $S$ as
follows:
if $i$ is included in the sketch for $b$ then
$r^{(b)}_k(I\setminus\{i\})=r^{(b)}_{k+1}(I)$.  Otherwise,
$r^{(b)}_k(I\setminus\{i\})=r^{(b)}_{k}(I)$.
These values are constant over $\Omega(i,r^{-i})$.
 We provide explicit expressions for $p(i,r^{-i})$ (Eq.~(\ref{coloc_p})), for $i\in
S(r)$.

\smallskip
\noindent {\bf Independent ranks} (independent bottom-$k$ sketches):
The probability over $\Omega(i,r^{-i})$ that $i$ is included in the
bottom-$k$ sketch of $b$ is ${\bf
F}_{w^{(b)}(i)}(r^{(b)}_{k}(I\setminus\{i\}))$. It is included in
$S(r')$ if and only if it is included for at least one of $b\in
\cW$. Since $r'^{(b)}(i)$ are independent,
\begin{equation}\label{colocind}
p(i,r^{-i})=1-\prod_{b\in\cW}(1-{\bf F}_{w^{(b)}(i)}(r^{(b)}_{k}(I\setminus\{i\})))\ .
\end{equation}

Specifically,\\
\begin{tabular}{ll}
For \EXP\ ranks:$\quad$ &
$p(i,r^{-i})=1- \exp\left(-\sum_{b\in\cW} w^{(b)}(i)r^{(b)}_{k}(I\setminus\{i\}))\right)$ \ . \\
For \IPPS\ ranks: &
$p(i,r^{-i})=1-\prod_{b\in\cW}(1-\min\{1,w^{(b)}(i)r^{(b)}_{k}(I\setminus\{i\})\})$.
\end{tabular}

\smallskip
\noindent {\bf Shared-seed consistent ranks} (coordinated bottom-$k$ sketches):
Item $i$ is included in the sketch of assignment $b$ for $r'\in
\Omega(i,r^{-i})$ if and only if $u(i)\leq {\bf
F}_{w^{(b)}(i)}(r^{(b)}_{k}(I\setminus\{i\}))$. The probability that
it is included for at least one of $b\in\cW$ is
\begin{equation}\label{colocssc}
p(i,r^{-i})=\max_{b\in\cW}\{{\bf F}_{w^{(b)}(i)}(r^{(b)}_{k}(I\setminus\{i\}))\}\ .
\end{equation}

Specifically,\\
\begin{tabular}{ll}
For \EXP\ ranks:$\quad$ &
$p(i,r^{-i})=1-\exp(-\max_{b\in\cW}\{w^{(b)}(i)r^{(b)}_{k}(I\setminus\{i\})\})$\ .\\
For \IPPS\ ranks: &
$p(i,r^{-i})=\min\left\{1,\max_{b\in\cW}\{w^{(b)}(i)r^{(b)}_{k}(I\setminus\{i\})\}\right\}$\ .
\end{tabular}

\smallskip
\noindent {\bf Independent-differences consistent ranks}
(coordinated bottom-$k$ sketches): Let $w^{(b_1)}(i)\leq \cdots \leq
w^{(b_h)}(i)$ be the entries of the weight vector of $i$. Recall
that $r^{(b_j)}(i)\leftarrow \min_{a=1}^{j} d_j$ where
$d_j\leftarrow \EXP[w^{(b_j)}(i)-w^{(b_{j-1})}(i)]$ (we define
$w^{(0)}(i)\equiv 0$ and $\EXP[0] \equiv 0$).

We also define $M_\ell=\max_{a=\ell}^h
r^{(b_a)}_{k}(I\setminus\{i\})$ ($\ell\in [h]$), and
 the event $A_j$ to consist of all rank assignments such
that $j$ is the smallest index for which $d_j \le M_j$. Clearly the
events $A_j$ are disjoint and $p(i,r^{-i})=\sum_{\ell=1}^h \pr[A_\ell]$.

The probabilities $\pr[A_\ell]$ can be computed using a linear pass
on the sorted weight vector of $i$ using the independence of
$d_\ell$'s as follows  {\scriptsize
\begin{eqnarray*}
\Pr[A_1] & = & \Pr[d_1\leq M_1]={\bf F}_{w^{(b_1)}(i)}(M_1)\ ;\\
\Pr[A_2] & = & \Pr[d_1 > M_1 \wedge d_2\leq M_2]
=  (1-{\bf F}_{w^{(b_1)}(i)}(M_1)){\bf F}_{w^{(b_2)}(i)-w^{(b_1)}(i)}(M_2)\ ;\\
& \ldots & \\
\Pr[A_\ell] & = & \Pr[\bigwedge_{a=1}^{\ell-1} (d_a > M_a) \wedge d_\ell \leq M_\ell] = \prod_{j=1}^{\ell-1} (1-{\bf F}_{w^{(b_{j})}(i)-w^{(b_{j-1})}(i)}(M_{j}))  \cdot \;\; {\bf
F}_{w^{(b_{\ell})}(i)-w^{(b_{\ell-1})}(i)}(M_{\ell})\ .
\end{eqnarray*}
}

\smallskip
\noindent
 {\bf Generic consistent rank assignments (coordinated
sketches)} Let $\cR\subset \cW$ be the set of assignments relevant
for $f$ and $d$. Let
$r_{k}^{(\min \cR)}(I\setminus\{i\})\equiv\min_{b\in\cR}
r_{k}^{(b)}(I\setminus\{i\})$.  We use a more restrictive selection
\begin{equation} \label{gen_consistent:eq}
i\in \cS^{*}(i)\quad \iff \quad  \min_{b\in\cR}r^{(b)}(i)\leq
r_{k}^{(\min \cR)}(I\setminus\{i\})\ .
\end{equation}
For all consistent rank assignments,
we have
$$p(i,r^{-i})={\bf F}_{w^{(\max \cR)}(i)}(r_{k}^{(\min \cR)}(I\setminus\{i\}))\ .$$

It is easy to see that $\cS^{*}$ satisfies the requirements of
the template estimator (Section~\ref{genest:sec}) for any
consistent ranks distribution.
While generic and simpler than the tailored derivations for shared-seed and independent differences, 
this estimator is weaker because $\cS^*$ is less inclusive
(immediate consequence of Lemma~\ref{incl:lemma}).

\ignore{: Consider a key $i$ and $\Omega(i,r^{-i})$.
The general estimators has
a smaller $p(i,r^{-i})$ than the estimator of (Eq.~\ref{coloc_p}).
\begin{eqnarray*}
\pr[\exists b\in\cW , r^{(b)}(i)<r^{(b)}_{k+1}] & \geq & \pr[\exists
b\in\cW , r^{(b)}(i)<r_{k+1}] \\
 & = & {\bf F}_{w^{(\max)}(i)}(r_{k+1}(\cW)).
\end{eqnarray*}
}

\section{Dispersed weights} \label{dispersed:sec}

Let $r$ be a rank assignment for $(I,\cW)$.  
As in the co-located model the summary $S(r)$ contains all keys in the
 bottom-$k$ sketches $s_k(I,r^{(b)})$ for $b\in \cW$.  But in the
dispersed weights model $w^{(b)}(i)$ (for $i\in I,b\in\cW$) is
included in $S(r)$ if and only if $i\in s_k(I,r^{(b)})$.
 Table~\ref{notation:table} summarizes the notation we use in this 
section.

\begin{table}
 \begin{tabular}{l|l}
\hline
{\bf notation} & {\bf explanation} \\
\hline
$w^{(\cR)}(i)$ and $r^{(\cR)}(i)$ & the weight and rank assignments restricted
to $\cR$.\\
\hline
$r^{(\min \cR)}_{h}(J)$=$\min_{b\in\cR} r^{(b)}_{h}(J)$ & the smallest $h$-smallest rank, over $\cR$ \\
$r^{(\ell^{th}\smallest\ \cR)}_{h}(J)$ & The $\ell^{th}$
smallest  value in  $\{r^{(b)}_{h}(J)| b\in\cR\}$\\
$r^{(\min \cR)}(i)=\min_{b\in\cR} r^{(b)}(i)$ & The min rank value of $i$ over $\cR$ \\
$r^{(\max \cR)}(i)=\max_{b\in\cR}r^{(b)}(i)$ & the max rank value of $i$ over $\cR$\\
\hline
$w^{(\max \cR)}(i)=\max_{b\in\cR} w^{(b)}(i)$ & Maximum weight $i$ assumes over $\cR$ \\
$w^{(\min \cR)}(i)=\min_{b\in\cR} w^{(b)}(i)$ & Minimum weight $i$ assumes over $\cR$ \\
$w^{(\ell^{\mbox{{\scriptsize th}}}\mbox{-} \largest\ \cR)}(i)$ &
$\ell$th largest weight $i$ assumes over $\cR$ (in $\{w^{(b)}(i)|b\in\cR\}$) \\
$w^{(\top \ell\ \cR)}(i)$ & Restriction of $w^{(\cR)}(i)$ to the top-$\ell$ assignments of $i$. \\
\hline
$b^{(\max \cR)}(i)=\arg\max_{b\in\cR} w^{(b)}(i)$ & The weight
assignment from $\cR$ which maximizes $i$'s weight. \\
$b^{(\min \cR)}(i)=\arg\min_{b\in\cR} w^{(b)}(i)$ & The weight
assignment from $\cR$ which minimizes $i$'s weight.\\
$b^{(\top \ell\ \cR)}(i)$ & Subset of $\cR$ containing assignments with top-$\ell$ weights for $i$\\
$b^{(\ell^{\mbox{{\scriptsize th}}}\mbox{-} \largest\ \cR)}(i)$ &
assignment $b\in \cR$ with $\ell$th largest weight $w^{(b)}(i)$ \\
\hline
\end{tabular}
 \caption{Notation table. $\cR\subset\cW$ is a subset of assignments, $i\in I$, is a key, $J\subset I$ is a subset of keys, $h\geq 1$ is an integer. When the dependency on $\cR$ is clear from
context, it is omitted.}\label{notation:table}
\end{table}

  An {\em aggregation} is specified by the pair $f,d$.
Assignments $\cR\subset\cW$ are
{\em relevant assignments} for an aggregation
if $f$ and $d$ depend only on $w^{(\cR)}$.
In the dispersed
weights model, samples taken for assignments not in $\cR$ do not
contain any useful information for estimating our aggregate.
We are therefore interested in
identifying a minimal such set $\cR$.

 We next characterize a class of
aggregates which includes the minimum ($f=w^{(\min \cR)}$), maximum ($f=w^{(\max\cR)}$) and quantiles over a set $\cR$ of assignments (for example,
$f(i)$ being the median, in this case the $4^{th}$ largest weight,
of $\{w^{(1)}(i)$, $w^{(2)}(i)$, $\ldots$, $w^{(7)}(i)\}$).

\begin{definition}
We say that an aggregation ($f$,$d$) is
 {\em $\top \ell$ dependent} if
\begin{eqnarray}
f(i) \equiv  f(w^{(\top \ell\ \cR)}(i),b^{(\top \ell\ \cR)}(i))&& \label{cond1}\\
d(i) \equiv  d(w^{(\top \ell\ \cR)}(i),b^{(\top \ell\ \cR)}(i))&& \label{cond2}\\
 w^{(\ell^{th}\mbox{-}\largest\ \cR)}(i)=0  \Rightarrow  d(i)f(i)=0\ .&&\label{cond3}
\end{eqnarray}
\end{definition}

Notice that functions $f$ and $d$ that satisfy (\ref{cond3}) for some
$\ell$ also satisfy (\ref{cond3}) for $\ell' < \ell$ but do not necessarily
satisfy (\ref{cond1}) and (\ref{cond2}) for $\ell' < \ell$.
Of special interest are the two extreme cases:
{\em $\max$-dependence}, when $\ell=1$, and {\em $\min$-dependence}, when $\ell=|\cR|$.
With $\min$-dependence,  (\ref{cond1}) and (\ref{cond2}) are redundant
(always hold)
and (\ref{cond3}) is $w^{(\min \cR)}(i)=0 \Rightarrow f(i)d(i)=0$.

 The aggregations specified by
$f(i)=w^{(\min \cR)}(i)$ and any predicate $d$ are
$\min$-dependent, but not $\top \ell$ dependent for any $\ell\not=|\cR|$.
The aggregations specified by
$f(i) = w^{(\max \cR)}(i)$ and any attribute-based predicate $d$ are $\max$-dependent, but not $\top \ell$ dependent for any $\ell\not=1$.
More generally, aggregations specified by 
$f(i)=w^{(\ell^{\mbox{{\scriptsize th}}}\mbox{-} \largest\ \cR)}(i)$
(the $\ell^{th}$ largest weight)
and attribute-based $d$ are $\top \ell$ dependent but not
$\top h$ dependent for any $h\not=\ell$.

 We use our template estimator to obtain
unbiased nonnegative
estimators for $\top \ell$ dependent aggregations over coordinated sketches($1\leq \ell \leq |\cR|$) and $\min$-dependent
aggregations over independent sketches.

We present two
such estimators which we name {\em s-set} and {\em l-set} and respectively denote $\cS^*_s$ and $\cS_l^*$ the
respective selections in the template estimator.
 We have that $\cS^*_l \supset \cS^*_s$ and moreover, $\cS^*_l$ is
a maximal set for which
we can determine the $\ell$ largest
weights. Hence, from Lemma~\ref{incl:lemma}, 
the $l$-set estimator dominates the $s$-set estimator. 
On the
other hand, s-set
estimators have a simple closed universal expression for all
coordinated sketches whereas we present a closed-form l-set estimators 
only for shared-seed coordinated sketches.

\subsection{$s$-set $\top \ell$ dependence} \label{sset:sec}

\noindent
{\bf $s$-set $\top \ell$ dependence estimator (coordinated sketches):}

\begin{center}
\noindent
\framebox{
\parbox[t]{5in}{
\noindent
$\forall i\in S$,\\
$\quad\bullet$
$\ \cR'(i) \leftarrow \left\{b\in\cR \mid
r^{(b)}(i)<r^{(\min \cR)}_{k}(I\setminus\{i \})\right\}$\\
$\quad\bullet$ {\bf if} $\quad |\cR'(i)| \geq \ell \quad\quad$
$\{*\quad S \in \cS_s^*(i) \quad *\}$\\
\begin{eqnarray*}
 w^{(\top\ell\ \cR)}(i) & \leftarrow & w^{(\top\ell\ \cR'(i))}(i)\\
 b^{(\top\ell\ \cR)}(i) & \leftarrow & b^{(\top\ell\ \cR'(i))}(i) \\
 f(i) & \leftarrow & f(w^{(\top\ell\ \cR)}(i),b^{(\top\ell\ \cR)}(i))\\
 d(i) & \leftarrow & d(w^{(\top\ell\ \cR)}(i),b^{(\top\ell\ \cR)}(i))\\
 p(i,r^{-i}) & \leftarrow & {\bf F}_{w^{(\ell^{th}\largest\ \cR)}(i)}(r^{(\min \cR)}_{k}(I\setminus \{i\}))\\
a^{f}(i) & \leftarrow & \frac{f(i)}{p(i,r^{-i})}
\end{eqnarray*}

\noindent
 {\bf Output}  $\sum_{i | S\in\cS_s^*(i) \wedge d(i)} a^{f}(i)$
}}
\end{center}

\medskip

For $f(i)= w^{(\max \cR)}(i)$  ($\ell=1$),
$$S\in \cS_s^{*}  \iff  \exists b\in\cR \mid
r^{(b)}(i)<r^{(\min \cR)}_{k}(I\setminus\{i\})$$
and we obtain the adjusted weights:
\begin{equation} \label{dispmax}
a^{(\max \cR)}(i)=\frac{w^{(\max \cR)}(i)}{{\bf
F}_{w^{(\max \cR)}(i)}(r^{(\min \cR)}_{k}(I\setminus\{i\}))}\ .
\end{equation}

\medskip
\noindent {\bf Correctness:}
Clearly when rank assignments are consistent or independent,
any $\ell$-dependent $f$ and $d$ 
satisfy the first requirement of the template estimator, namely, for
any $i$ with $f(i)d(i)>0$ and any fixed assignment on ranks to 
$I\setminus \{i\}$, there is positive probability that $S\in \cS_s^*(i)$.
The $\ell$ largest weights of $i$ are all positive and there is a 
positive probability that all respective assignments have
maximum rank at most $r^{(\min \cR)}_{k}(I\setminus\{i\})$.

We now establish that the requirements of the template estimator are 
satisfied and that it is correctly applied.
\begin{lemma} \label{scschar:lemma} Let $r$ be a consistent rank assignment.
\begin{itemize}
\item[{\bf (i)}]
  If $S\in \cS_s^*(i)$,
\begin{eqnarray*}
w^{(\top\ell\ \cR)}(i) & = & w^{(\top\ell\ \cR'(i))}(i)\ , \\
b^{(\top\ell\ \cR)}(i) & = & b^{(\top\ell\ \cR'(i))}(i)\ .
\end{eqnarray*}
\item[{\bf (ii)}] The computation of $p(i,r^{-i})$ is correct.
\end{itemize}
\end{lemma}
\begin{proof}
{\bf (i)}: We have $S\in\cS_s^*(i)$  if and only if
$|\cR'(i)|\geq \ell$.
Therefore, there is some $\ell'\geq \ell$ such that
$b^{(\ell'^{th}\largest\ \cR)}(i) \in \cR'(i)$.
It suffices to show that for all $h\leq \ell'$,
$b^{(h^{th}\largest\ \cR)}(i) \in \cR'(i)$.
From consistency of ranks, $w^{(b_1)}(i)> w^{(b_2)}(i)$ if and only if
$r^{(b_1)}(i)< r^{(b_2)}(i)$.
Therefore, $r^{(b_1)}(i)< r^{(\min \cR)}_{k}(I\setminus\{i\})$
implies $r^{(b_2)}(i)< r^{(\min \cR)}_{k}(I\setminus\{i\})$.
Equivalently, $b_1\in \cR'(i)$ implies
$b_2\in \cR'(i)$.

\noindent
{\bf (ii)}:
 The value $r^{(\min \cR)}_{k}(I\setminus\{i\})$ depends
on the rank values of all keys other than $i$ and is the same for all
assignments $r'\in \Omega(i,r^{-i})$.
It can always be computed from $S(r)$ since $r^{(b)}_{k}(I\setminus\{i\})$
can be determined for all $b\in\cR$ regardless if $i\in s_k(I,r^{(b)})$ or not
($r^{(b)}_{k}(I\setminus\{i\})$ is the $(k+1)$th smallest rank value
if $i\in s_k(I,r^{(b)})$ and is the $k$th smallest rank value otherwise).
By the definition of the template estimator $p(i,r^{-i})$ should be
 the probability, conditioned on $\Omega(i,r^{-i})$,
that for  at least $\ell$ assignments $b\in \cR$ we have
$b\in \cR'(i)$, that is,
$r'^{(b)}(i)<r_{k}^{(\min \cR)}(I\setminus i)$.
Using  part (i), this is equivalent to
the condition that for all
$b\in b^{(\top\ell\ \cR)}(i)$,
$r'^{(b)}(i)<r_{k}^{(\min \cR)}(I\setminus i)$.
From consistency of ranks, this condition holds if and only if
the $\ell$th smallest rank in $r^{(\cR)}(i)$,  associated
 with the
$\ell^{th}$ largest weight,  is less than
$r^{(\min \cR)}_{k}(I\setminus\{i\})$.
This probability is exactly
${\bf F}_{w^{(\ell^{th}\largest\ \cR)}(i)}(r^{(\min \cR)}_{k}(I\setminus \{i\}))$ and the lemma follows.
\end{proof}

\begin{lemma} \label{scsk:coro}
For $\ell=1$ ($\max$-dependence), $|\{i | S\in\cS_s^{*}(i)\}|\geq k-1$ (at least $k-1$ keys obtain nonnegative adjusted weights).
\end{lemma}
\begin{proof}

Let $b$ be such that $r^{(b)}_{k}(I)$ is minimized.
each one of the  $k-1$ smallest-rank keys in
$s_k(I,r^{(b)})$ must have
$S\in \cS_s^{*}(i)$.
\end{proof}

\ignore{
If $f$ has the form $w^{(b_1)}(i)+\max\{w^{(b_2)}(i),w^{(b_3)}(i)\}$,
we can estimate $w^{(b_1)}(i)$ using $s_k(I,r^{b})$
and apply the above for $\max\{w^{(b_2)}(i),w^{(b_3)}(i)\}$, using
relevant sets $\cR=\{b_1,b_2,b_3\}$.

}

\subsubsection{Min-dependence s-set estimator.} \label{sset_mindep:sec}
The s-set estimator has 
a particularly simple formulation when $\ell=|\cR|$ (min-dependence).
The expression for $p_s(i,r^{-i})$ below holds for any rank distribution and
can be computed also for independent ranks.

{\bf Min-dependence s-set estimator:}

\begin{center}
{\small \noindent \framebox{
\parbox[t]{5in}{
\noindent $\bullet\quad$
$S\in \cS_s^{*}(i)  \iff \forall {b\in\cR}, r^{(b)}(i)<r^{(\min \cR)}_{k+1}(I)$ \\
\noindent $\bullet\quad$ if $S\in S_s^{*}(i),
$~\quad$ p_s(i,r^{-i})\leftarrow \pr[\forall b\in\cR ,
r'^{(b)}(i)<r^{(\min \cR)}_{k+1}(I)\mid r'\in\Omega(i,r^{-i})]$ }}}
\end{center}

 We have $S\in \cS_s^{*}(i)$ if $i$ is included in all
$|\cR|$ sketches with rank value that is at most $r^{(\min \cR)}_{k+1}(I)$.
For coordinated sketches,
we have that for all $b\in
\cR$, $r^{(b)}(i)<r^{(\min \cR)}_{k+1}(I)$ if and only if
$r^{(\max \cR)}(i)<r^{(\min \cR)}_{k+1}(I)$.
 Therefore,
\begin{eqnarray*}
p_s(i,r^{-i}) & = & \Pr[r'^{(\max \cR)}(i)<r^{(\min \cR)}_{k+1}(I)\mid  r'\in\Omega(i,r^{-i})]\\
 & = & {\bf F}_{w^{(\min \cR)}(i)}(r^{(\min \cR)}_{k+1}(I))
\end{eqnarray*}

 For independent sketches, for each $b\in
\cR$, the events $r^{(b)}(i)<r^{(\min \cR)}_{k+1}(I)$ are independent.
Therefore,
\begin{equation*}
p_s(i,r^{-i})  =  \prod_{b\in \cR} {\bf F}_{w^{(b)}(i)}(r^{(\min \cR)}_{k+1}(I))
\end{equation*}

For coordinated sketches and
$f(i)=w^{(\min \cR)}(i)$,
\begin{equation}
a_s^{(\min \cR)}(i)=\frac{w^{(\min \cR)}(i)}{{\bf F}_{w^{(\min
\cR)}(i)}(r^{(\min \cR)}_{k+1}(I))}\ ,
\end{equation}
(when $S\in\cS_s^*(i)$, $a_s^{(\min \cR)}(i)=0$ otherwise).

\subsection{$l$-set $\top \ell$ dependence} \label{lset:sec}

The $l$-set estimator which we now present is the tightest template
estimator for $f(i)=w^{(\ell^{th} \largest\ \cR)}(i)$:
$S\in \cS_l^*(i)$ if and only if the sample (with $k-1$) 
contains sufficient information to determine
$w^{(\top\ell\; \cR)}$ and $b^{(\top\ell\; \cR)}$.  The latter is
a necessary condition in the template and therefore $\cS_l^*$ is the most inclusive possible selection (Lemma~\ref{incl:lemma}).

 When $\ell=1$ (max-dependence), the l-set and s-set estimators are 
the same.
 When $\ell=|\cR|$ (min-dependence), $S\in\cS_l^{*}(i)$ 
when $i$ is
included in all $|\cR|$ bottom-$k$  sketches. For $\ell$-dependence,
$S\in \cS_l^{*}(i)$ if the top $\ell$ weights
of $i$ are included in $S$ and we have upper bound on
all other weights that are no more than $w^{(\ell^{th} \largest \cR)}(i)$.
We can obtain such upper bounds when the seeds $u^{(b)}(i)$ are readily
available for all $b\in \cR$ and $i\in S(r)$.
Under these conditions we
can compute $f(i)$ and $d(i)$ when $S\in \cS_l^*(i)$ and we will
show that we can also compute $p(i,r^{-i})$.

Consider sketches where the seeds $u^{(b)}(i)$
are available for all $i\in S(r)$ and $b\in \cR$.
This {\em known seeds} requirement holds for shared-seed coordinated sketches
(since $u^{(b)}(i) \equiv u(i)$ and $u(i)$ is
available for all $i\in S(r)$) and can be explicitly made for independent
sketches.

\medskip
\noindent
{\bf $\top \ell$ dependence $l$-set estimator:}
 
\begin{center}
{\small
\noindent
\framebox{
\parbox[t]{6.2in}{
\noindent $\bullet$ {\bf For } $i\in S$: \\
$~\quad\bullet$ $\cR'(i) \leftarrow \left\{b\in\cR \mid r^{(b)}(i)<r^{(b)}_{k+1}(I)\right\}$\\
$~\quad\quad\bullet$ {\bf If} $|\cR'(i)| \geq \ell$ {\bf and }
$\forall b\in \cR\setminus b^{(\top\ell\ \cR'(i))}(i)$,
$u^{(b)}(i)< {\bf F}_{w^{(\ell^{th} \largest \cR'(i))}(i)}(r^{(b)}_{k}(I\setminus \{i\}))$\\
/* $S\in \cS^*_l(i)$ */
$~\quad\quad\quad$\begin{eqnarray*}
 w^{(\top \ell\; \cR)}(i) \, &  \leftarrow & \, w^{(\top \ell \; \cR'(i))}(i), \\
 b^{(\top \ell \; \cR)}(i) \, & \leftarrow & \, b^{(\top \ell \; \cR'(i))}(i),\\
  f(i) & \leftarrow & f(w^{(\top\ell\ \cR)}(i),b^{(\top\ell\ \cR)}(i))\\
 d(i) & \leftarrow & d(w^{(\top\ell\ \cR)}(i),b^{(\top\ell\ \cR)}(i))\\
 p_\ell(i,r^{-i})\, & \leftarrow & \, \pr[\forall b\in
b^{(\top \ell\ \cR)}(i), r'^{(b)}(i)<r^{(b)}_{k+1}(I)\, \wedge \\ 
&& \,
\forall b\in R\setminus b^{(\top\ell\ \cR)}(i), u^{(b)}(i)< {\bf
F}_{w^{(\ell^{th} \largest\ \cR)}(i)}(r^{(b)}_{k}(I\setminus
\{i\})) \mid r'\in\Omega(i,r^{-i})] \\
a^{f}(i) & \leftarrow & \frac{f(i)}{p(i,r^{-i})}
\end{eqnarray*}
}
} 
} 
\end{center}

For
 shared-seed coordinated sketches
\begin{equation} \label{pp_l:eq}
p_\ell(i,r^{-i})=\min\{\min_{b\in b^{(\top\ell\ \cR)(i)}} {\bf F}_{w^{(b)}(i)}(r^{(b)}_{k+1}(I)),
\min_{b\not\in b^{(\top\ell\ \cR)}} {\bf F}_{w^{(\ell^{th}\largest\ \cR)}(i)}(r^{(b)}_{k+1}(I))\}
\end{equation}

 For independent ranks with known seeds:
\begin{equation} \label{pp_l_ind:eq}
p_\ell(i,r^{-i})=\prod_{b\in  b^{(\top\ell\ \cR)(i)}} {\bf
F}_{w^{(b)}(i)}(r^{(b)}_{k+1}(I)) \prod_{b\not\in  b^{(\top\ell\
\cR)(i)}} {\bf F}_{w^{(\ell^{th}\largest)}(i)}(r^{(b)}_{k+1}(I)) \ .
\end{equation}

\noindent
The {\bf Min-dependence l-set estimator}
 has the simpler form
(The l-set estimator uses the seed values $u_i$ only
when $l < |\cR|$):
\begin{center}
{\small
\noindent
\framebox{
\parbox[t]{4.7in}{
\noindent
$\bullet\quad$
$S\in \cS_l^{*}(i) ~\quad\ \iff  ~\quad\  \forall b\in\cR,  r^{(b)}(i)<r^{(b)}_{k+1}(I)$ \\
$\bullet\quad$ if $S\in \cS_l^{*}(i),\quad\ p_\ell(i,r^{-i})\leftarrow
\pr[\forall b\in\cR , r'^{(b)}(i)<r^{(b)}_{k+1}(I)\mid
r'\in\Omega(i,r^{-i})]$ }}}
\end{center}

$p_\ell(i,r^{-i})$  for shared-seed consistent ranks is:
\begin{equation} \label{p_l:eq}
p_\ell(i,r^{-i})=\min_{b\in \cR} {\bf F}_{w^{(b)}(i)}(r^{(b)}_{k+1}(I)) \
.
\end{equation}
For \EXP\ ranks,
$$p_\ell(i,r^{-i})=1-\exp(-\min_{b\in\cR} w^{(b)}(i)r^{(b)}_{k+1}(I))$$ and
for \IPPS\ ranks,
$p_\ell(i,r^{-i})=\min\{1,\min_{b\in\cR}\{w^{(b)}(i)r^{(b)}_{k+1}(I)
\}\}$.
\ignore{ For independent-differences consistent ranks,
    $p_\ell(i,r^{-i})$ is expressed as a
simultaneous bound on all prefix-sums of a set of independent
exponentially-distributed random variables.  We omit the details.}
  
 For independent sketches: 
\begin{equation} \label{p_l_ind:eq}
p_\ell(i,r^{-i})=\prod_{b\in \cR} {\bf F}_{w^{(b)}(i)}(r^{(b)}_{k+1}(I))\ .
\end{equation}

 By contrasting (\ref{pp_l:eq}) and  (\ref{pp_l_ind:eq})  or (\ref{p_l:eq}) and  (\ref{p_l_ind:eq})   we
can see that the respective  inclusion probability can be
exponentially smaller (in $|\cR|$) for independent sketches than
with coordinated sketches.  Since the variance $\var[a(i)]$ is
proportional to $(\frac{1}{p_\ell(i,r^{-i})}-1)$, we can have
exponentially larger variance.

 \noindent {\bf Correctness:}
 It is easy to see that the first requirement of the template estimator
(Section~\ref{genest:sec})
is satisfied: for both consistent and independent ranks, for
$\ell$-dependent $f$ and $d$, and $i$ such that 
$f(i)d(i)>0$ has $\pr[S\in \cS_l^*] > 0$.
Furthermore, when $S\in\cS_l(i)$, 
the top-$\ell$ weights $w^{(\top\ell\ \cR)}(i)$ are available from $S(r)$ and
therefore $f$ and $d$ can be evaluated.
Lastly, it is easy to see that the computation of the inclusion probability
is correct and that it depends only on the projection of $r$ on
$I\setminus\{i\}$.  The value $r^{(b)}_k(I\setminus\{i\})$ can always be determined from $S$: it is equal to $r^{(b)}_{k+1}(I)$ when $i$ is in the sketch of $b$ and to $r^{(b)}_{k}(I)$ otherwise.

Let $a_l^{(\min \cR)}(i)$ be the adjusted weight for
$f(i)=w^{(\min \cR)}(i)$ of the
 l-set estimator
using shared-seed consistent ranks, and let
$a_{ind}^{(\min \cR)}(i)$ be the adjusted weight for
$f(i)=w^{(\min \cR)}(i)$ of the  l-set estimator
using independent ranks.

\ignore{
  Coordination is a critical ingredient in the derivation of
these bounds.  For independent sketches the bounds are very
weak.  To
qualitatively understand this, note
that if there are $|\cW|$ weight functions,
variance increases like a power in $|\cW|$ even if weight assignments
are identical.  If inclusion probabilities
are $p^{(s)}(i)$ then we need to use joint one
$\prod_s p^{(s)}(i)$.  so the variance is $(1/\prod_s p^{(s)}(i)-1)w^{\max}(i)$
instead of $(1/p^{(s)}(i)-1)w^{(s)}(i)$.
unless $p^{(s)}(i)$ are close to $1$.
}

\subsection{The Range ($L_1$ difference).}

  The template estimator is not applicable to range estimates.  This is because there are weight vectors (say $w(i)=(1,0)$) where the range is positive
($w^{(L_1)}=1$ in this case) and there is $0$ probability to determine 
it from the sample $S$ (intuitively, because we can never be ``sure'' that the second value is $0$).

 Fortunately, we can use the relation
 $w^{(L_1 \cR)}(i)=w^{(\max \cR)}(i)-w^{(\min \cR)}(i)$ with the
$\ell$-dependence estimators and get the estimator
\begin{equation}
a^{(L_1 \cR)}(i)  =  a^{(\max \cR)}(i)-a^{(\min \cR)}(i)
\end{equation}

We use $a^{(\max \cR)}(i)$ for the max-dependence l-set (same as s-set) 
estimator and $a^{(\min \cR)}(i)$ when the statement applies to
both the s-set and l-set min-dependence 
estimators.  When specifics is warrented, we use 
$a_s^{(L_1 \cR)}(i) = a^{(\max \cR)}(i)-a_s^{(\min \cR)}(i)$ and
$a_l^{(L_1 \cR)}(i)  = 
a^{(\max \cR)}(i)-a_l^{(\min \cR)}(i)$.

  $a^{(L_1 \cR)}(i)$ is obviously an unbiased estimate of
$w^{(L_1 \cR)}$, because it is the difference of two
unbiased estimators.  We show that for
a consistent $r$,  it is also nonnegative.


We use the notation $p^{(\max \cR)}(i,r^{-i})$, $p_s^{(\min \cR)}(i,r^{-i})$,
and $p_l^{(\min \cR)}(i,r^{-i})$ for the respective inclusion
probabilities (and $p^{(\min \cR)}$ when the statement applies to both 
the respective s-set and l-set estimators).

 We first establish the following lemma.
\begin{lemma} \label{maxminprobratio}
For consistent $r$ with \IPPS\ or \EXP\ ranks, $\cR$, $k$, and $i\in I$.
$$
\frac{p^{(\max \cR)}(i,r^{-i})}{p^{(\min \cR)}(i,r^{-i})}\le
\frac{w^{(\max \cR)}(i)}{w^{(\min \cR)}(i)} \ .$$
 \end{lemma}
\begin{proof}
Since, $p_s^{(\min \cR)}(i,r^{-i}) \leq p_\ell^{(\min \cR)}(i,r^{-i})$, it
suffices to establish the inequality for $p_s^{(\min \cR)}(i,r^{-i})$.

%
%
%

 For \IPPS\ ranks it suffices to show that for any $\tau$
$$\frac{\min\{1,\tau w^{(\max \cR)}(i)\}}{\min\{1,\tau w^{(\min \cR)}(i)\}}\leq \frac{w^{(\max \cR)}(i)}{w^{(\min \cR)}(i)}\ .$$
This is clear in case the numerator of the left hand side is $\tau
w^{(\max \cR)}(i)$ and the denominator is $\tau w^{(\min \cR)}(i)$.
Otherwise, since $\tau w^{(\max \cR)}(i) \ge \tau w^{(\min \cR)}(i)$
the numerator is $1 < \tau w^{(\max \cR)}(i)$ so the inequality must
also hold.

 For \EXP\ ranks, we need to show that for any
 $\tau$,
$$\frac{1-\exp(-\tau w^{(\max \cR)}(i))}{1-\exp(-\tau w^{(\min \cR)}(i))}\leq \frac{w^{(\max \cR)}(i)}{w^{(\min \cR)}(i)}\ .$$
Taking $r =w^{(\max \cR)}(i)/w^{(\min \cR)}(i)$ and $\gamma =
\exp(-\tau w^{(\min \cR)}(i))$, this  follows since for any $r\geq
1$ and $0<\gamma<1$, $\frac{1-\gamma^r}{1-\gamma}\leq r$.
\end{proof}

\begin{lemma}
For consistent $r$ with \IPPS\ or \EXP\ ranks,
$\forall i\in I$, $a^{(L_1 \cR)}(i)\geq 0$.
\end{lemma}
\begin{proof}
It suffices to show that $a^{(\min \cR)}(i) \leq a^{(\max \cR)}(i)$.
We first observe that $a^{(\min \cR)}(i)>0$ implies
$a^{(\max \cR)}(i)>0$.  If $a^{(\max \cR)}(i)>0$ and
$a^{(\min \cR)}(i)=0$ we are done. Otherwise, the claim follows
using Lemma~\ref{maxminprobratio}.
\end{proof}

\ignore{  
\subsection{Variance of s-set and l-set estimators}
The l-set estimators
have at most the variance of the s-set estimators:
\begin{lemma} \label{sversusl}
For any weight function $f$ and $i\in I$,
$$\var[a_l^{(f)}(i)]\leq \var[a_s^{(f)}(i)]$$
\end{lemma}
\begin{proof}
Since $S_s^{*}(r)\subset S_l^{*}(r)$, it follows from
Lemma~\ref{incl:lemma} that the l-set estimator has at most the
variance  of the s-set estimator (since $p_s(i,r^{-i}) \leq
p_\ell(i,r^{-i})$).
\end{proof}
}

\section{Variance properties} \label{varbounds:sec}

\subsection{Covariances}

  We conjecture that the \rc\ estimators we presented have zero covariances.
\ignore{ That is, for all $i\not=j \in I$, $\E[a^{(f)}(i)a^{(f)}(j)]=
f(i)f(j)$.}
This conjecture is consistent with empirical observations and with
properties of related \rc\ estimators~\cite{bottomk:VLDB2008,CK:sigmetrics09}.
With zero covariances, the variance $\var[a^{(f)}(J)]$ is
the sum over $i\in J$ of the per-key variances $\var[a^{(f)}(i)]$.
Hence, if two adjusted-weights
estimators $a_1$ and $a_2$ have $\var[a_1(i)]\geq \var[a_2(i)]$ for all $i\in I$, then the relations holds for all $J\subset I$.

\begin{conjecture} \label{cov:thm}
All our estimators for colocated or dispersed summaries have zero
covariances:  For all $i\not=j \in I$, $\E[a^{(f)}(i)a^{(f)}(j)]=
f(i)f(j)$.
\end{conjecture}

\subsection{Variance bounds}

  We use the notation $t^{(f)}_k(i)$ for the \rc\ $f$-adjusted weights
assigned by  an \rc\ estimators applied to a bottom-$k$ sketch of
$(I,f)$. We also write $t^{(w^{(b)})}_k(i)$ as $t^{(b)}_k(i)$ for short.

  We measure the variance of an adjusted weight assignment $a$ using
$\sigv[a]=\sum_{i\in I}\var[a(i)]$.  To establish variance relation
between two estimators, it suffices to establish it for
each key $i$.
 Furthermore, if the
estimators are defined with respect to the same distribution of rank assignments
then it suffices to establish variance relation with respect
to some $\Omega(i,r^{-i})$.
(Since these subspaces partition $\Omega$ and our estimators are
unbiased on each subspace).

 The variance of adjusted $f$-weights $a^{(f)}(i)$ for $i\in I$ are
\begin{equation} \label{vari:eq}
\var_{\Omega(i,r^{-i})}[a^{(f)}(i)]=f(i)^2\left(\frac{1}{p(i,r^{-i})}-1 \right)\ .
\end{equation}

\smallskip \noindent {\bf Colocated single-assignment estimators.}
 We show that our single-assignment
inclusive estimators for co-located summaries (independent or
coordinated) dominate plain RC estimators based on a single bottom-$k$ sketch.
\begin{lemma}
 For $b\in\cW$ and $i\in I$, let $a^{(b)}(i)$ be the
 adjusted weights for co-located summaries
computed by our estimator (using $S^{*}(r)\equiv S(r)$ and inclusion probabilities
(\ref{coloc_p})).
Then, $\var[a^{(b)}(i)]\leq \var[t^{(b)}(i)]$.
\end{lemma}
\begin{proof}
Consider  applying the generic estimator with
$S^{*}(r)$ containing all keys $i$ with $r^{(b)}(i)<r^{(b)}_{k+1}(I)$.
This estimator assigns to $i$ an adjusted weight of $0$ if
$r^{(b)}(i)>r^{(b)}_{k+1}(I)$ and an adjusted weight of
$w^{(b)}(i)/{\bf F}_{w^{(b)}(i)}(r^{(b)}_{k+1}(I))$ otherwise.
This is the same adjusted weights as assigned by
the \rc\ bottom-$k$ estimator if we apply it using the rank assignment
to $(I,w^{(b)})$ obtained by restricting $r$ (that is using $r^{(b)}$).
The lemma now follows from Lemma~\ref{incl:lemma}.

A direct proof:  It suffices to establish the variance
relation for a particular subspace $\Omega(i,r^{-i})$ considering the restriction of $r^{(b)}$ of
$r$ as the rank assignment for $(I,w^{(b)})$.
In $\Omega(i,r^{-i})$, the variance of $t^{(b)}_k(i)$  is
$$\var_{\Omega(i,r^{-i})}[t^{(b)}_k(i)]=w^{(b)}(i)^2(1/{\bf F}_{w^{(b)}(i)}(r^{(b)}_{k+1}(I)) -1)\ .$$
From (\ref{coloc_p}), $p(i,r^{-i})\geq {\bf F}_{w^{(b)}(i)}(r^{(b)}_{k+1}(I))$.
Therefore,
$$\var_{\Omega(i,r^{-i})}[a^{(b)}(i)]=w^{(b)}(i)^2(1/p(i,r^{-i}) -1)\leq \var_{\Omega(i,r^{-i})}[t^{(b)}_k(i)]\ .$$
\end{proof}

\noindent
{\bf Approximation quality of multiple-assignment estimators.}
 The quality of the estimate depends on the relation between $f$
and the weight assignment(s) with respect to which the weighted sampling
is performed. We refer to these assignments as {\em primary}.
Variance is minimized when $f(i)$ are the primary weights but
often $f$ must be {\em secondary}:
$f$ may not be known at the time of sampling,
the number of different functions $f$ that are of interest can
be large -- to estimate all pairwise similarities we need ${|\cW|
\choose 2}$ different ``weight-assignments''.
 For dispersed weights, even if known apriori,
weighted samples with respect to some multiple-assignment $f$
cannot, generally, be computed in a scalable way.
 We bound the variance of our $\min$, $\max$, and $L_1$ estimators.

\smallskip \noindent {\bf Colocated $\min$, $\max$, and $L_1$ estimators.}
 We bound the variance of inclusive
estimators for $\min$, $\max$, and $L_1$ using the variance
of inclusive estimators for the respective primary weight
assignments.
\begin{lemma} \label{colocmultivar}
For $f\in\{\max \cR, \min \cR, L_1 \cR\}$, let
$a^{(f)}(i)$ be
the adjusted $w^{(f)}$-weights for co-located summaries computed by our estimator (using $S^{*}(r)\equiv S(r)$ and inclusion probabilities
(\ref{coloc_p})).
\begin{eqnarray*}
\var[a^{(\min \cR)}(i)] & = &  \min_{b\in\cR}\var[a^{(b)}(i)]\ , \\
\var[a^{(\max \cR)}(i)] & = &  \max_{b\in\cR}\var[a^{(b)}(i)]\ , \\
\var[a^{(L_1 \cR)}(i)] & \leq & \var[a^{(\max \cR)}(i)]\ .
\end{eqnarray*}
\end{lemma}
\notinproc{
\begin{proof}
 It suffices to establish this relation in a subspace $\Omega(i,r^{-i})$.
All inclusive estimators share the same inclusion probabilities
$p(i,r^{-i})$ and the variance is as in Equation (\ref{vari:eq}).  The
proof is immediate from the definitions, substituting
$w^{(\min \cR)}(i)=\min_{b\in\cR} w^{(b)}(i)$,
$w^{(\max \cR)}(i)=\max_{b\in\cR} w^{(b)(i)}$,  and $w^{(L_1
\cR)}(i)=w^{(\max \cR)}(i)-w^{(\min \cR)}(i)$.
\end{proof}
} 
The following relations are an immediate corollary of Lemma~\ref{colocmultivar}:

$$
\SV[a^{(\min \cR)}]  \leq  \min_{b\in\cR}\SV[a^{(b)}]\ , \;\;
\SV[a^{(\max \cR)}]  \leq  \max_{b\in\cR}\SV[a^{(b)}]\ , $$
$$
\SV[a^{(L_1 \cR)}]  \leq  \SV[a^{(\max \cR)}]\leq
\max_{b\in\cR}\SV[a^{(b)}]\ .
$$

\smallskip \noindent {\bf Relative variance bound for $\max$: }
For both the dispersed and the colocated models,
we show that the
variance of the $\max$ estimator is at most that
of an estimator applied to a weighted sample taken with
$\max$ being the primary weight.
More precisely,
$a^{(\max \cR)}(i)$ has at most the
variance of an \rc\ estimator applied to a bottom-$k$ sketch of
$(I,w^{(\max \cR)})$ (obtained with respect to the same ${\bf f}_w$
($w>0$)). Hence, the relative variance bounds of single-assignment
bottom-$k$ sketch estimators are
applicable~\cite{bottomk07:ds,bottomk:VLDB2008,DLT:jacm07}.

\begin{lemma}
Let $t_k^{(\max \cR)}(i)$ be the adjusted weights of the \rc\
estimator applied to a bottom-$k$ sketch of $(I,w^{(\max \cR)})$.
For any $i\in I$, $\var[a^{(\max \cR)}(i)]\leq
\var[t_k^{(\max \cR)}(i)]$.
\end{lemma}
\notinproc{
\begin{proof}
By Lemma~\ref{ranksformax} for consistent ranks
$r^{(\min \cR)}=\min_{b\in \cR} r^{(b)}(i)$ is a valid rank
assignment for $(I,w^{(\max \cR)})$ (using the same rank
distributions). So it follows that
 \rc\ adjusted weights with respect to
$w^{(\max \cR)}$ can be stated as a redundant application of the template
estimator where $S\in \cS_1^*(i)$ when $i$ is one of the $k$ least
ranked keys with respect to $r^{(\min \cR)}$.

From Lemma~\ref{scsk:coro}, the selection $\cS^{*}$ is such that
$S\in \cS^*(i)$ for the $k-1$ least ranked keys (or more)
with respect to
$r^{(\min \cR)}$. Hence, $\cS^*_1\subset \cS^{*}$.  Applying
Lemma~\ref{incl:lemma}, we obtain that for all $i\in I$,
$\var[a^{(\max \cR)}(i)]\leq \var[t_k^{(\max \cR)}(i)]$.
\end{proof}
}

\ignore{
{\bf  tailored proof}
 Since $\ell>k$,
the inclusion probability using
the $k$th smallest rank
is at least that when using the $\ell$th smallest rank:
$$p_k(i)={\bf F}_{w^{\max}(i)}(r_k(I\setminus\{i\}))\leq
{\bf F}_{w^{\max}(i)}(r_\ell(I\setminus\{i\}))=p_{\ell}(i)$$
The variance conditioned on this subspace is
$$\var[a^{\max}(i)]=(w^{(\max)})^2(1/p_{\ell}-1)\leq
\var[a(i)]=(w^{(\max)})^2(1/p_{k}-1)\ .$$
}

\noindent
{\bf Dispersed model $\min$ and $L_1$ estimators.}
We bound the absolute variance of  our $w^{(\min \cR)}$ estimator
in terms of the variance of $w^{(b)}$-estimators for $b\in \cR$.
Let $t_k^{(b)}$ be \rc\ adjusted $w^{(b)}$-weights using the
bottom-$k$ sketch with ranks $r^{(b)}$.
\begin{lemma} \label{disvarmin}
For shared-seed consistent $r$, for all $i\in I$,
\begin{equation*}
\var[a_\ell^{(\min \cR)}(i)] \leq \max_{b\in \cR}\var[t^{(b)}(i)]
\end{equation*}
\end{lemma}
\notinproc{
\begin{proof}
Fixing $i$ and $\Omega(i,r^{-i})$, for shared seed
$p^{(\min \cR)}_\ell(i,r^{-i})=\min_{b\in \cR} {\bf
F}_{w^{(b)}(i)}(r^{(b)}_{k+1}(I))$.

Let $b'$ be such that $p^{(\min \cR)}_\ell(i,r^{-i})={\bf
F}_{w^{(b')}(i)}(r^{(b')}_{k+1}(I))$. We have that
{\tiny
\begin{equation*}
 w^{(\min \cR)}(i)^2\left(\frac{1}{p_\ell^{(\min \cR)}(i,r^{-i})}-1 \right) \leq  w^{(b')}(i)^2 \left( \frac{1}{{\bf F}_{w^{(b')}(i)}(r^{(b')}_{k+1}(I))}-1 \right)\ .
\end{equation*}
}
 From this follows that
{\tiny
\begin{equation*}
 w^{(\min \cR)}(i)^2\left(\frac{1}{p_\ell^{(\min \cR)}(i,r^{-i})}-1 \right) \leq \max_{b\in \cR} w^{(b)}(i)^2\left( \frac{1}{{\bf F}_{w^{(b)}(i)}(r^{(b)}_{k+1}(I))}-1 \right)\ .
\end{equation*}
}
 which is equivalent to the statement of the lemma.
\end{proof}
}

It follows from this lemma that there exists a $b\in \cR$ such that
$\sigv[a^{(\min \cR)}]\leq \sum_{b\in\cR} \sigv[a^{(b)}]$.

\begin{lemma} \label{disvarL1}
For consistent $r$, for all $i\in I$,
\begin{equation*}
\var[a^{(L_1 \cR)}(i)] \leq  \var[a^{(\min \cR)}(i)] + \var[a^{(\max \cR)}(i)] \end{equation*}
\end{lemma}
\notinproc{
\begin{proof}
Fixing $i$ and $\Omega(i,r^{-i})$,
With probability $p^{(\min \cR)}(i,r^{-i})$,\\
$a^{(L_1\cR)}=w^{(\max \cR)}(i)/p^{(\max \cR)}(i,r^{-i})-
w^{(\min \cR)}(i)/p^{(\min \cR)}(i,r^{-i})$. With probability
$p^{(\max \cR)}(i,r^{-i})-p^{(\min \cR)}(i,r^{-i})$,\\
$a^{(L_1\cR)}=w^{(\max \cR)}(i)/p^{(\max \cR)}(i,r^{-i})$.

 We have $\var[a^{(L_1\cR)}(i)]=\E[a^{(L_1\cR)}(i)^2]-(w^{(\max \cR)}(i)-w^{(\min \cR)}(i))^2$.
Substituting in the above we obtain
{\scriptsize
\begin{eqnarray*}
\lefteqn{\var[a^{(L_1\cR)}(i)]=}\\ &  & w^{(\max \cR)}(i)^2(\frac{1}{p^{(\max \cR)}(i,r^{-i})}-1)+
w^{(\min \cR)}(i)^2(\frac{1}{p^{(\min \cR)}(i,r^{-i})}-1)\\
&& -2 w^{(\max \cR)}(i) w^{(\min \cR)}(i)(\frac{1}{p^{(\max \cR)}(i,r^{-i})}-1)\\
& = & \var[a^{(\max \cR)}(i)] + \var[a^{(\min \cR)}(i)]\\
&& -2 w^{(\max \cR)}(i)
w^{(\min \cR)}(i)(\frac{1}{p^{(\max \cR)}(i,r^{-i})}-1)
\end{eqnarray*}
}
\end{proof}
} In particular we get $\sigv[a^{(L_1
\cR)}]\leq \sigv[a^{(\min \cR)}] + \sigv[a^{(\max \cR)}]$.

\ignore{
We consider
$a^{\max(\cR)}()$ ($w^{\max(\cR)}$-adjusted weights),
$a^{\min(\cR)\mbox{-s}}$ and $a^{\min(\cR)\mbox{-l}}$
($w^{\min(\cR)}$-adjusted weights based on $S^{*\mbox{-s}}$ and
$S^{*\mbox{-l}}$),
$a^{L_1(\cR)\mbox{-s}}(i)=a^{\max(\cR)}(i)-a^{\min(\cR)\mbox{-s}}(i)$ and
$a^{L_1(\cR)\mbox{-l}}(i)=a^{\max}(i)-a^{\min(\cR)\mbox{-l}}(i)$ ($w^{L_1(\cR)}$-adjusted weights).

 We can also show using Lemma~\ref{maxminprobratio} that
 $\var[a^{(\min \cR)}(i)]\leq (w^{(\min \cR) (i)}/w^{(\max \cR)}(i))\var[a^{(\max \cR)}(i)]+w^{(\min \cR)}(i)(w^{(\max \cR)}(i)-w^{(\min \cR)}(i))$.

\begin{lemma}
For each $i\in I$,
\begin{eqnarray*}
\var[a^{(\max \cR)}(i)] & \leq & \max_b\var[a^{(b)}(i)] \\
\var[a_\ell^{(\min \cR)}(i)] & \leq & \var[a_s^{(\min \cR)}(i)]  \leq  \max_b\var[a^{(b)}(i)] \\
\var[a_\ell^{L_1\cR}(i)] & \leq & \var[a_s^{L_1\cR}(i)] \leq
\var[a^{(\max \cR)}(i)]
\end{eqnarray*}
\end{lemma}
\begin{proof}
 ************  to complete/clean up **********
*** ??  ****  check again first two lines for correct versions.

\end{proof}
}

\ignore{
{\bf ***** Conjecture that justifies use of $\SV$ *****}
\begin{lemma}
For the estimators we considered, for all $J\subset I$,
$$\sum_{i\in J} \var[a_{k-1}(i)]\leq \var[\sum_{i\in J}a_{k}(i)] \leq
\sum_{i\in J} \var[a_{k}(i)]$$
\end{lemma}

}

\section{Evaluation} \label{eval:sec}

  We evaluate the performance of our estimators on summaries, of
independent and coordinated sketches,  produced for
the colocated and the dispersed data models.

\subsection{Datasets}

\noindent
  $\bullet$ {\bf IP dataset1:}
  A set of about $9.2\times 10^6$ IP packets
from a gateway router.  For each packet we have
source and destination
IP addresses (srcIP and destIP), source and destination ports (srcPort and destPort),  protocol,  and total bytes.

\smallskip
\noindent $\diamond$ {\em Colocated data:} Packets were aggregated
by each of the following keys.

\smallskip
\noindent {\bf keys:} 4tuples (srcIP, destIP, srcPort and destPort)
($1.09\times 10^6$ distinct keys). \\
{\bf Weight assignments:} number of bytes ($4.25\times 10^9$ total),
number of packets ($9.2\times 10^6$ total), and uniform (weight of 1
for each key).

\smallskip
\noindent {\bf keys:} destIP ($3.76\times 10^4$ unique destinations)
{\bf Weight assignments:} number of bytes ($4.25\times 10^9$ total),
number of packets ($9.2\times 10^6$ total), number of distinct
4-tuples  ($1.09\times 10^6$ total), and uniform (weight of 1 for
each key).

\smallskip
\noindent $\diamond$ {\em Dispersed data:}
We partitioned the packet
stream into two consecutive sets with the same number of packets
($4.6\times 10^6$) in each.  We refer to the first set as {\em
period1} and to the second set as {\em period2}. For each
period packets were aggregated by keys.
As keys we used the destIP or a pair consisting of both the srcIP and the
destIP. We considered three attributes for each key, namely,
total number of bytes, number of packets, or the number of distinct 4tuples with
that key. For each attribute we got two weight assignments
$w^{(1)}$ and $w^{(2)}$
one for each period.
(See
Table~\ref{data:properties}).

\begin{table*}[htbp]
\centerline{ {\scriptsize
\begin{tabular}{|c|ccccc|}
\hline
key, weight & $\sum_i w^{(1)}(i)$ & $\sum_i w^{(2)}(i)$ & $\sum_i w^{(\max \{1,2\})}(i)$ & $\sum_i w^{(\min \{1,2\})}(i)$ & $\sum_i w^{(L_1 \{1,2\})}(i)$ \\
\hline
destIP, 4tuple & $5.42\times 10^5$ & $5.54\times 10^5$ & $7.47\times 10^5$ & $3.49\times 10^5$ & $3.98\times 10^5$ \\
destIP, bytes & $2.08\times 10^9$ & $2.17\times 10^9$ & $3.26\times 10^9$ & $9.96\times 10^8$ & $2.26\times 10^9$ \\
srcIP$+$destIP, packets & $4.61\times 10^6$ &  $4.61\times 10^6$  &  $7.61\times 10^6$ &  $1.61\times 10^6$ & $6.00\times 10^6$ \\
srcIP$+$destIP, bytes & $2.08\times 10^9$ & $2.17\times 10^9$ & $3.49\times 10^9$ & $7.65\times 10^8$ & $2.72\times 10^9$\\
\hline
\end{tabular}
}} \caption{IP dataset1 \label{data:properties}}
\end{table*}

\medskip
\noindent
$\bullet$ {\bf IP dataset2:}  IP packet trace from
an IP router  during August 1, 2008.
Packets were partitioned to one hour
time periods.

\noindent $\diamond$ {\em Colocated data:} {\bf keys:} destIP or
4tuples. {\bf weight assignments:} bytes, packets, IP flows, and
uniform.

 We used the packet stream of Hour3 which has
 $1.73\times 10^5$ distinct destIPs, $1.87\times 10^{10}$ total bytes,
$4.93\times 10^7$ packets, $1.30\times 10^6$ distinct flows, and
$0.94 \times 10^5$ distinct 4tuples.

\smallskip
\noindent
$\diamond$ {\em Dispersed data:}
The packets in each hour were aggregated into different weight
assignments.
We used keys that are destIP or 4tuples and weights that are
corresponding bytes.
We thus obtained a weight assignment $w^{(h)}$ for each hour.

 The following table
summarizes some properties of the data for the first 4 hours
and for the sets of hours $\cR=\{1,2\}$ and $\cR=\{1,2,3,4\}$.
The table lists the
number of distinct keys (destIP or 4tuples) and total bytes
$\sum_i w^{(h)}(i)$ for each hour or set of hours.\\
\smallskip
\centerline{
{\scriptsize
\begin{tabular}{|c|rrrr|rr|}
\hline
  hours         &  $1$  &  $2$   &   $3$   &  $4$  & $\{1,2\}$  & $\{1,2,3,4\}$\\
\hline
 destIP ($\times 10^5$) &  $2.17$ & $2.96$ & $1.73$ & $1.76$ & $3.41$ & $3.61$ \\
 4tuples ($\times 10^6$) &  $1.05$ & $1.17$ & $0.94$ & $0.99$ & $2.10$ & $3.74$ \\
 bytes ($\times 10^{10}$)& 2.00 & 1.84 & 1.87 & 1.81 & 3.84 & 7.52   \\
\hline
\end{tabular}
}}

 The following table lists,
for destIP and 4tuple keys, the sums
$\sum_i w^{(\min \cR)}(i)$, $\sum_i w^{(\max \cR)}(i)$, and $\sum_i w^{(L_1 \cR)}(i)$, for $\cR=\{1,2\}$ and $\cR=\{1,2,3,4\}$.\\
\smallskip
\centerline{
{\scriptsize
\begin{tabular}{|r|r r| r r|}
\hline
key &\multicolumn{2}{c|}{destIP} & \multicolumn{2}{c|}{4tuple} \\
$\cR$   & $\{1,2\}$ & $\{1,2,3,4\}$ & $\{1,2\}$ & $\{1,2,3,4\}$ \\
\hline
  $\min$ ($\times 10^{10}$)  & $1.51$ & $1.33$  & $0.86$  & $0.82$  \\
  $\max$ ($\times 10^{10}$) & $2.33$ & $3.02$  & $2.99$  & $4.92$  \\
  $L_1$ ($\times 10^{10}$) & $0.83$ & $1.69$ & $2.13$ & $4.11$ \\
\hline
\end{tabular}
}}

\medskip
\noindent
$\bullet$ {\bf Netflix Prize Data~\cite{netflix}}.  The dataset
contains dated ratings of $1.77 \times 10^4$ movies.
We used all
2005 ratings ($5.33\times 10^7$).
Each key corresponds to a movie and we used 12 weight assignments
$b\in \{1\ldots 12\}$ that corresponds to months.  The weight
$w^{(b)}(i)$ is the number of ratings of movie $i$ in month $b$.
(See Table~\ref{netflix:tab} for more details.)

\begin{table*}[htbp]
{\scriptsize \centerline{
\begin{tabular}{|l|rrrrrrrrrrrr|rrr|}
\hline
months &  $1$ & $2$ & $3$ & $4$ & $5$ & $6$ & $7$ & $8$ & $9$ & $10$ & $11$ & $12$ & $1$,$2$ & $1$-$6$ & $1$-$12$ \\
\hline
 movies ($\times 10^4$) & $1.54$ & $1.58$ & $1.61$ & $1.64 $ & $1.66 $ & $1.68 $ & $1.70 $ & $1.73 $ & $1.73 $ & $1.77 $ & $1.73 $ & $1.73 $ & 1.60 & $1.71$ & $1.77$ \\
ratings ($\times 10^6$) & $4.70$  & $4.10$ & $4.31$ & $4.16$ & $4.39$ & $5.30$ & $4.95$ & $5.26$ & $4.91$ & $5.16$ & $3.61$ & $2.41$ & $8.80$ & $27.0$ & $53.3$ \\
$\min$ ($\times 10^6$) & &&&&&&&&&&&& 3.72  & 2.97  &  1.68 \\
$\max$ ($\times 10^6$) & &&&&&&&&&&&& 5.08  & 6.79  &  7.95 \\
$L_1$  ($\times 10^6$) & &&&&&&&&&&&& 1.35  & 3.82  &  6.27 \\
\hline
\end{tabular}
} \caption{Netflix data set.   Distinct movies (number of movies
with at least one rating)  and total number of ratings  for each
month ($1,\ldots,12$) in 2005 and for periods $\cR=\{1,2\}$,
$\cR=\{1,\ldots,6\}$, and $\cR=\{1,\ldots,12\}$. For these periods,
we also show $\sum_i w^{(\min \cR)}(i)$, $\sum_i w^{(\max \cR)}(i)$,
and $\sum_i w^{(L_1 \cR)}(i)$. \vspace{-0.3cm}
  \label{netflix:tab}} }
\end{table*}


\medskip
\noindent $\bullet$ {\bf Stocks data:} Data set contains daily data
for about 8.9K ticker symbols, for October 2008 (23 trading days).
Daily data of each ticker had 5 price attributes (open, high, low,
close, adjusted\_close) and volume traded. Table~\ref{stock:table1}
lists totals of these weights for each trading day.
\begin{table*}[htbp]
\centerline{ {\tiny
\begin{tabular}{c}
\begin{tabular}{c|rrrrrrrrrrrr}
 & $1$ & $2$ & $3$ & $4$ & $5$ & $6$ & $7$ & $8$ & $9$ & $10$ & $11$ & $12$ \\
\hline
open & $1.81$ & $1.80$ & $1.75$ & $1.68$ & $1.65$ & $1.55$ & $1.56$ & $1.42$ & $1.50$ & $1.61$ & $1.54$ & $1.47$  \\
high & $1.85$ & $1.83$ & $1.81$ & $1.72$ & $1.70$ & $1.63$ & $1.61$ & $1.54$ & $1.61$ & $1.67$ & $1.57$ & $1.53$ \\
low & $1.78$ & $1.73$ & $1.70$ & $1.57$ & $1.57$ & $1.50$ & $1.45$ & $1.33$ & $1.46$ & $1.52$ & $1.45$ & $1.40$  \\
close & $1.82$ & $1.75$ & $1.72$ & $1.65$ & $1.59$ & $1.56$ & $1.48$ & $1.46$ & $1.58$ & $1.57$ & $1.47$ & $1.50$ \\
adj\_close & $1.81$ & $1.74$ & $1.72$ & $1.64$ & $1.58$ & $1.55$ & $1.47$ & $1.45$ & $1.57$ & $1.56$ & $1.46$ & $1.49$ \\
volume  & $1.52$ & $1.66$ & $1.82$ & $2.26$ & $1.96$ & $2.44$ &
$2.10$ & $3.14$ & $1.93$ & $2.22$ & $1.80$ & $2.27$
\end{tabular} \\
\\
\begin{tabular}{c|rrrrrrrrrrr}
  & $13$ & $14$ & $15$ & $16$ & $17$ & $18$ & $19$ & $20$ & $21$ & $22$ & $23$ \\
\hline
open   & $1.48$ & $1.52$ & $1.52$ & $1.48$ & $1.45$ & $1.37$ & $1.38$ & $1.38$ & $1.42$ & $1.46$ & $1.47$ \\
high   & $1.57$ & $1.57$ & $1.56$ & $1.52$ & $1.49$ & $1.44$ & $1.43$ & $1.45$ & $1.49$ & $1.50$ & $1.54$\\
low    & $1.44$ & $1.49$ & $1.49$ & $1.42$ & $1.38$ & $1.34$ & $1.34$ & $1.33$ & $1.39$ & $1.42$ & $1.44$ \\
close  & $1.50$ & $1.55$ & $1.51$ & $1.45$ & $1.44$ & $1.40$ & $1.36$ & $1.42$ & $1.44$ & $1.48$ & $1.51$\\
adj\_close  & $1.50$ & $1.54$ & $1.51$ & $1.44$ & $1.43$ & $1.39$ & $1.36$ & $1.42$ & $1.43$ & $1.47$ & $1.50$\\
volume & $1.84$ & $1.42$ & $1.43$ & $1.73$ & $2.05$ & $1.84$ &
$1.55$ & $1.99$ & $1.96$ & $1.71$ & $1.75$
\end{tabular}
\end{tabular}
} } \caption{Daily totals for 23 trading days in October, 2008.
Prices (open, high, low, close, adjusted\_close) are $\times 10^5$.
Volumes are in multiples of $10^{10}$.\label{stock:table1}}
\end{table*}

The ticker prices are
highly correlated both in terms of same attribute over different days
and the different price attributes in a given day.
The correlation is much stronger than for the volume attribute
or weight assignments used in the IP datasets.
At least 93\% of stocks had
positive volume each day and virtually all had positive (high, low, close, adjusted\_close) prices for the duration. This contrasts
the IP datasets, where it is much more likely for keys (destIPs or 4tuples)
to have zero weights in subsequent assignments.

\smallskip
\noindent
$\diamond$ {\em Colocated data:} {\bf keys:} ticker symbols;
{\bf weight assignments:} the six numeric attributes:
open, high, low, close, adjusted\_close, and volume in a given trading day.

\smallskip
\noindent $\diamond$ {\em Dispersed data}: {\bf keys:} ticker
symbols; {\bf weight assignments:} daily (high or volume) values for
each trading days.

 For multiple-assignment aggregates evaluation,
we used the first 2,5,10,15,23 trading days of October:
$\cR=\{1,2\}$ (October 1-2), $\cR=\{1,\ldots,5\}$ (October 1-7), $\cR=\{1,\ldots,10\}$ (October 1-14), $\cR=\{1,\ldots,15\}$ (October 1-21), $\cR=\{1,\ldots,23\}$ (October 1-31).
 The following table lists
$\sum_i w^{(\min \cR)}(i)$, $\sum_i w^{(\max \cR)}(i)$, and $\sum_i
w^{(L_1 \cR)}(i)$ for these sets of trading days.

\smallskip
\centerline{
{\tiny
\begin{tabular}{c|rrrrr|rrrrr|}
\cline{2-11}
 & \multicolumn{5}{|c|}{high ($\times 10^{5}$)} & \multicolumn{5}{|c|}{volume ($\times 10^{10}$)} \\
  & $1$-$2$  & $1$-$5$ & $1$-$10$ & $1$-$15$ & $1$-$23$   & $1$-$2$  & $1$-$5$ & $1$-$10$ & $1$-$15$ & $1$-$23$ \\
\hline
\multicolumn{1}{|c|}{$\min$} & $1.82$ & $1.67$ & $1.48$ & $1.44$ & $1.33$ & $1.34$ & $1.33$ & $1.30$ & $1.15$& $1.13$\\
\multicolumn{1}{|c|}{$\max$} & $1.87$ & $1.89$ & $1.92$ & $1.92$ & $1.94$ & $1.80$ & $2.54$ & $3.50$ & $3.59$& $3.77$\\
\multicolumn{1}{|c|}{$L_1$} & $0.05$ & $0.22$ & $0.44$ & $0.49$ & $0.61$ & $0.41$ & $1.20$ & $2.20$ & $2.43$& $2.64$\\
\hline
\end{tabular}
}}

\subsection{Dispersed data.}

  We evaluate our $w^{(\min \cR)}$, $w^{(\max \cR)}$, and
$w^{(L_1 \cR)}$ estimators as defined in
Section~\ref{dispersed:sec}: $a^{(\max \cR)}$, $a_s^{(\min \cR)}$,
$a_l^{(\min \cR)}$, $a_s^{(L_1 \cR)}$, and $a_l^{(L_1 \cR)}$ for
coordinated sketches and $a_{ind}^{(\min \cR)}$ for independent
sketches.

 We used shared-seed coordinated sketches and show results for
the \IPPS\ ranks (see Section~\ref{prelim:sec}).
Results for \EXP\ ranks were similar.

We measure performance using the absolute $\SV[a^{(f)}]$ and
normalized {\small $\nSV[a^{(f)}] \equiv \SV[a^{(f)}]/(\sum_{i\in
I} f(i))^2$} sums of per-key variances (as discussed in
Section~\ref{prelim:sec}), which we approximate by averaging square
errors over multiple (25-200) runs of the sampling algorithm.

\smallskip
\noindent
{\bf Coordinated versus Independent sketches.}
 We compare the
$w^{(\min \cR)}$ estimators $a_\ell^{(\min \cR)}$ (coordinated
sketches) and $a_{ind}^{(\min \cR)}$ (independent
sketches).\footnote{We consider $w^{(\min \cR)}$ because the estimator is applicable for independent sketches with unknown seeds.  
There are no nonnegative estimators for $w^{(\max \cR)}$
and $w^{(L_1 \cR)}$ estimators for independent sketches with unknown seeds.}

\onlyinproc{
\begin{figure*}[htbp]
\centerline{
\begin{tabular}{ccc}
\epsfig{figure=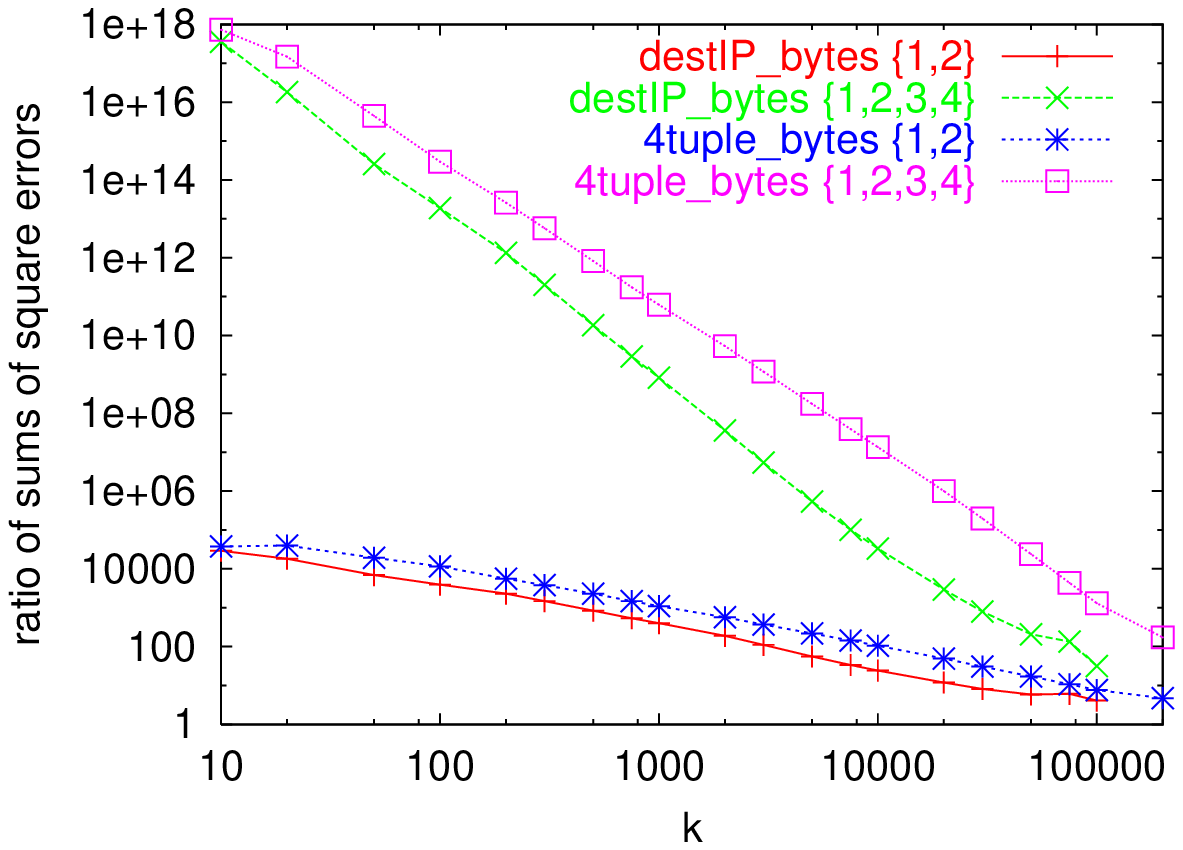,width=0.32\textwidth}
&
\epsfig{figure=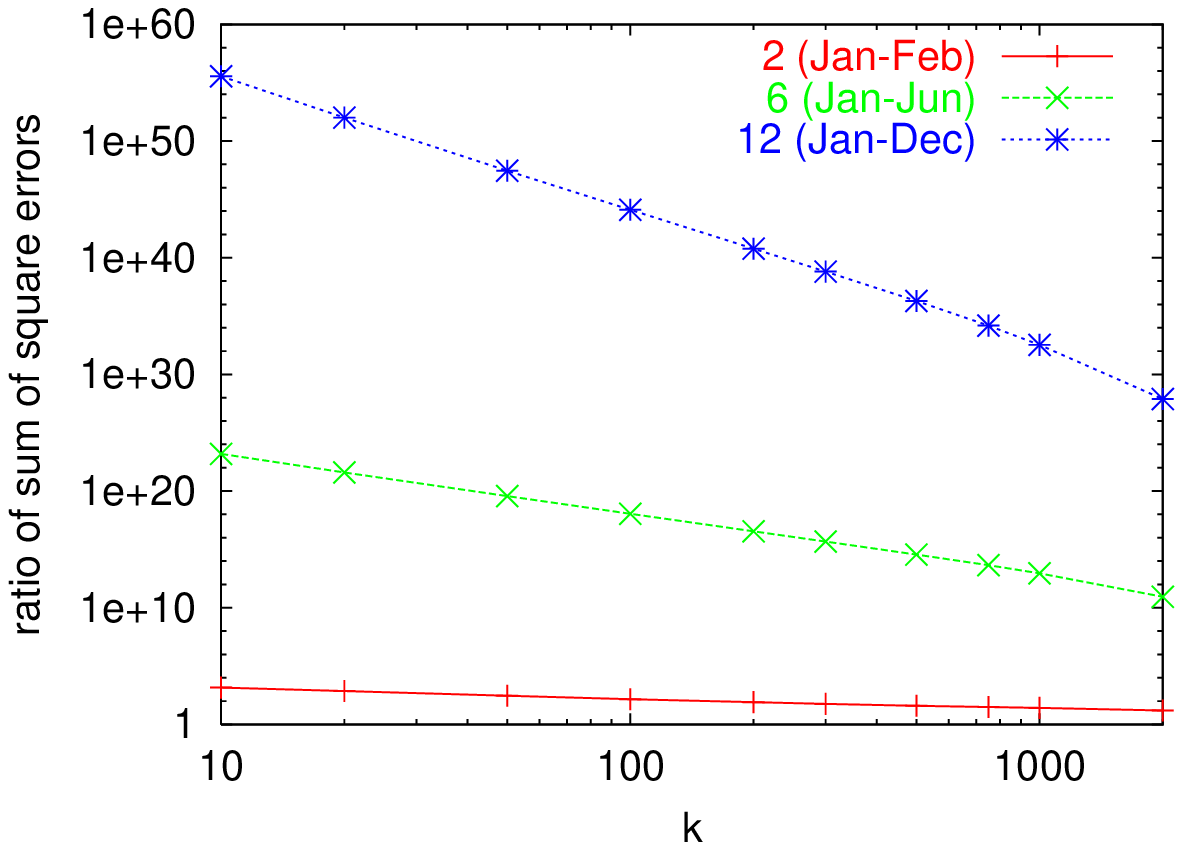,width=0.32\textwidth}
&
\epsfig{figure=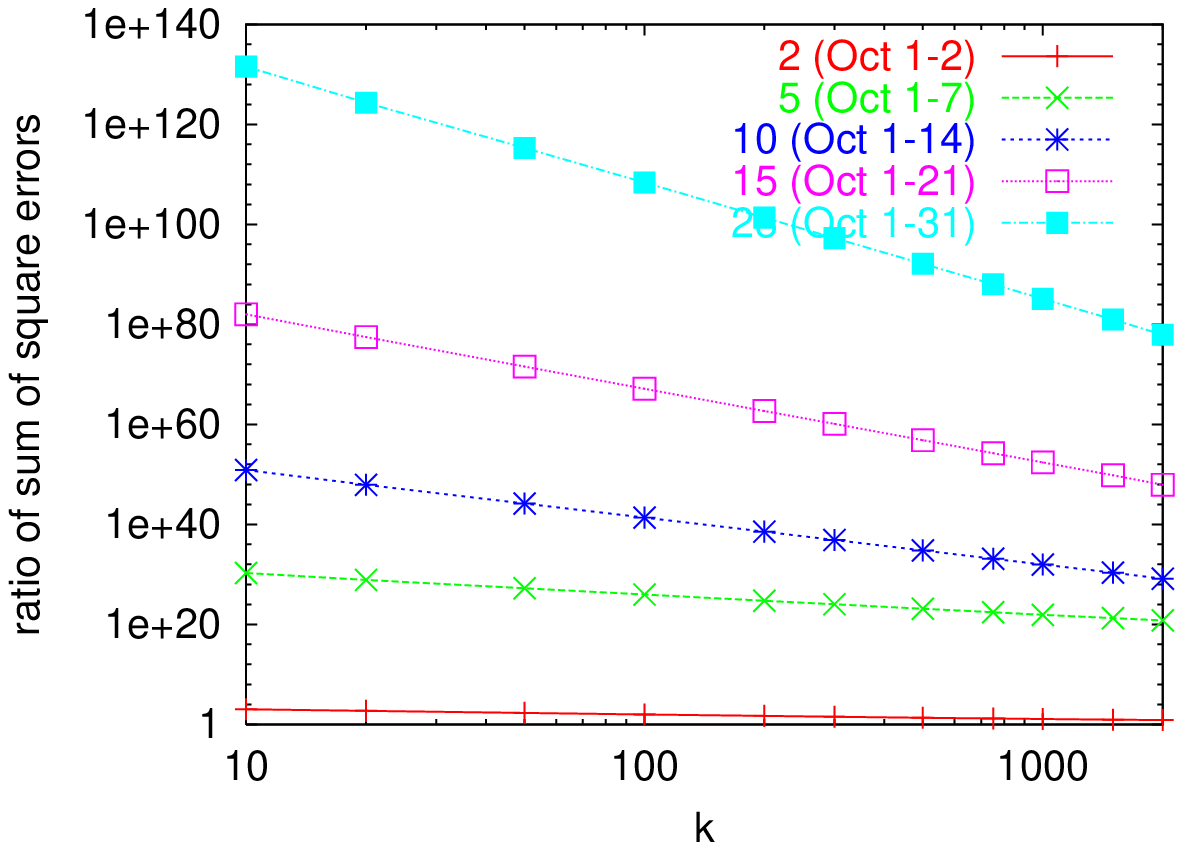,width=0.32\textwidth}
\end{tabular}
} \caption{Top:  IP dataset2 (left),  Netflix data set  (middle),
stocks dataset high-values  (right). Variance ratio of
$w^{(\min \cR)}$ estimators for independent and coordinated sketches
$\SV[a_{ind}^{(\min \cR)}]/\SV[a_l^{(\min \cR)}]$.}
\label{ind_coord_ratios_selected:fig}
\end{figure*}
} \notinproc{
\begin{figure*}[htbp]
\centerline{
\begin{tabular}{ccc}
\epsfig{figure=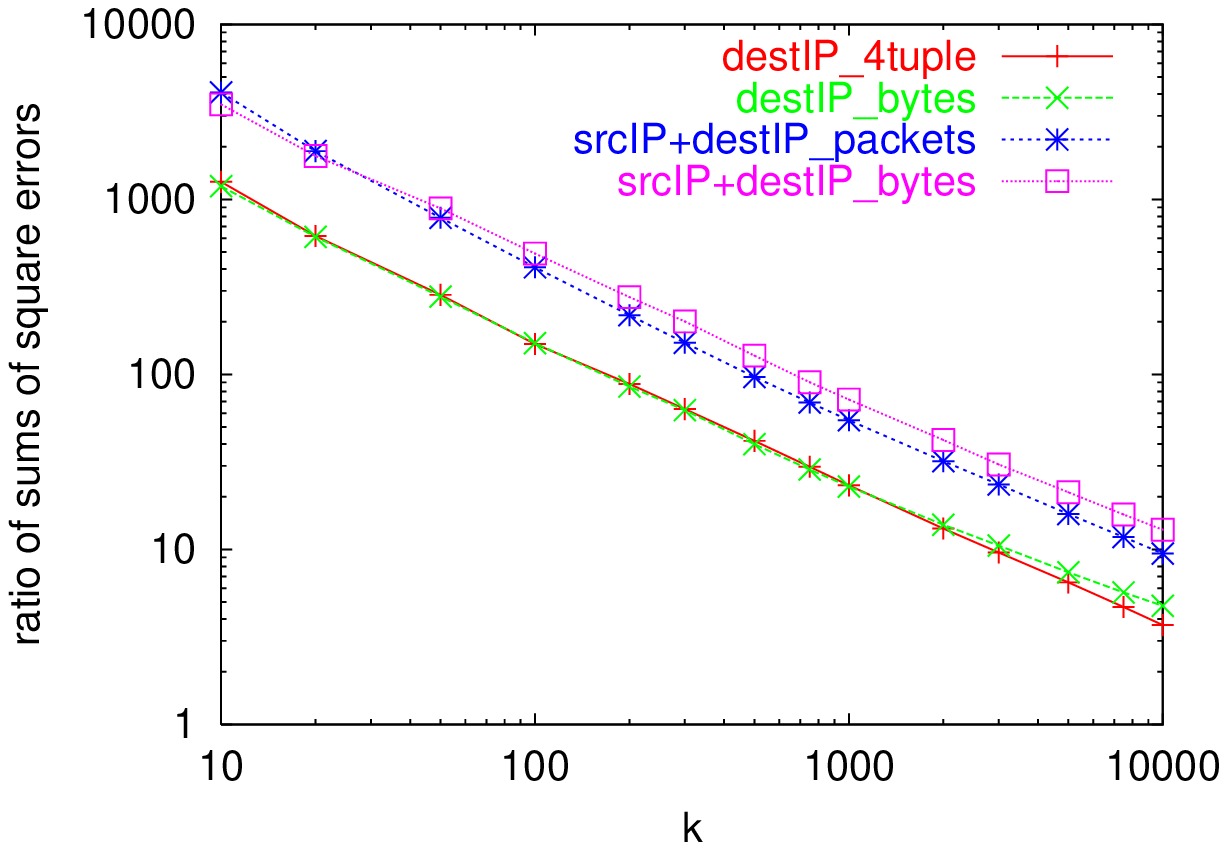,width=0.32\textwidth}
&
\epsfig{figure=multi_code/resultsL1/nfcapd20080801_ind_coord_ratio.eps,width=0.32\textwidth}
&
\epsfig{figure=multi_code/resultsL1/min_netflix2005_movie_req_p_month_ind_coord_ratio_r200.eps,width=0.32\textwidth} \\
\epsfig{figure=multi_code/resultsL1/min_stocks_high_200810_ind_coord_ratio_r50.eps,width=0.32\textwidth}
&
\epsfig{figure=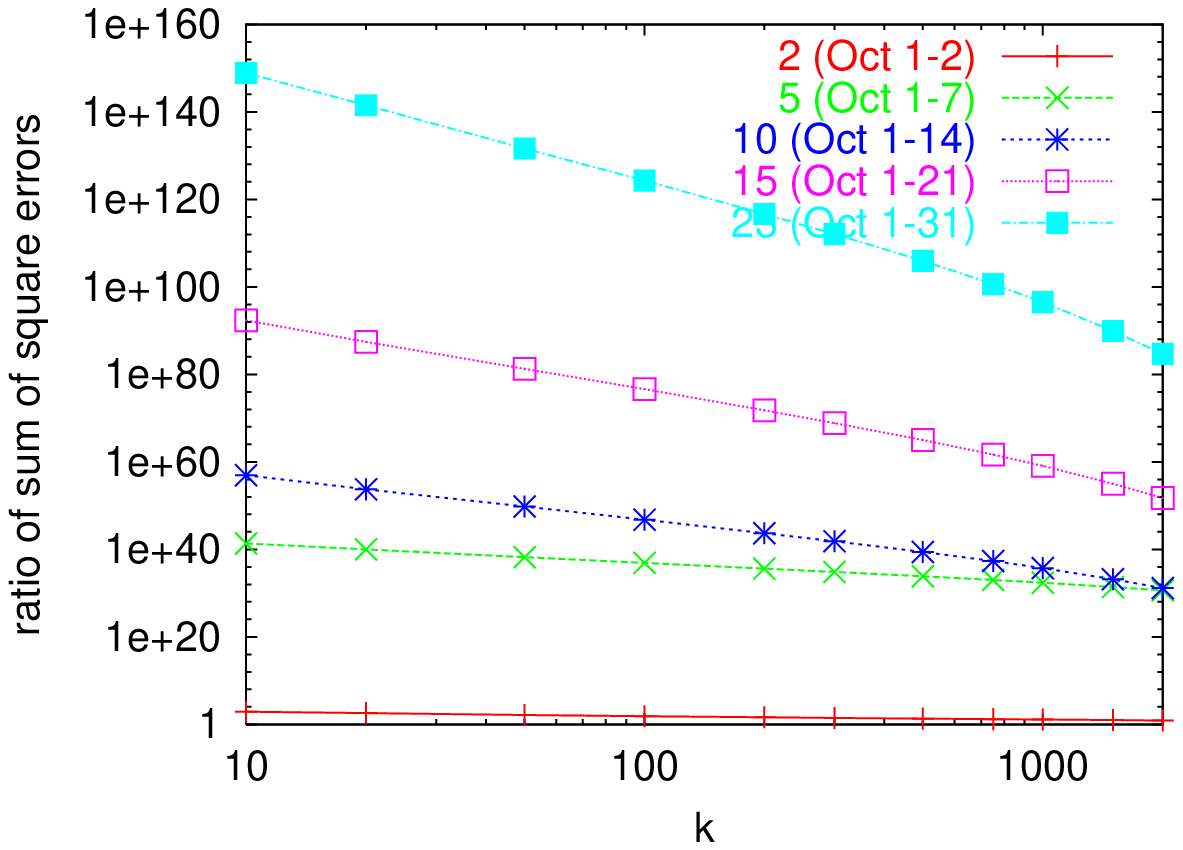,width=0.32\textwidth}
&
\end{tabular}
} \caption{Top:  IP dataset1 (left), IP dataset2 (middle), Netflix
data set (right). Bottom: stocks dataset high values (left),  stocks
dataset volume values (middle). Ratio of $w^{(\min \cR)}$ estimators
for independent and coordinated sketches
$\SV[a_{ind}^{(\min \cR)}]/\SV[a_\ell^{(\min \cR)}]$.}
\label{ind_coord_ratios:fig}
\end{figure*}
}

\notinproc{Figure~\ref{ind_coord_ratios:fig}}\onlyinproc{Figure~\ref{ind_coord_ratios_selected:fig}}
shows the ratio $\SV[a_{ind}^{(\min \cR)}]/\SV[a_\ell^{(\min \cR)}]$
as a function of $k$ for our datasets.  Across data sets, the
 variance of the independent-sketches
estimator is significantly larger, up to many
orders of magnitude, than the variance of coordinated-sketches estimators.
The ratio decreases with $k$ but remains significant even when
the sample size exceeds 10\% of the number of keys.

The ratio increases with the number of weight assignments: On the
Netflix data set, the ratio is 1-3 orders of magnitude for 2 months
(assignments) and 10-40 orders of magnitude for 6-12 months
(assignments).  On IP dataset 2, the gap is 1-5 orders of magnitude
for 2 assignments (hours) and 2-18 orders of magnitude for 4
assignments.  On the stocks data set, the gap is 1-3 orders of
magnitude for 2 assignments and reaches 150 orders of magnitude.  This
agrees with the exponential decrease of the inclusion probability with
the number of assignments for independent sketches (see
Section~\ref{lset:sec}).  These ratios demonstrate the estimation
power provided by coordination.

\smallskip
\noindent
{\bf Weighted versus unweighted coordinated sketches.}
We compare the performance of our estimators  to
known estimators applicable to
unweighted coordinated sketches
(coordinated sketches for uniform and global
weights~\cite{CK:sigmetrics09}).
To apply these methods, all positive weights were replaced by unit weights.
Because of the skewed nature
of the weight distribution, the ``unweighted'' estimators performed
poorly with variance being orders of magnitude larger.

\ignore{
\begin{figure} [htbp]
\centerline{\begin{tabular}{cc}
\epsfig{figure=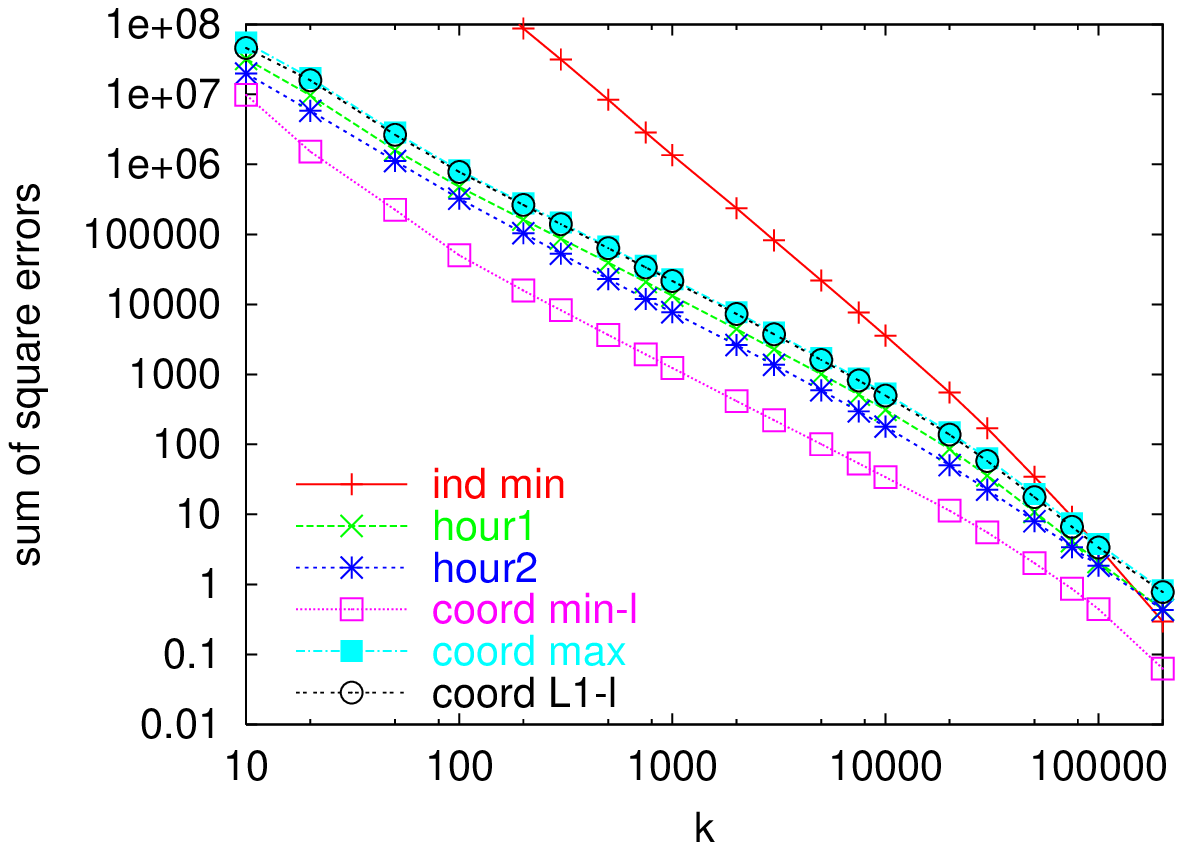,width=0.24\textwidth} &
\epsfig{figure=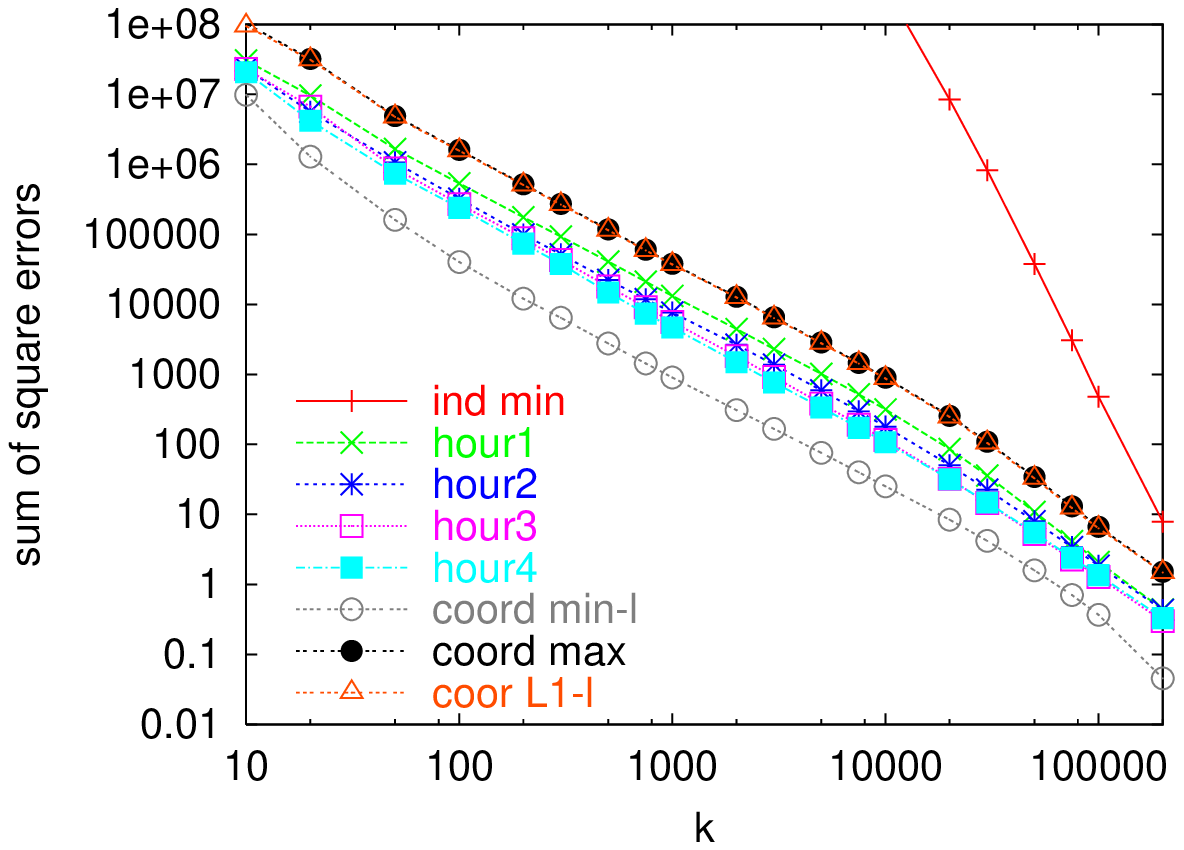,width=0.24\textwidth} \\
\epsfig{figure=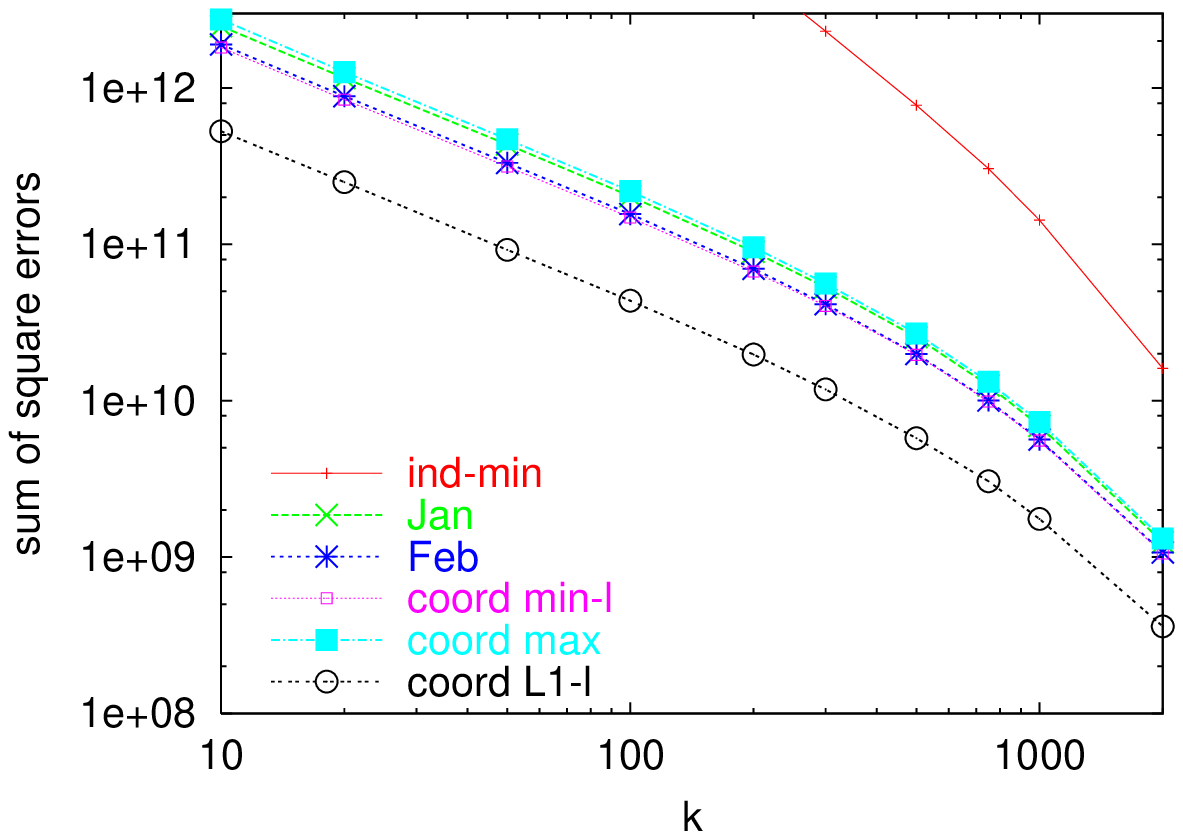,width=0.24\textwidth} &
\epsfig{figure=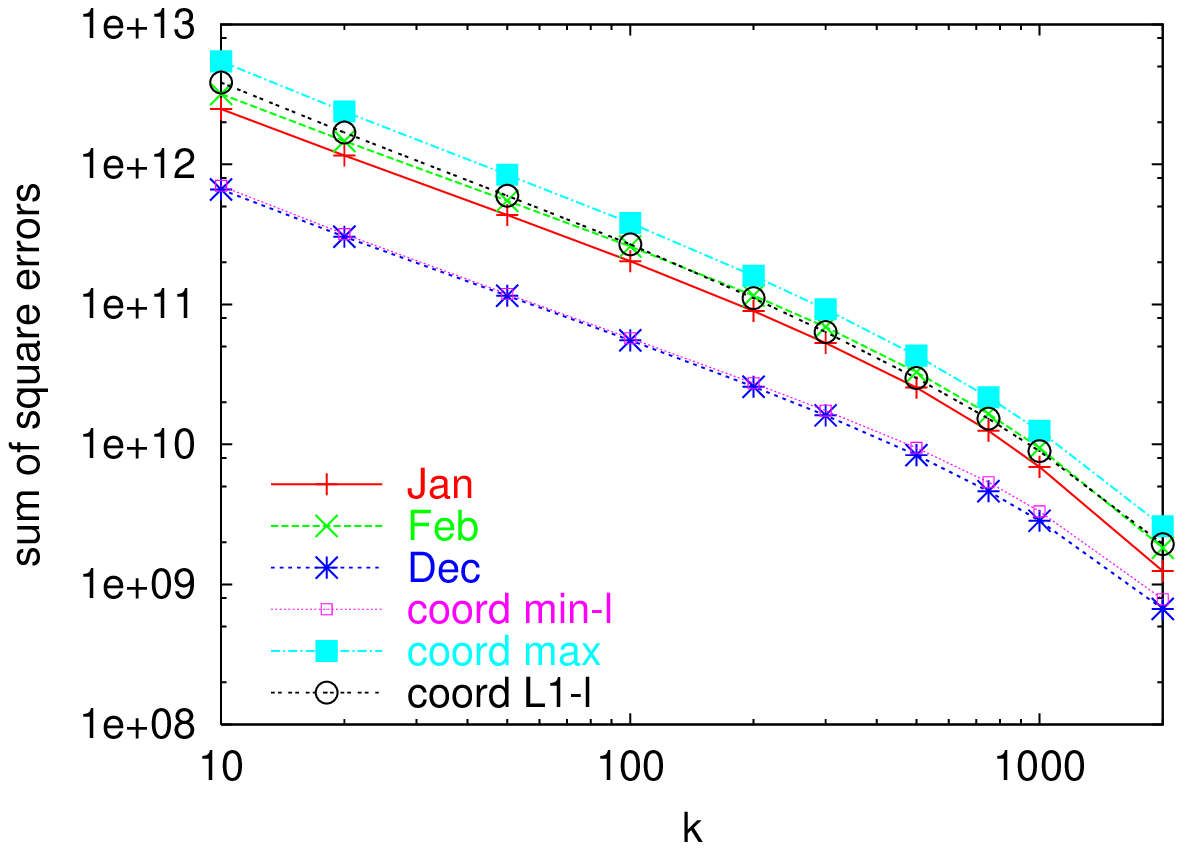,width=0.24\textwidth} \\
\epsfig{figure=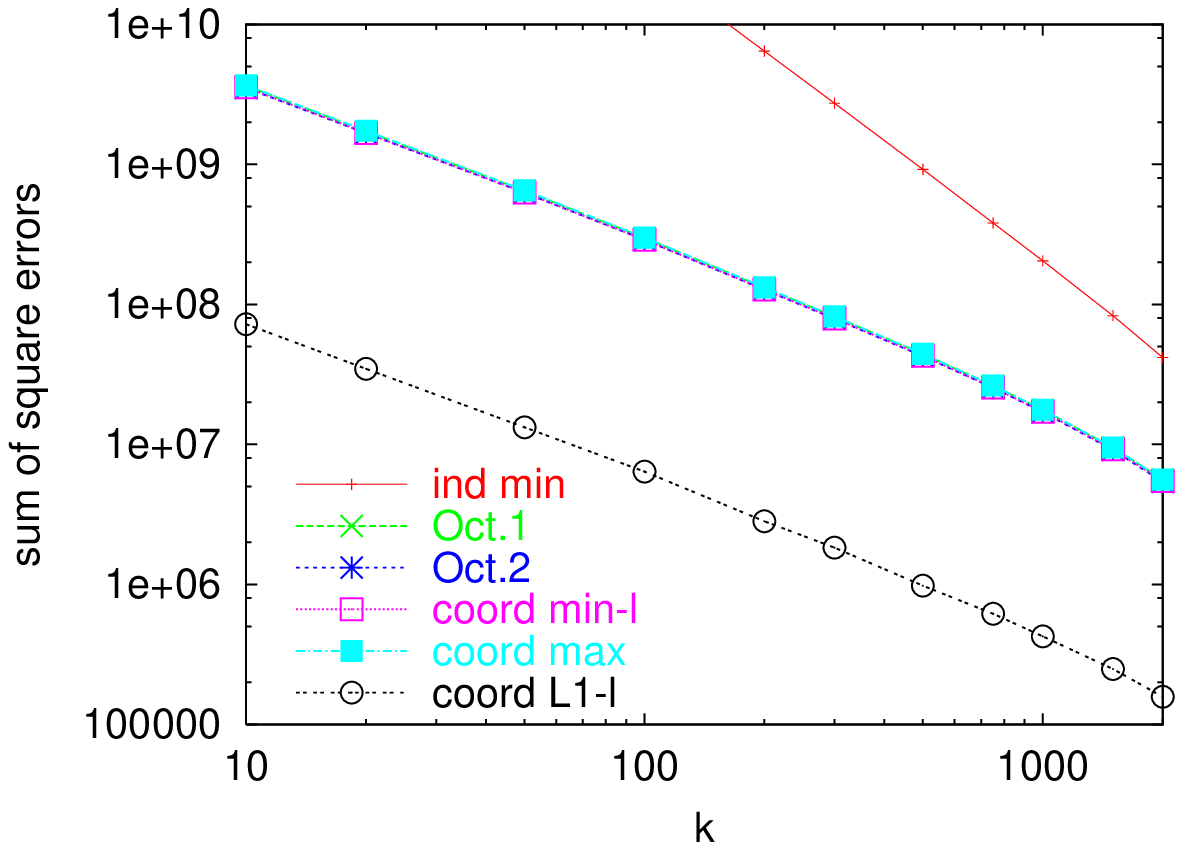,width=0.24\textwidth} &
\epsfig{figure=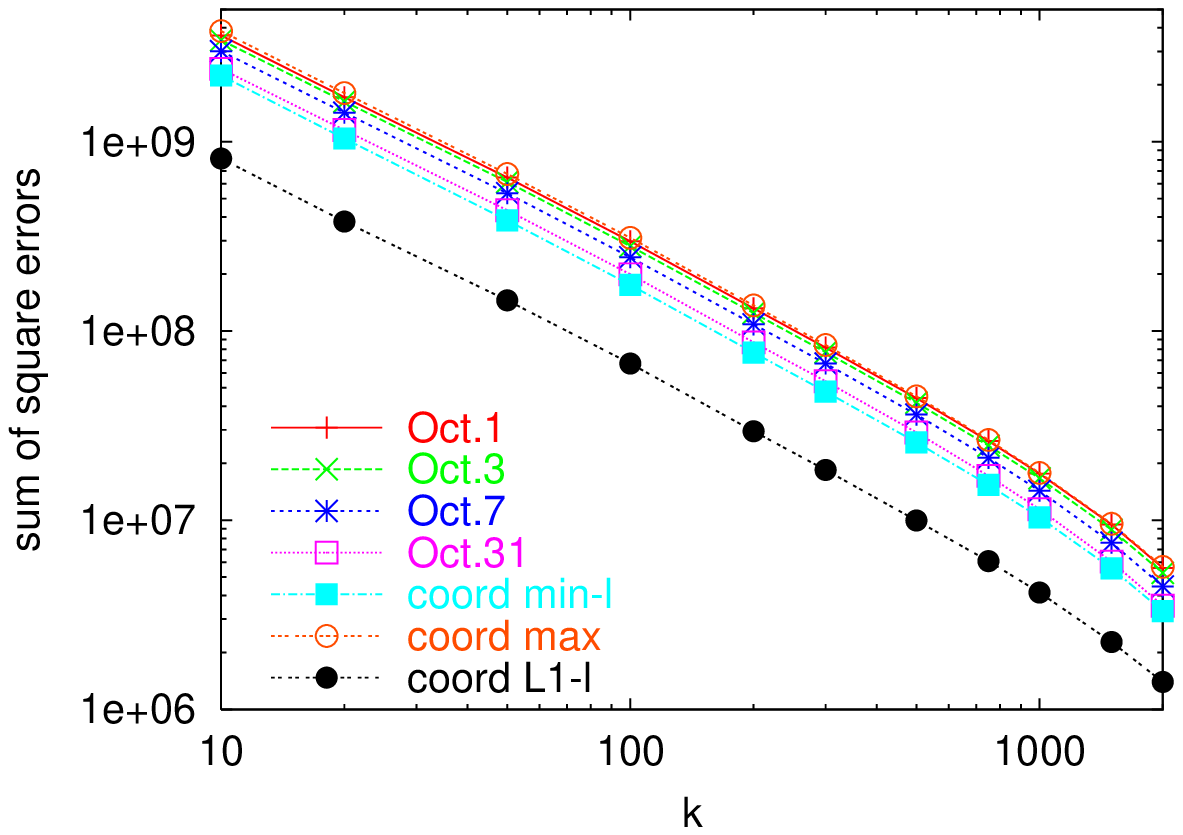,width=0.24\textwidth}
\end{tabular}
}
\caption{Top row: IP dataset2 key$=$4tuple  weight$=$bytes hours$=\{1,2\}$; IP dataset2 key$=$4tuple  weight$=$bytes hours$=\{1,2,3,4\}$. Middle row:
Netflix data set $\cR=\{1,2\}$,  $\cR=\{1,\ldots,12\}$.
Bottom row:
Stock dataset, high values: $\cR=\{1,2\}$ (October 1-2, 2008), $\cR=\{1,\ldots,23\}$ (all trading days in October, 2008).
\vspace*{-0.4cm}
\label{selected_sv:fig}}
\end{figure}
}

\smallskip
\noindent {\bf Variance  of multiple-assignment estimators.}
We relate the variance of our $w^{(\min \cR)}$,
$w^{(\max \cR)}$, and $w^{(L_1 \cR)}$ and the variance of the
optimal single-assignment estimators $a^{(b)}$ for the respective
individual weight assignments $w^{(b)}$ ($b\in\cR$).\notinproc{\footnote{For
\IPPS\ ranks, $a^{(b)}$ are essentially optimal as they minimize
$\SV[a^{(b)}]$ (and $\nSV[a^{(b)}]$) modulo a single
sample~\cite{DLT:jacm07,Szegedy:stoc06}.}}
Because the variance of $a_{ind}^{(\min \cR)}$ was typically many orders
of magnitude worse, we include it only when it fit in the scale
of the plot.
  The single-assignment estimators
$a^{(b)}$ are identical for independent and coordinated sketches
(constructed with the same $k$ and rank functions family), and
hence are shown once.

\notinproc{
\begin{figure*}[htbp]
\centerline{\begin{tabular}{cc}
\epsfig{figure=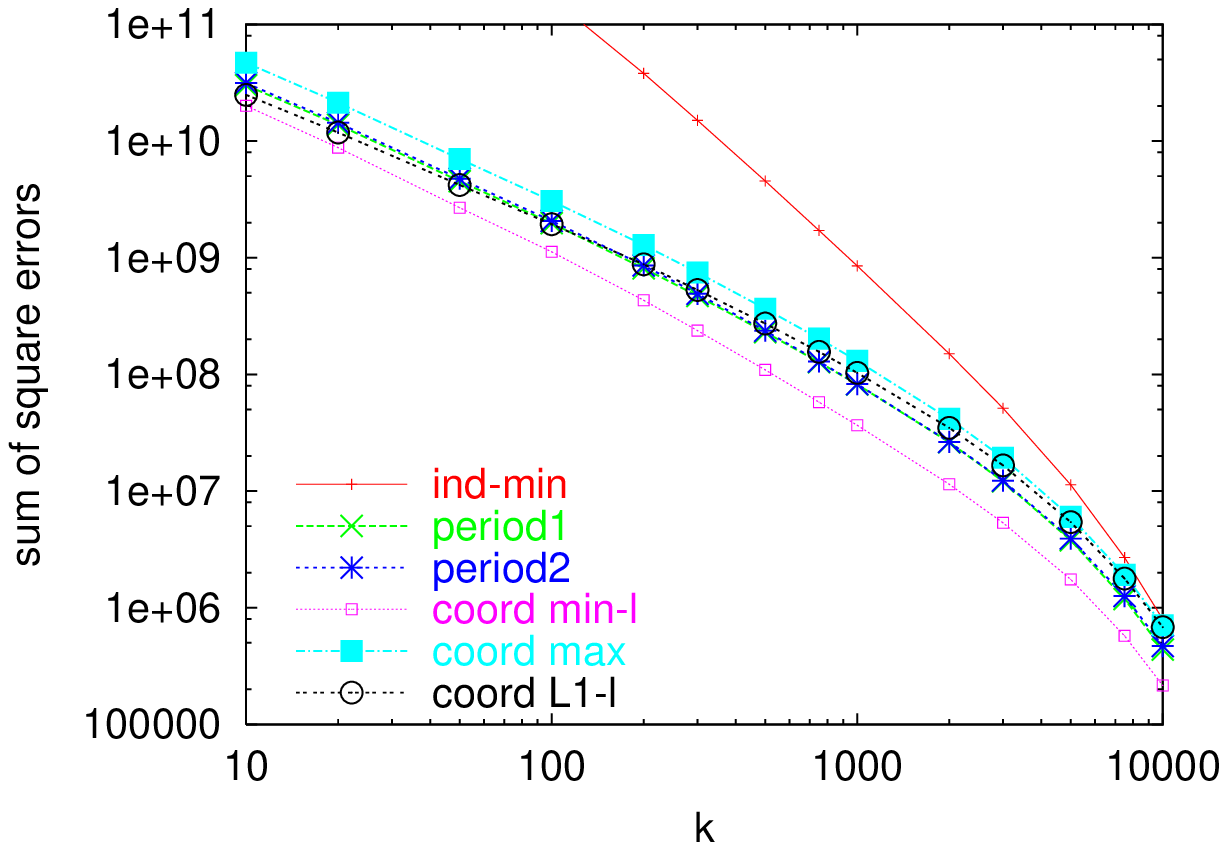,width=0.45\textwidth} &
\epsfig{figure=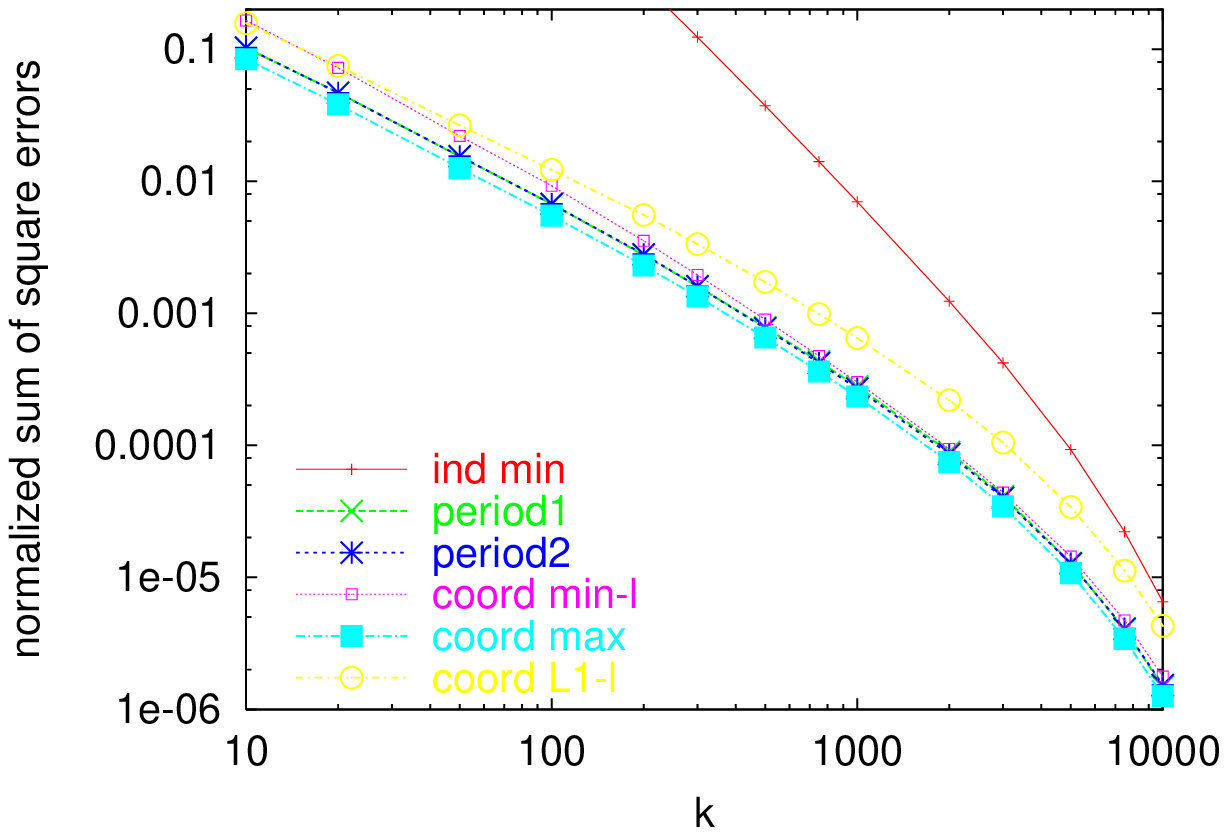,width=0.45\textwidth} \\
\epsfig{figure=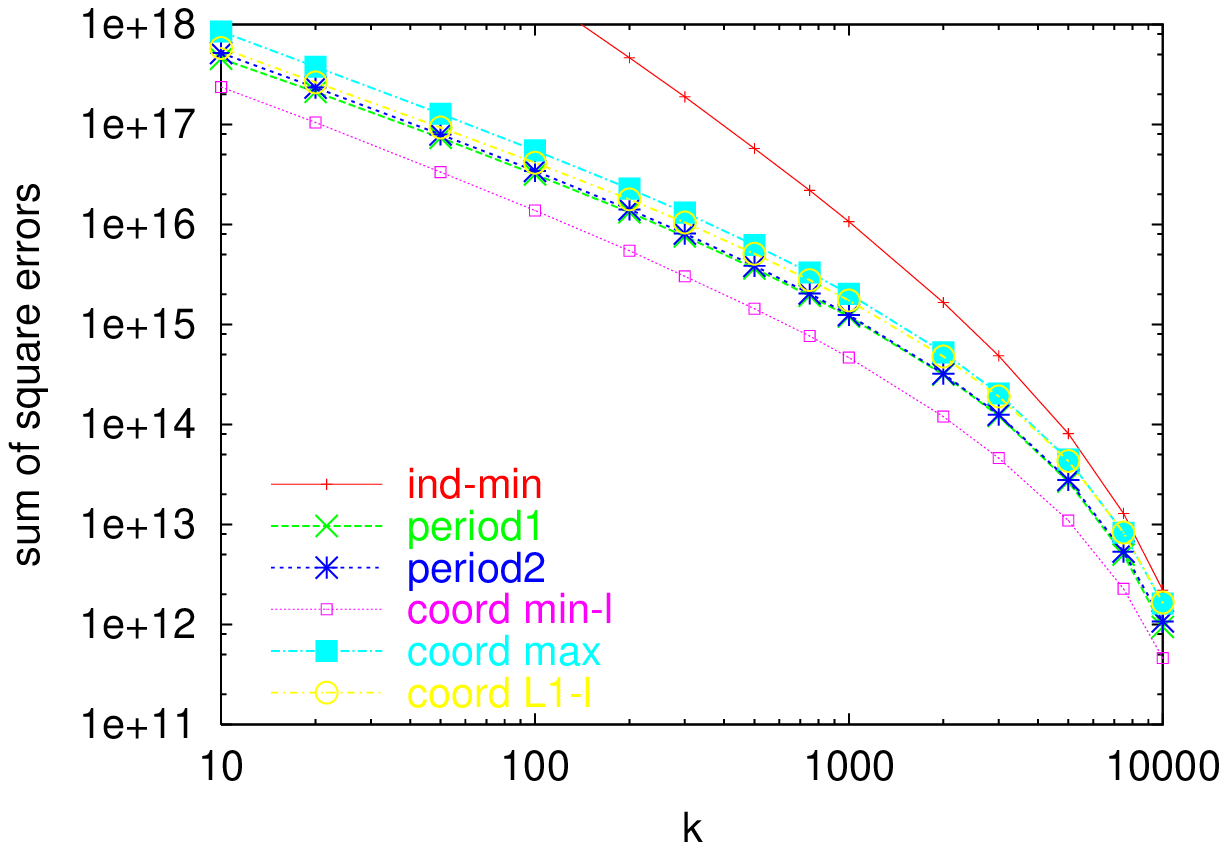,width=0.45\textwidth} &
\epsfig{figure=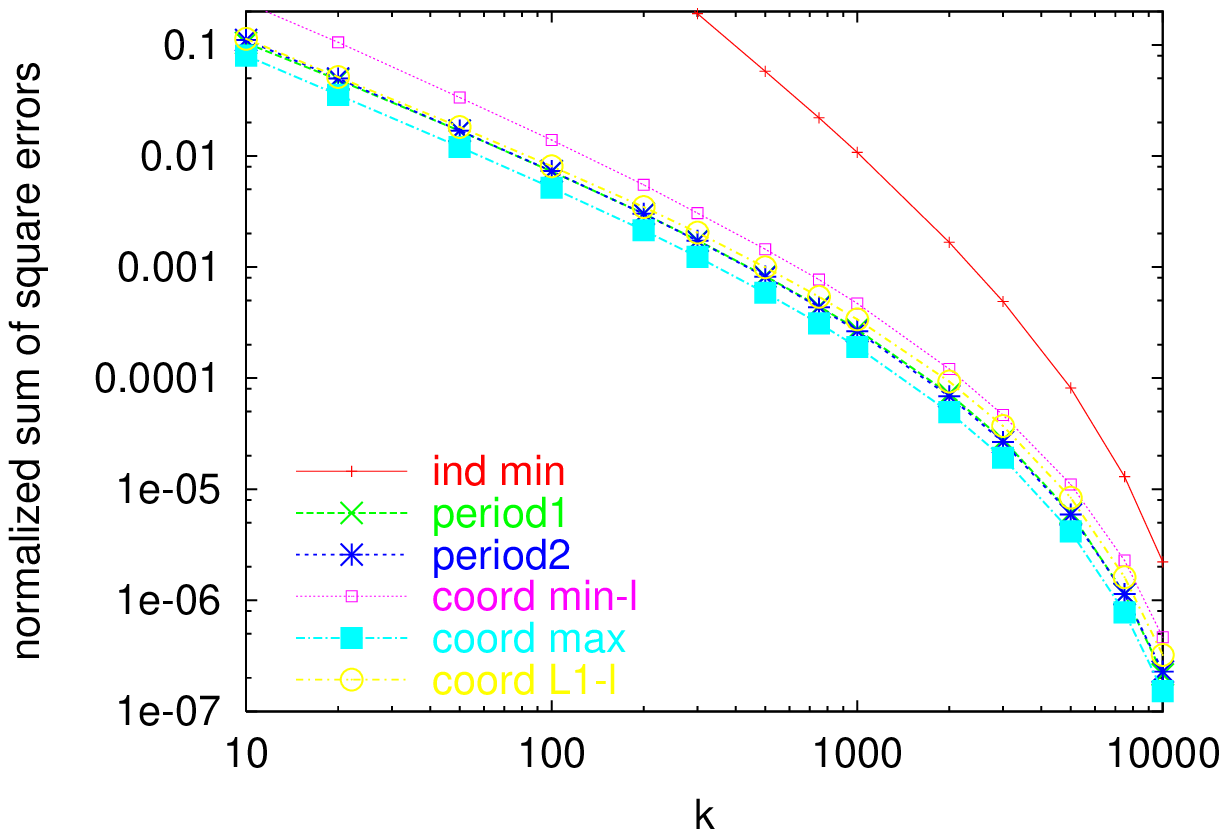,width=0.45\textwidth}\\
\epsfig{figure=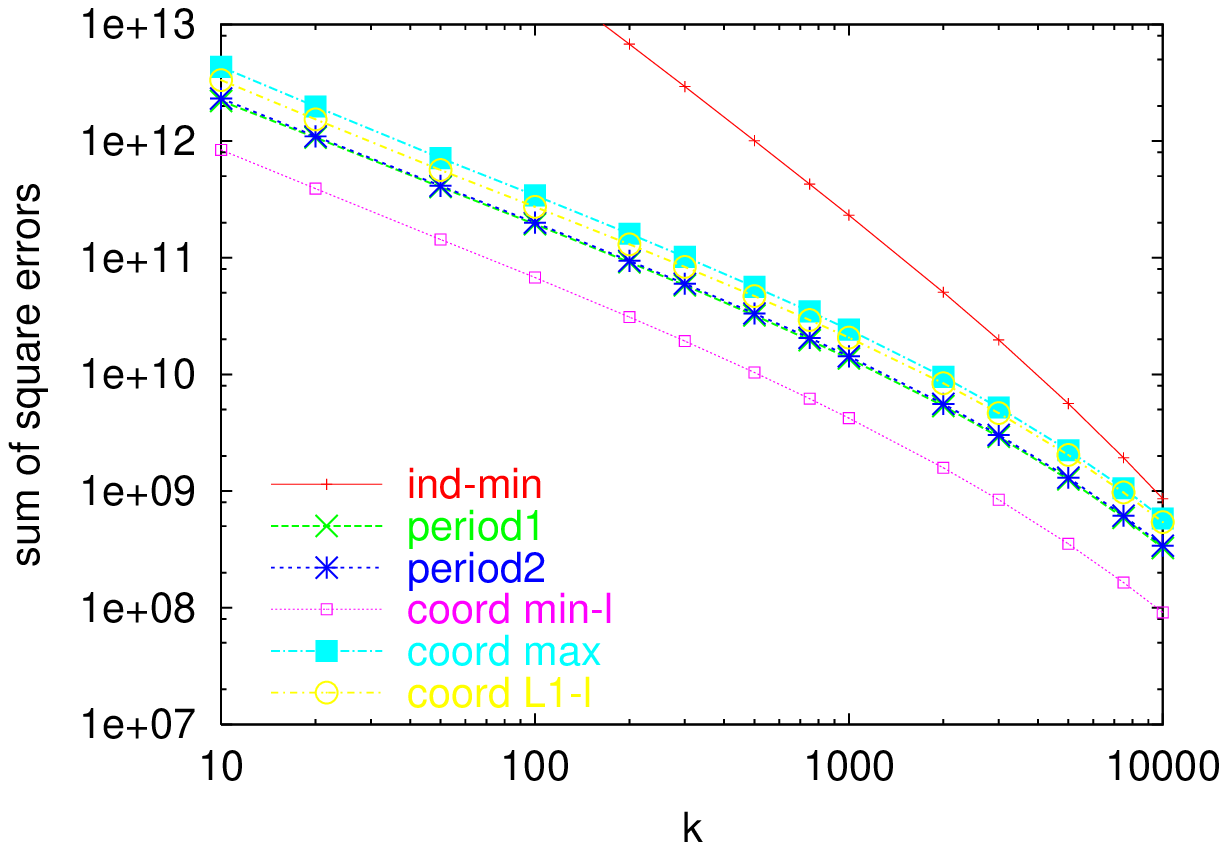,width=0.45\textwidth} &
\epsfig{figure=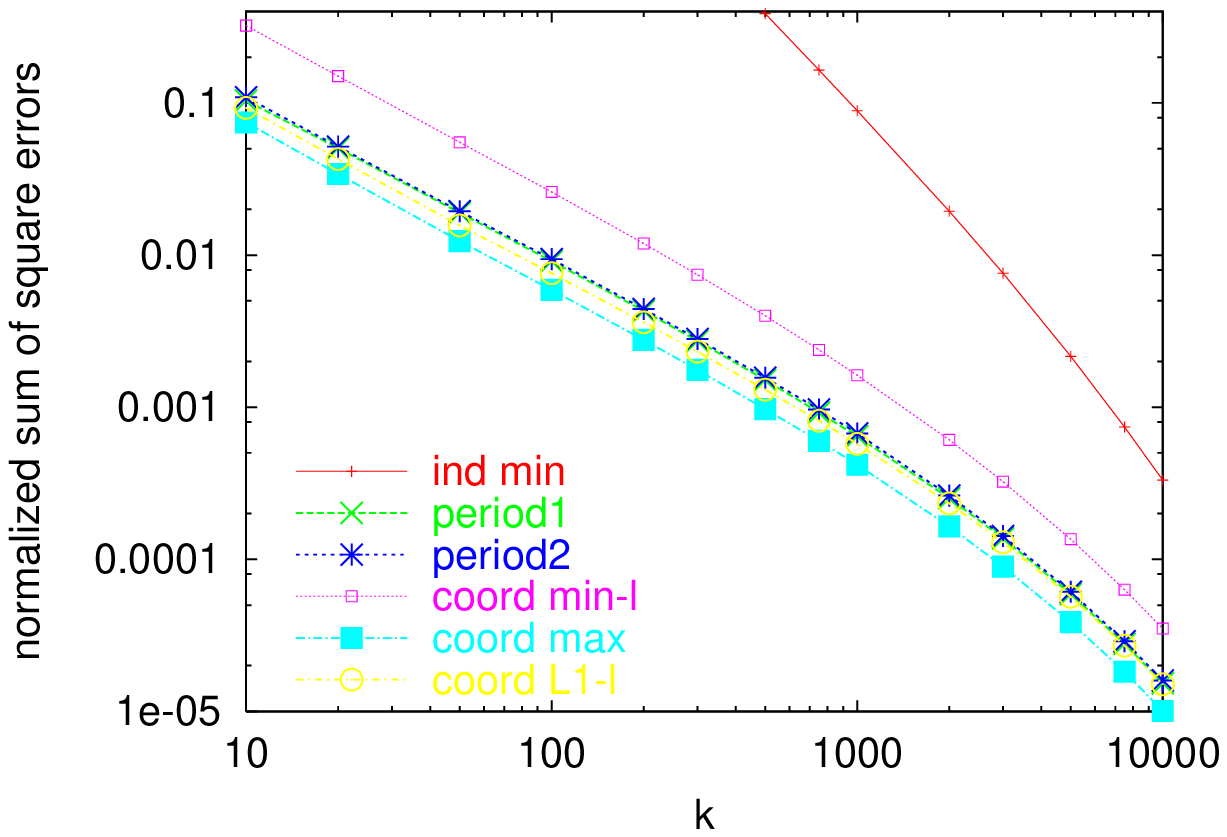,width=0.45\textwidth} \\
\epsfig{figure=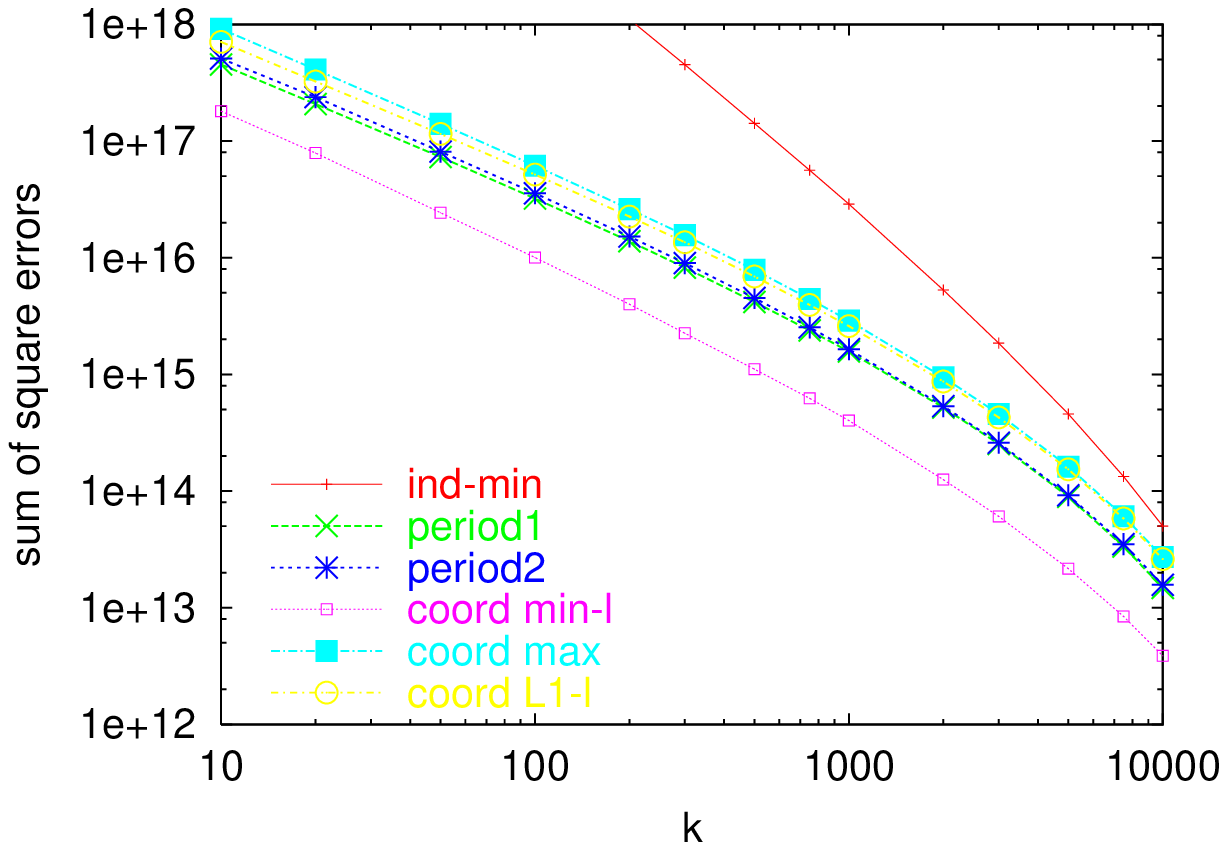,width=0.45\textwidth} &
\epsfig{figure=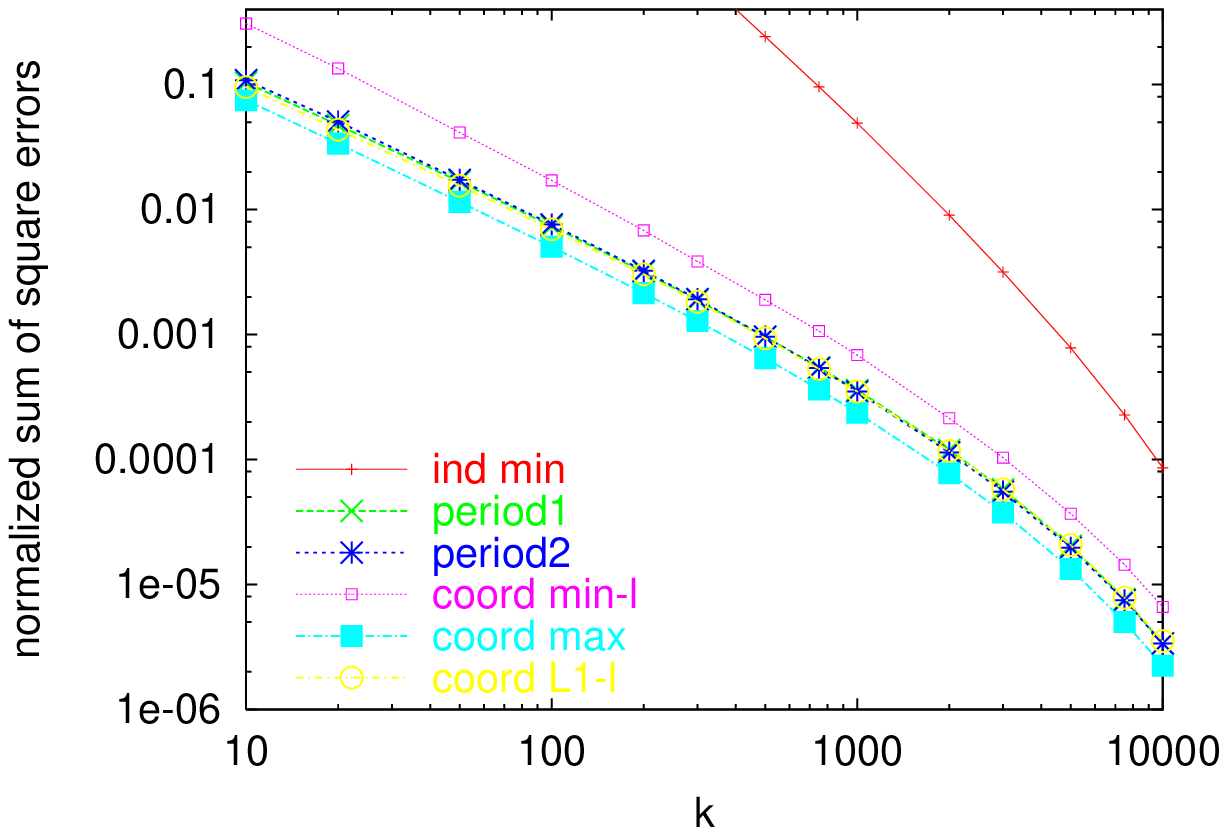,width=0.45\textwidth}
\end{tabular}
}
\caption{IP dataset1. Sum of square errors. left: absolute, right: normalized.  Top: key$=$destIP  weight$=$4tuple\_count, second row: key$=$destIP  weight$=$bytes.  Third row: key$=$srcIP$+$destIP, weight$=$packets.  Fourth row: key$=$srcIP$+$destIP, weight$=$bytes
\label{tperiods1:fig}}
\end{figure*}

\begin{figure*}[htbp]
\centerline{\begin{tabular}{cc}
\epsfig{figure=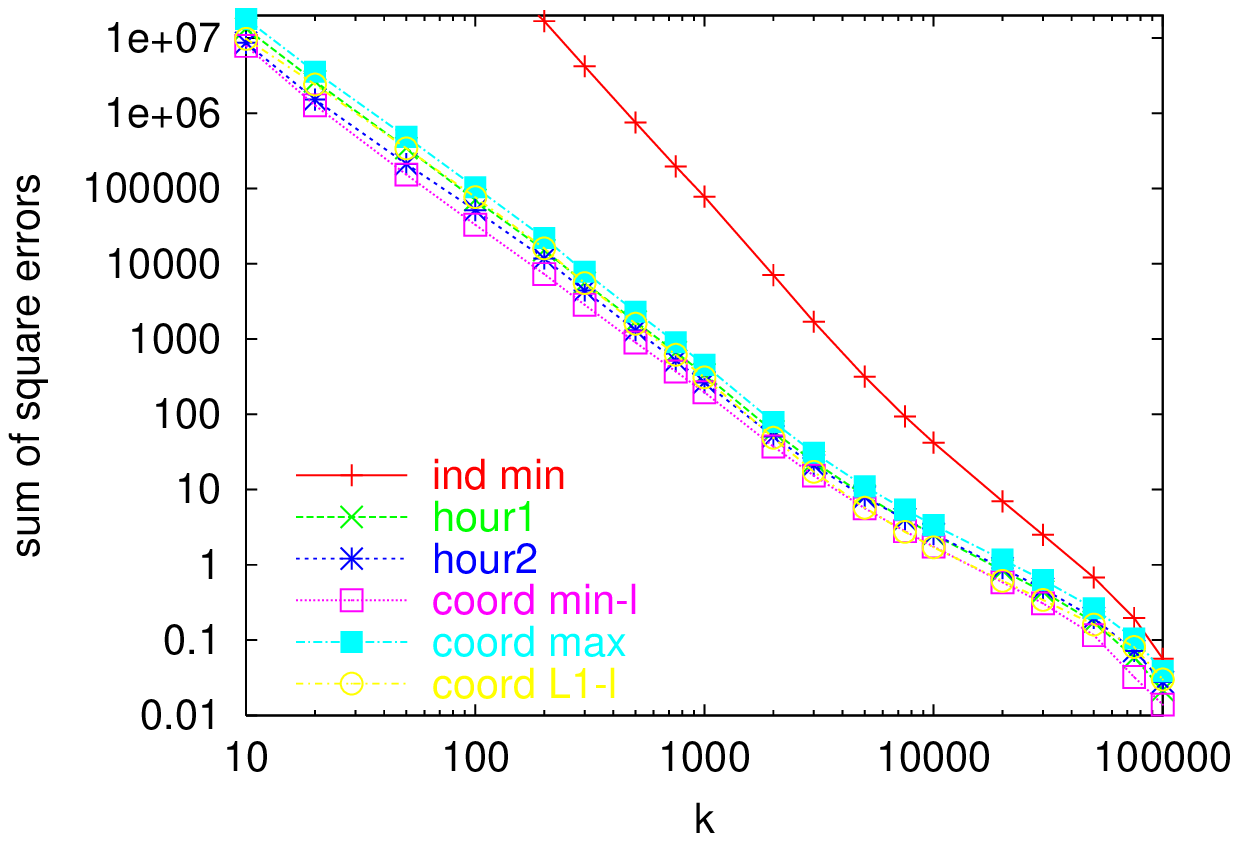,width=0.45\textwidth} &
\epsfig{figure=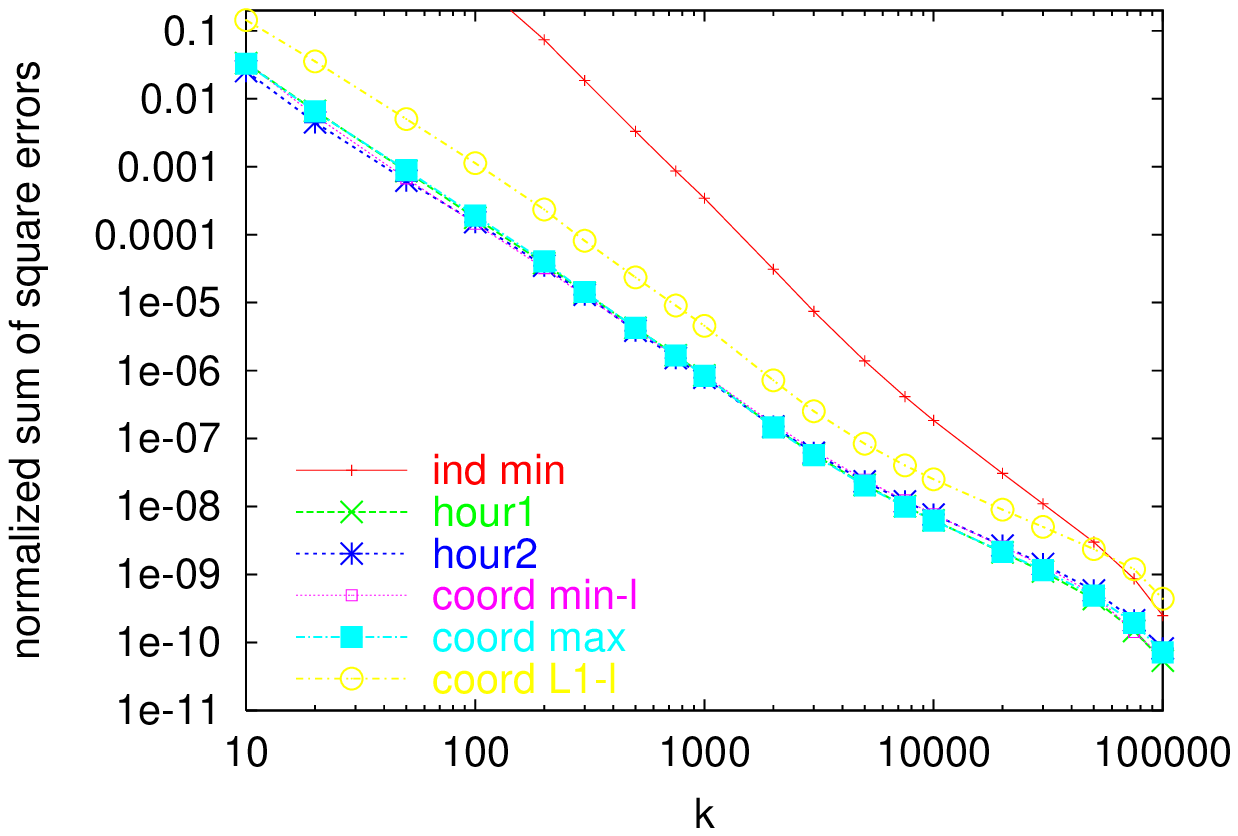,width=0.45\textwidth} \\
\epsfig{figure=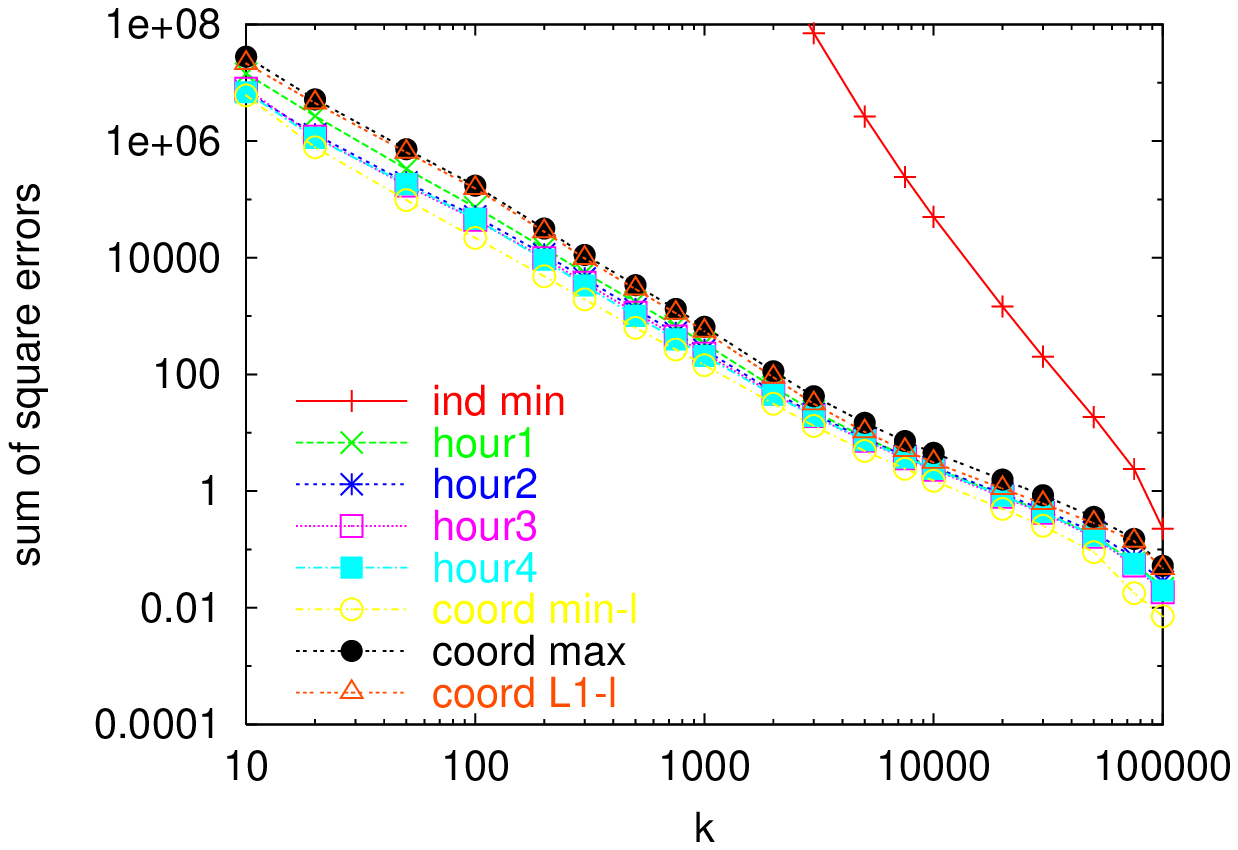,width=0.45\textwidth} &
\epsfig{figure=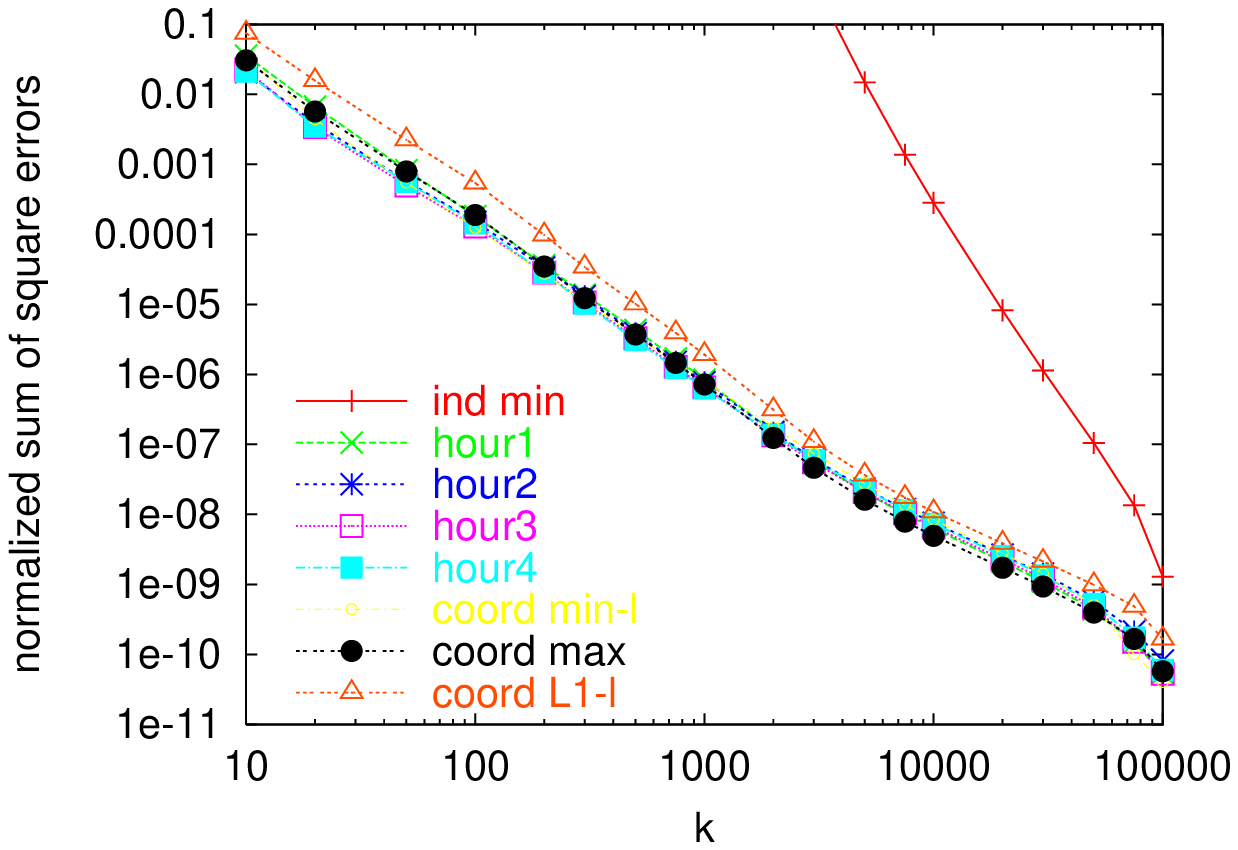,width=0.45\textwidth} \\
\epsfig{figure=multi_code/resultsL1/L1_4tuple_bytes_nfcapd20080801.4tuple.time12_r25.eps,width=0.45\textwidth} &
\epsfig{figure=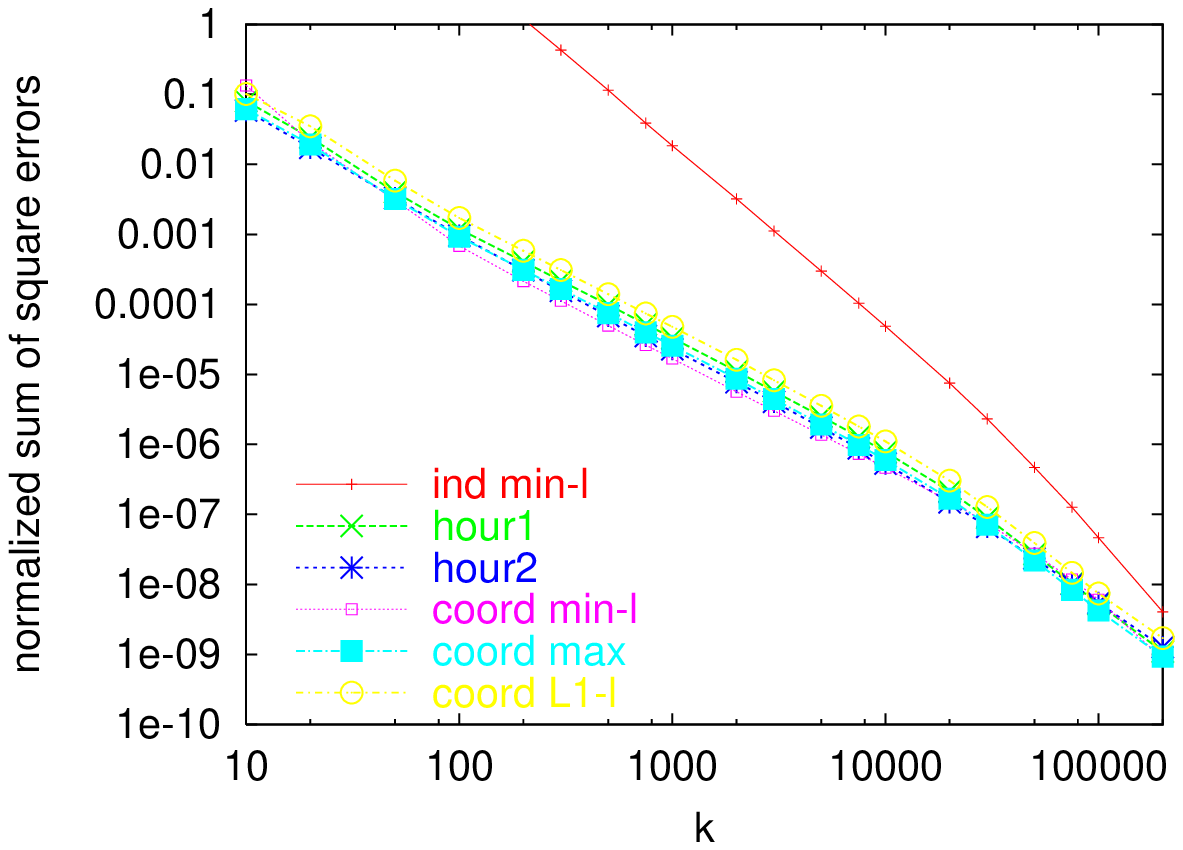,width=0.45\textwidth} \\
\epsfig{figure=multi_code/resultsL1/L1_4tuple_bytes_nfcapd20080801.4tuple.time1234_r10.eps,width=0.45\textwidth} &
\epsfig{figure=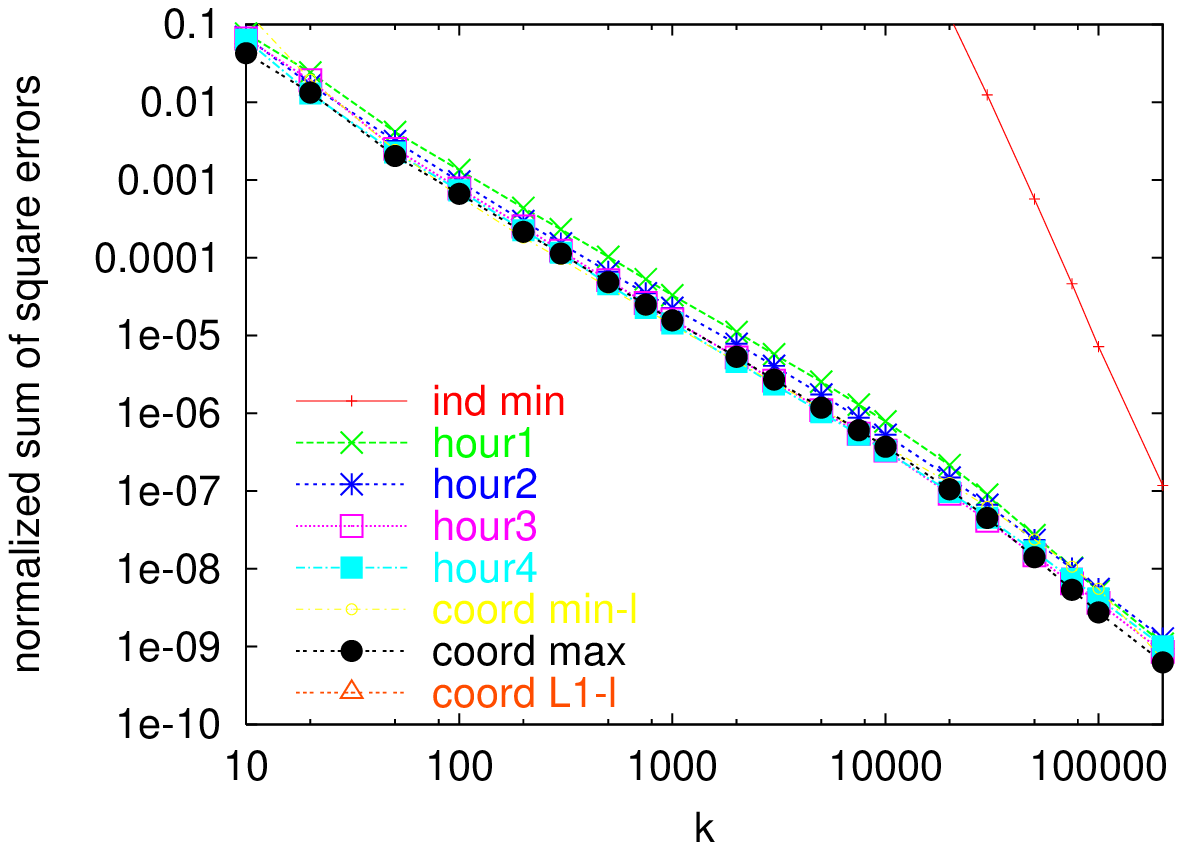,width=0.45\textwidth}
\end{tabular}
}
\caption{IP dataset2, sum of square errors. Left: absolute, Right: normalized.  Top to bottom: key$=$destIP  weight$=$bytes hours$=\{1,2\}$; key$=$destIP  weight$=$bytes hours$=\{1,2,3,4\}$; key$=$4tuple  weight$=$bytes hours$=\{1,2\}$; key$=$4tuple  weight$=$bytes hours$=\{1,2,3,4\}$.}
\label{tperiods2:fig}
\end{figure*}

\begin{figure*}[htbp]
\centerline{\begin{tabular}{cc}
\epsfig{figure=multi_code/resultsL1/L1_netflix2005_movie_req_p_month_m1_2_r200.eps,width=0.45\textwidth} &
\epsfig{figure=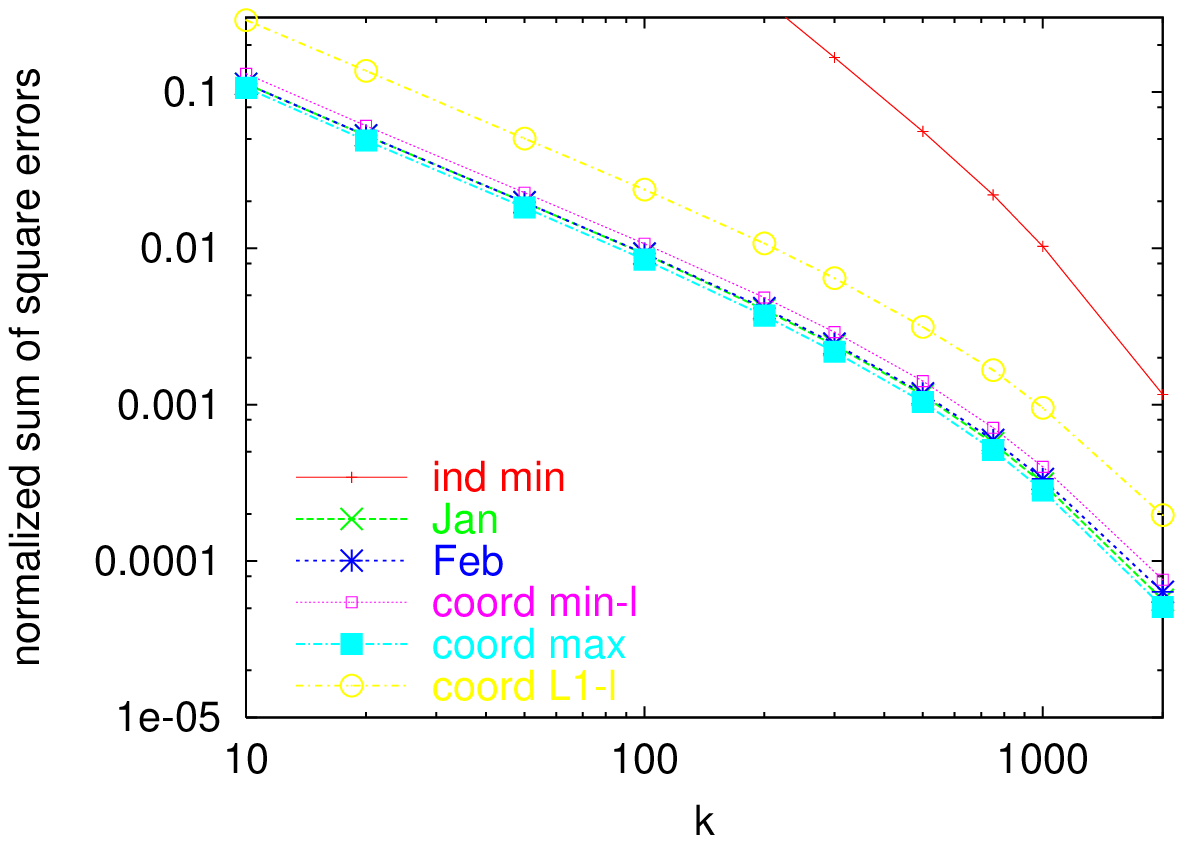,width=0.45\textwidth} \\
\epsfig{figure=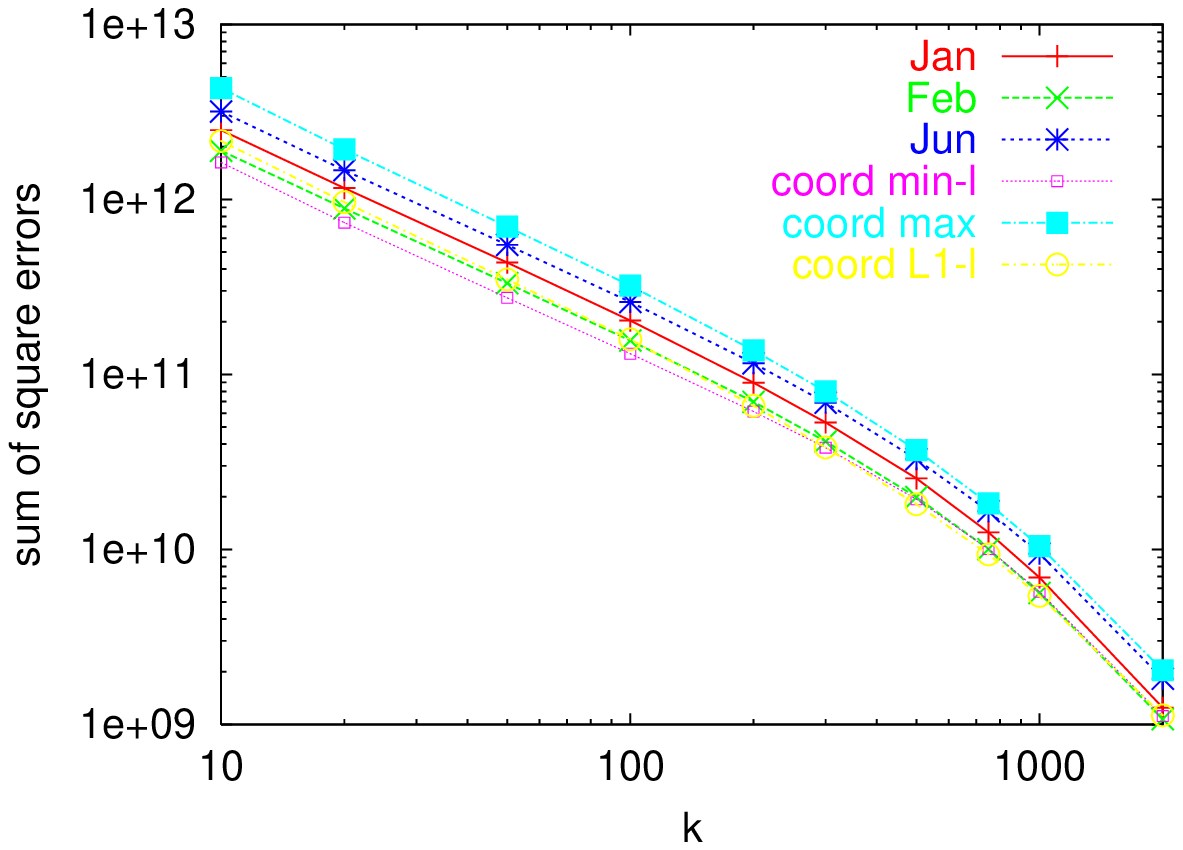,width=0.45\textwidth} &
\epsfig{figure=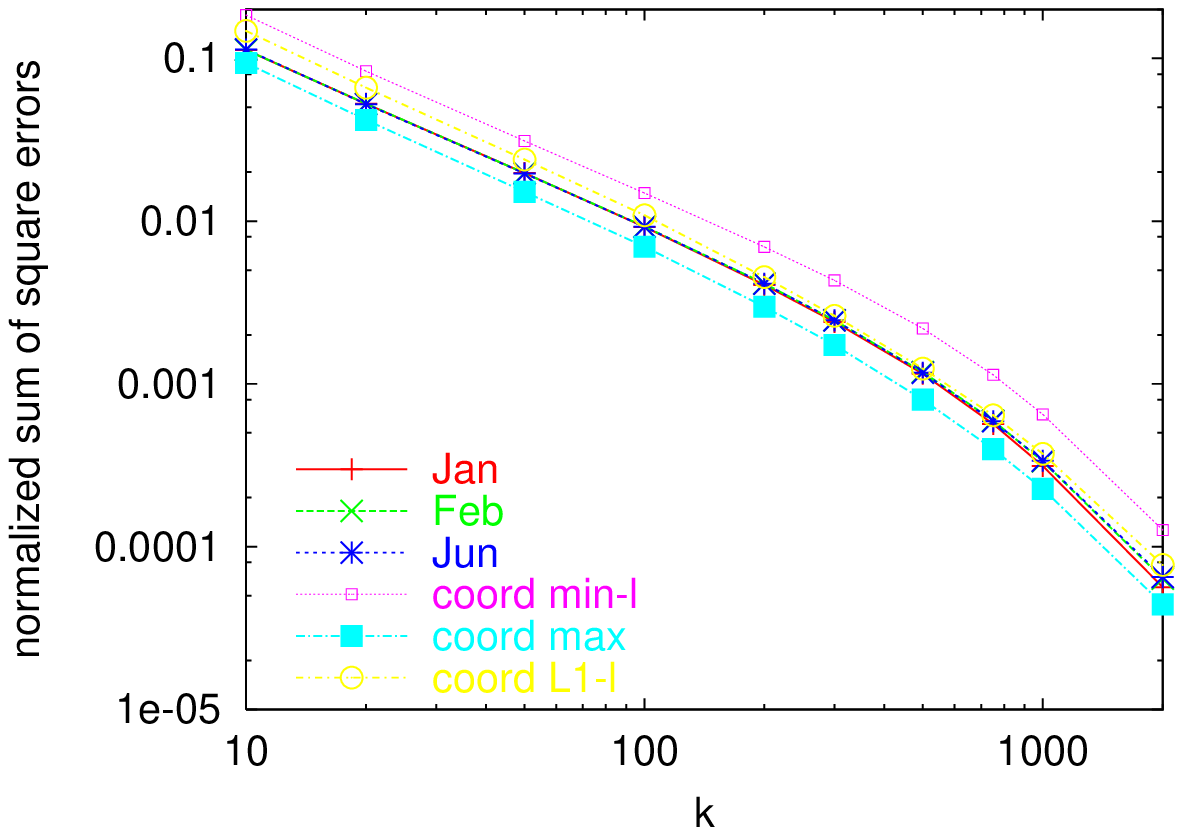,width=0.45\textwidth} \\
\epsfig{figure=multi_code/resultsL1/L1_netflix2005_movie_req_p_month_r200.eps,width=0.45\textwidth} &
\epsfig{figure=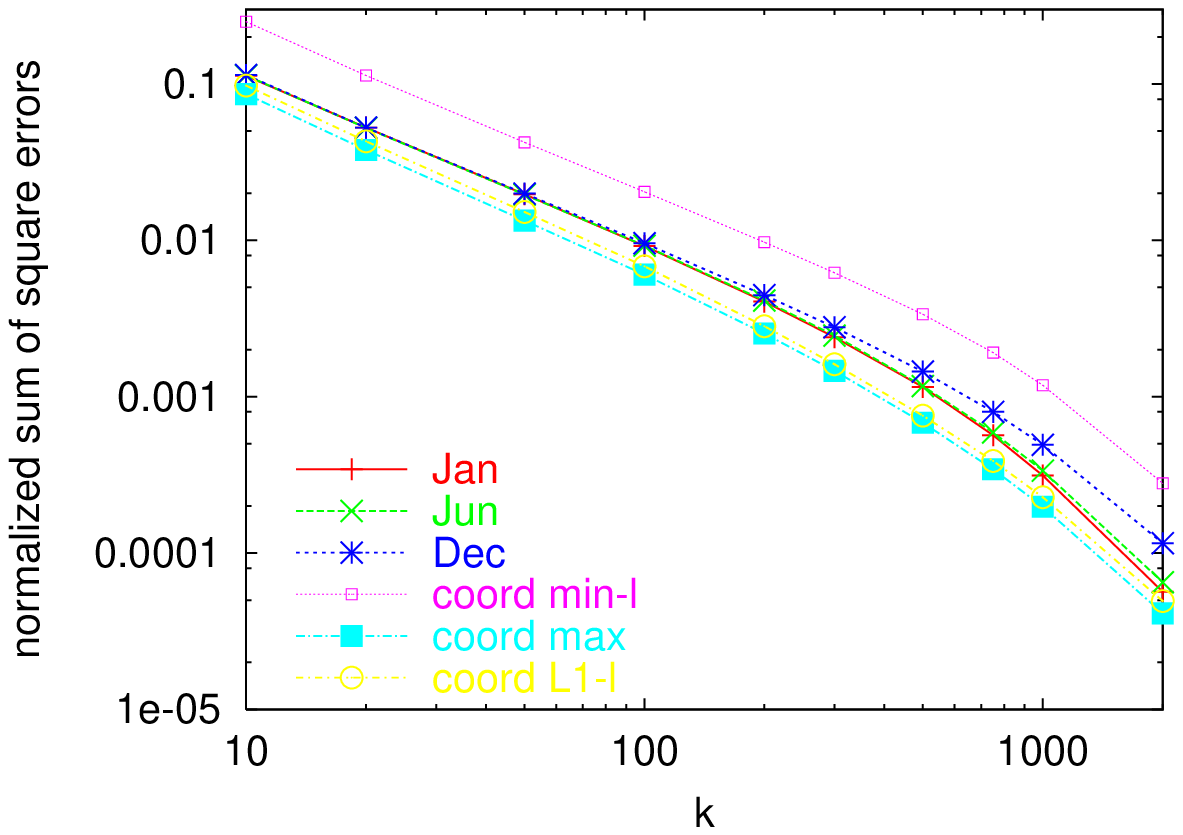,width=0.45\textwidth}
\end{tabular}
}
\caption{Netflix data set $\cR=\{1,2\}$ (top), $\cR=\{1,\ldots,6\}$ (middle), $\cR=\{1,\ldots,12\}$ (bottom).  $\SV$ (left) and $\nSV$ (right).}
\label{netflix_sv:fig}
\end{figure*}

\begin{figure*}[htbp]
\centerline{\begin{tabular}{ccc}
\epsfig{figure=multi_code/resultsL1/L1_stocks_high_200810_2_r50,width=0.32\textwidth} &
\epsfig{figure=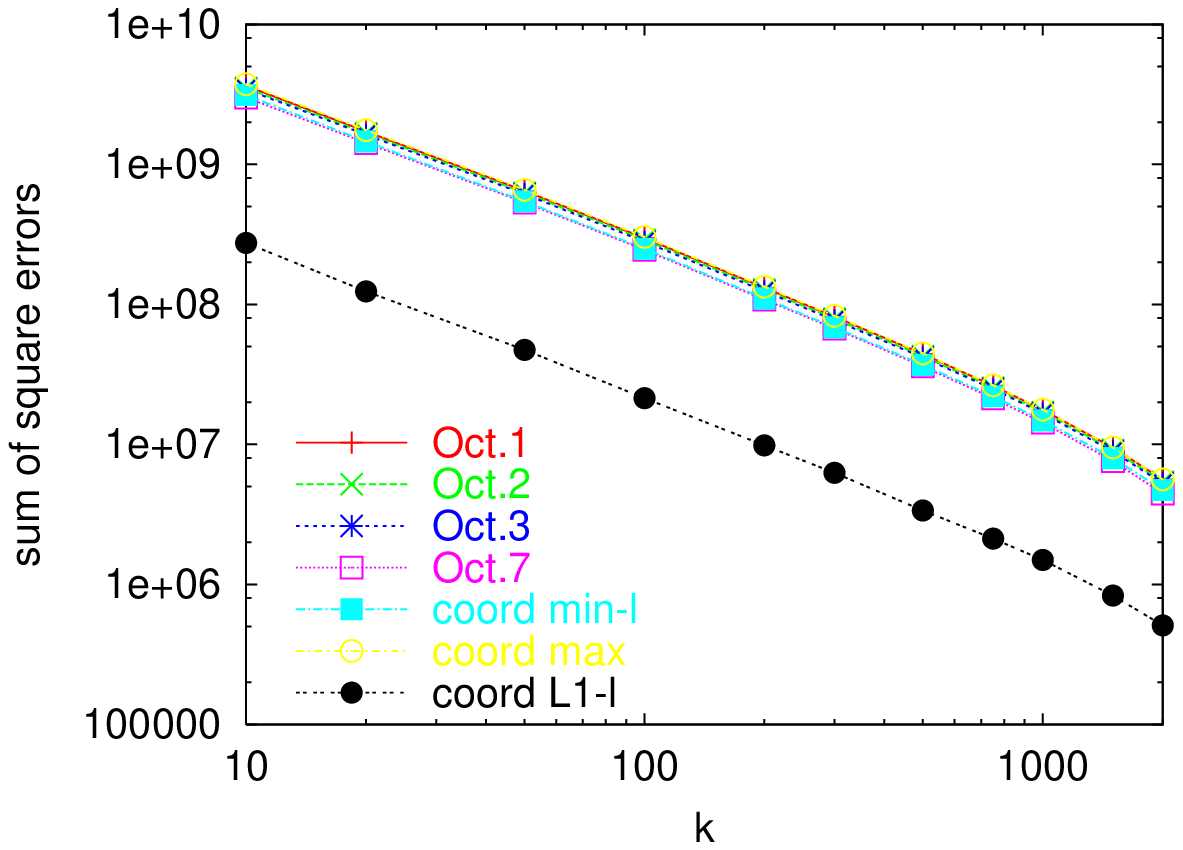,width=0.32\textwidth} &
\epsfig{figure=multi_code/resultsL1/L1_stocks_high_200810_23_r50,width=0.32\textwidth} \\
\epsfig{figure=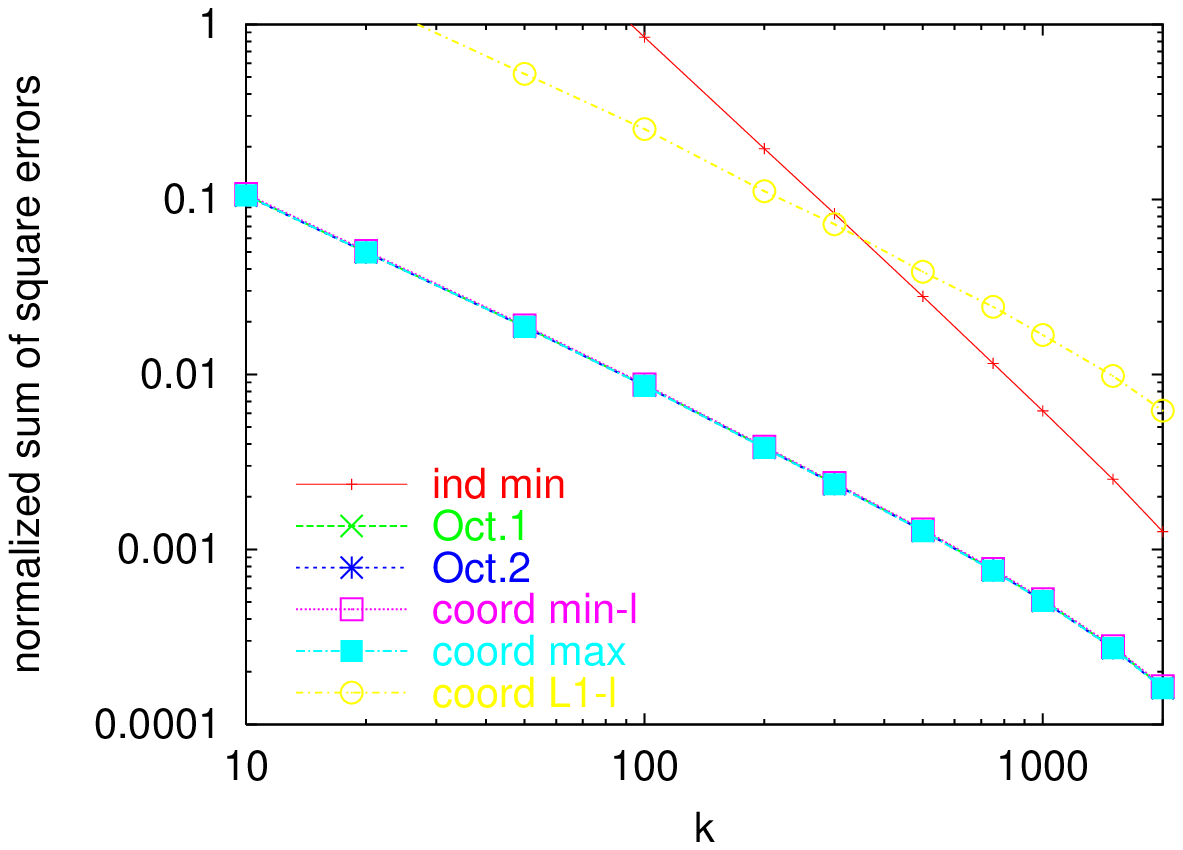,width=0.32\textwidth} &
\epsfig{figure=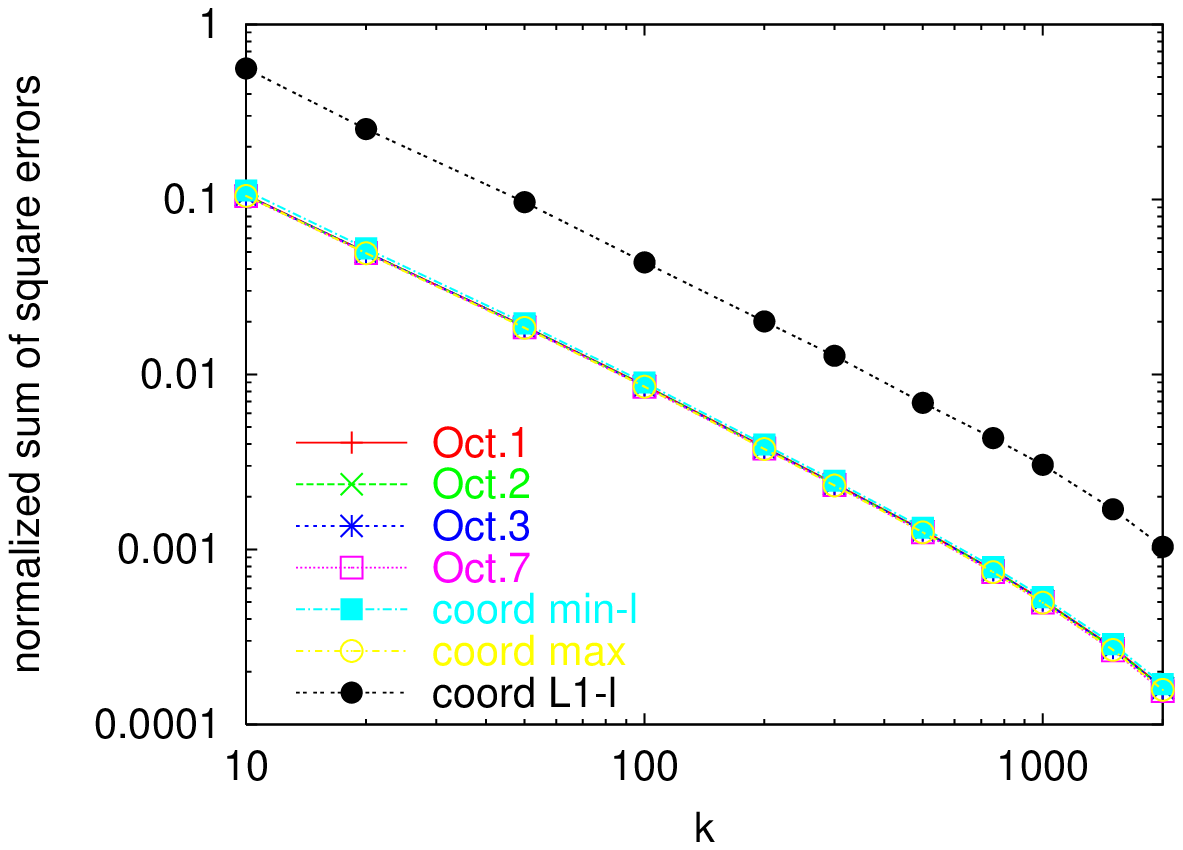,width=0.32\textwidth} &
\epsfig{figure=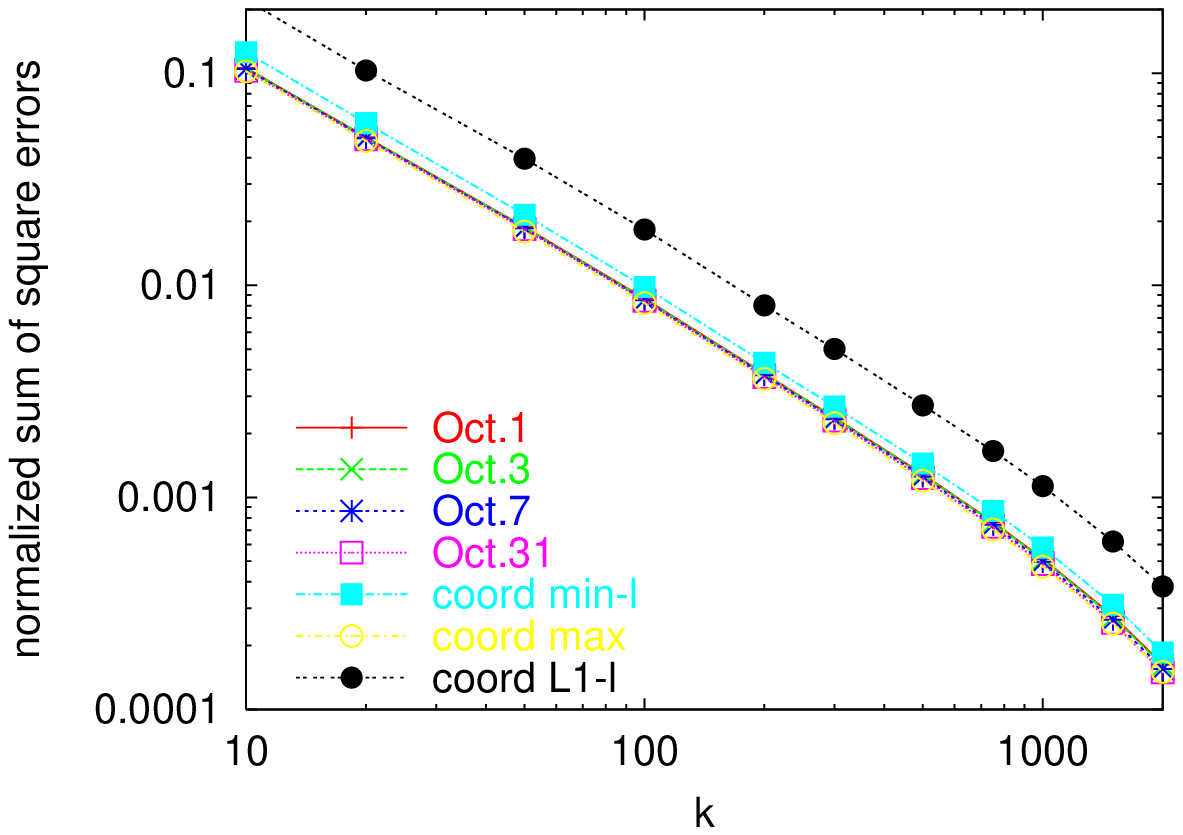,width=0.32\textwidth} \\
\epsfig{figure=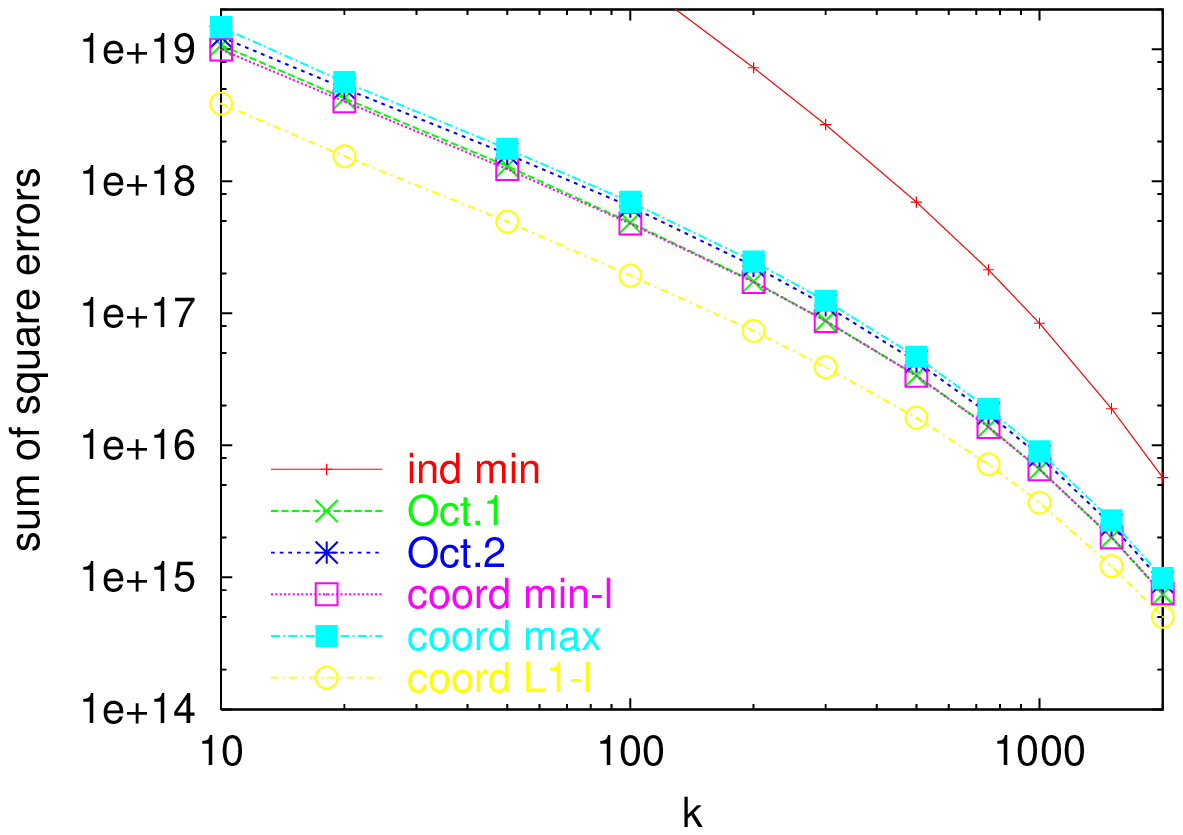,width=0.32\textwidth} &
\epsfig{figure=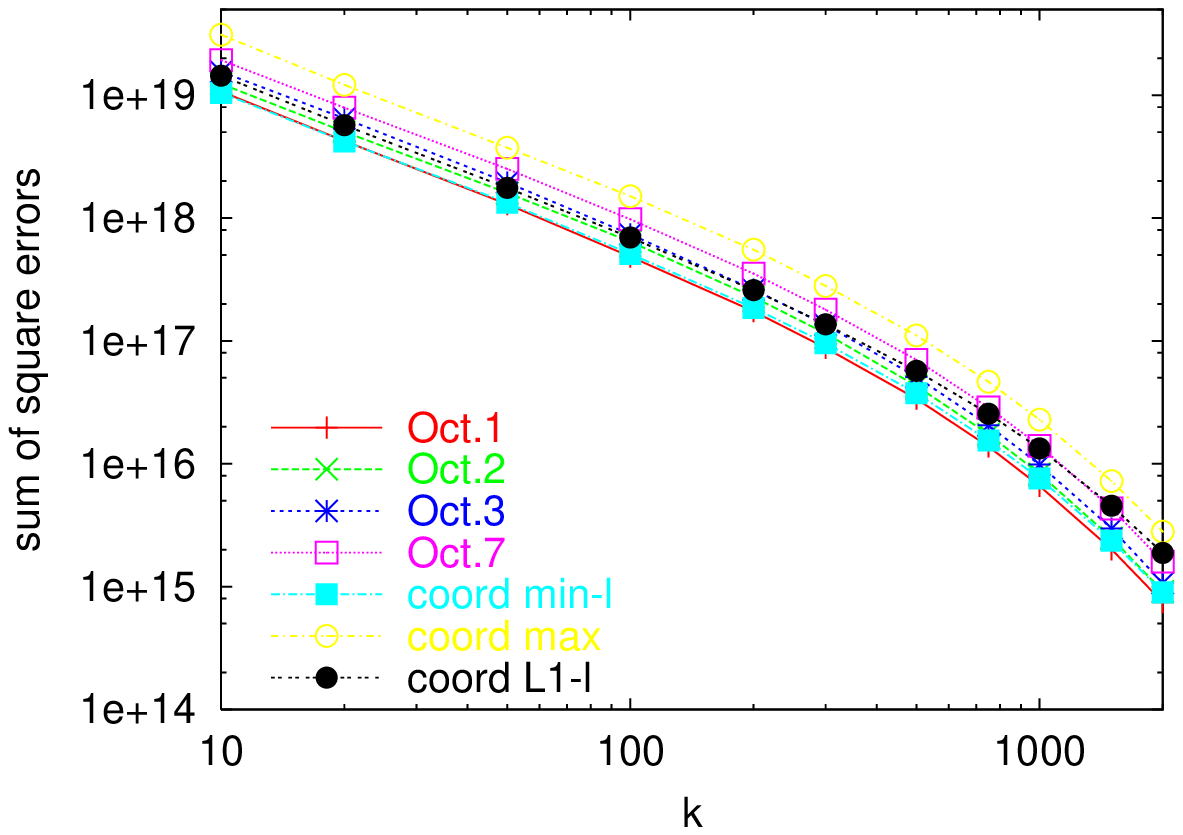,width=0.32\textwidth} &
\epsfig{figure=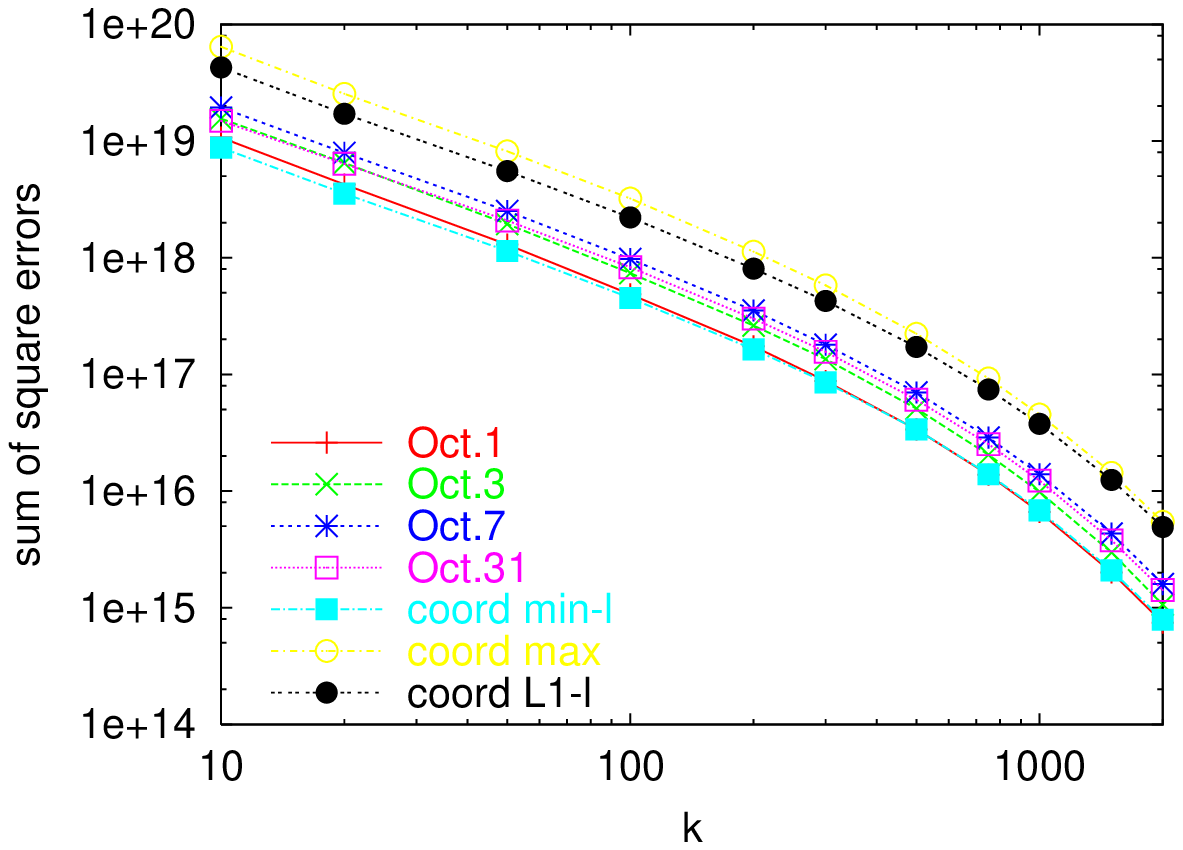,width=0.32\textwidth} \\
\epsfig{figure=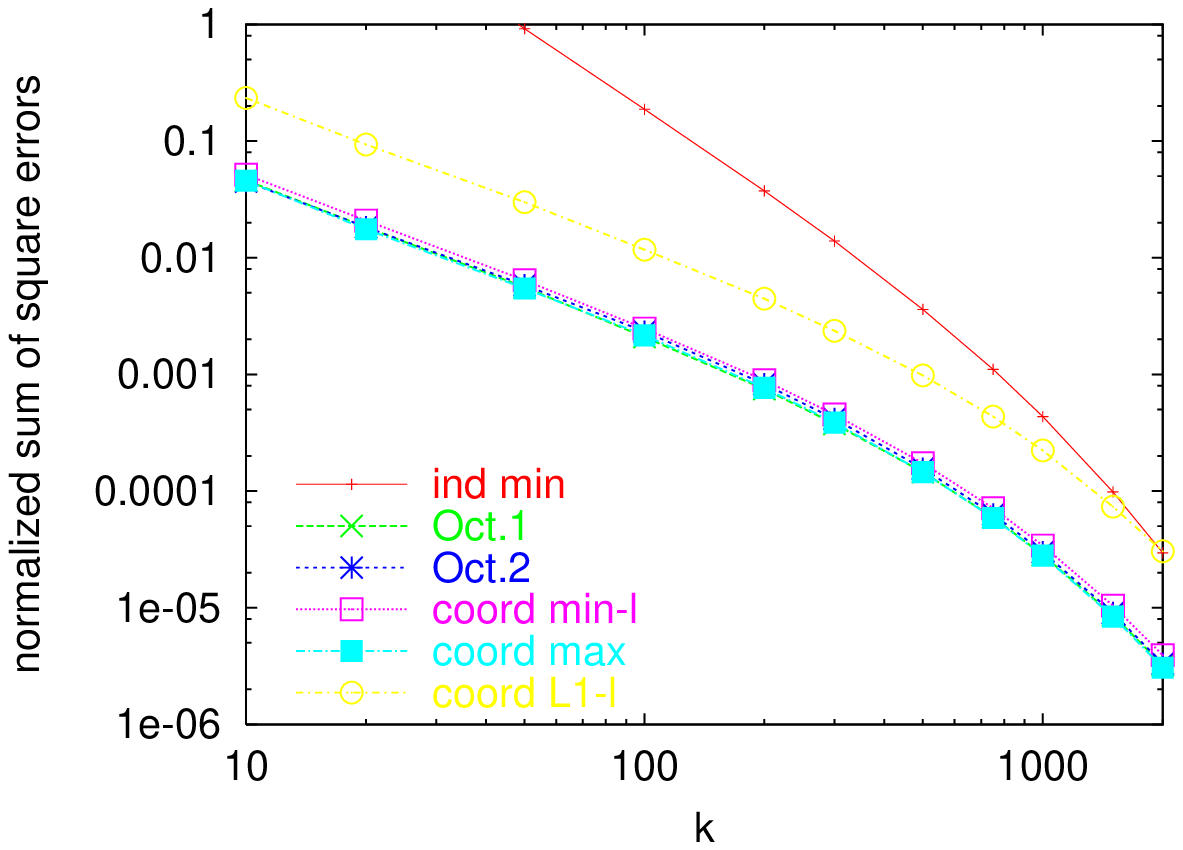,width=0.32\textwidth} &
\epsfig{figure=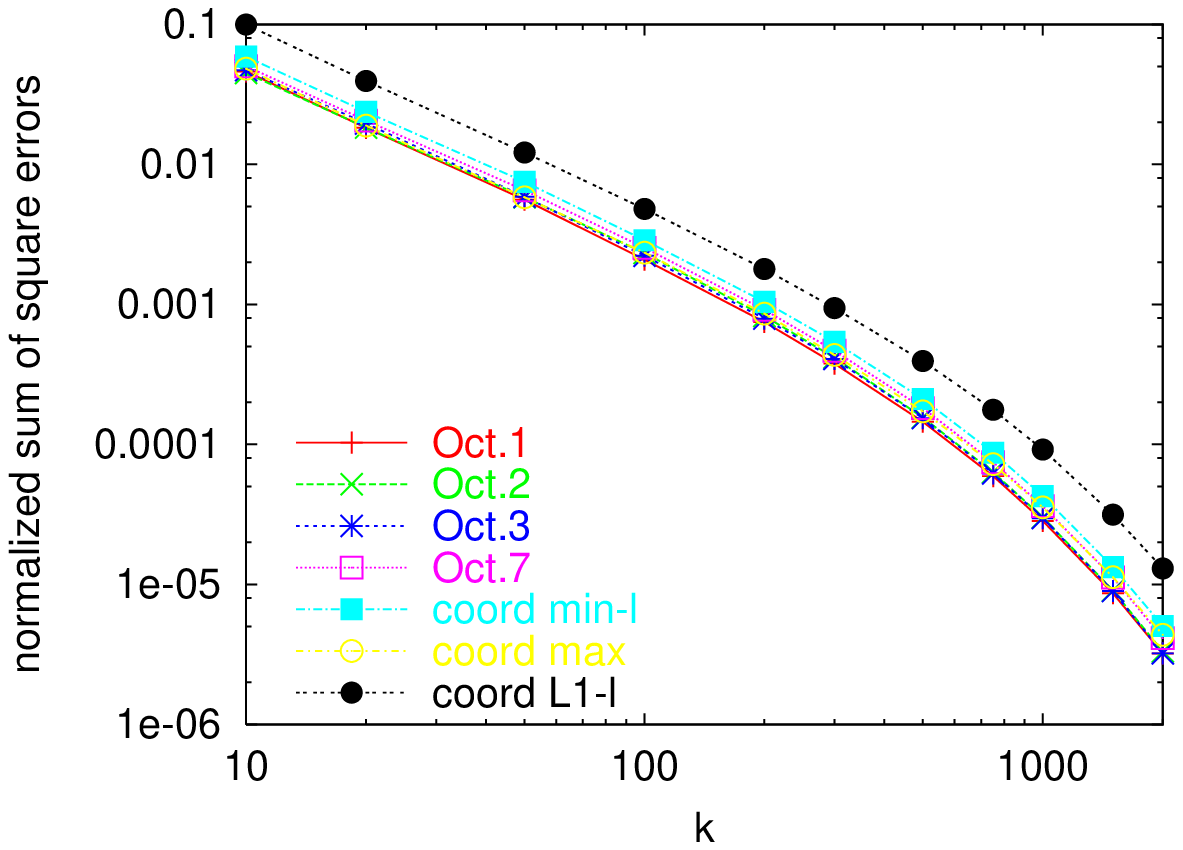,width=0.32\textwidth} &
\epsfig{figure=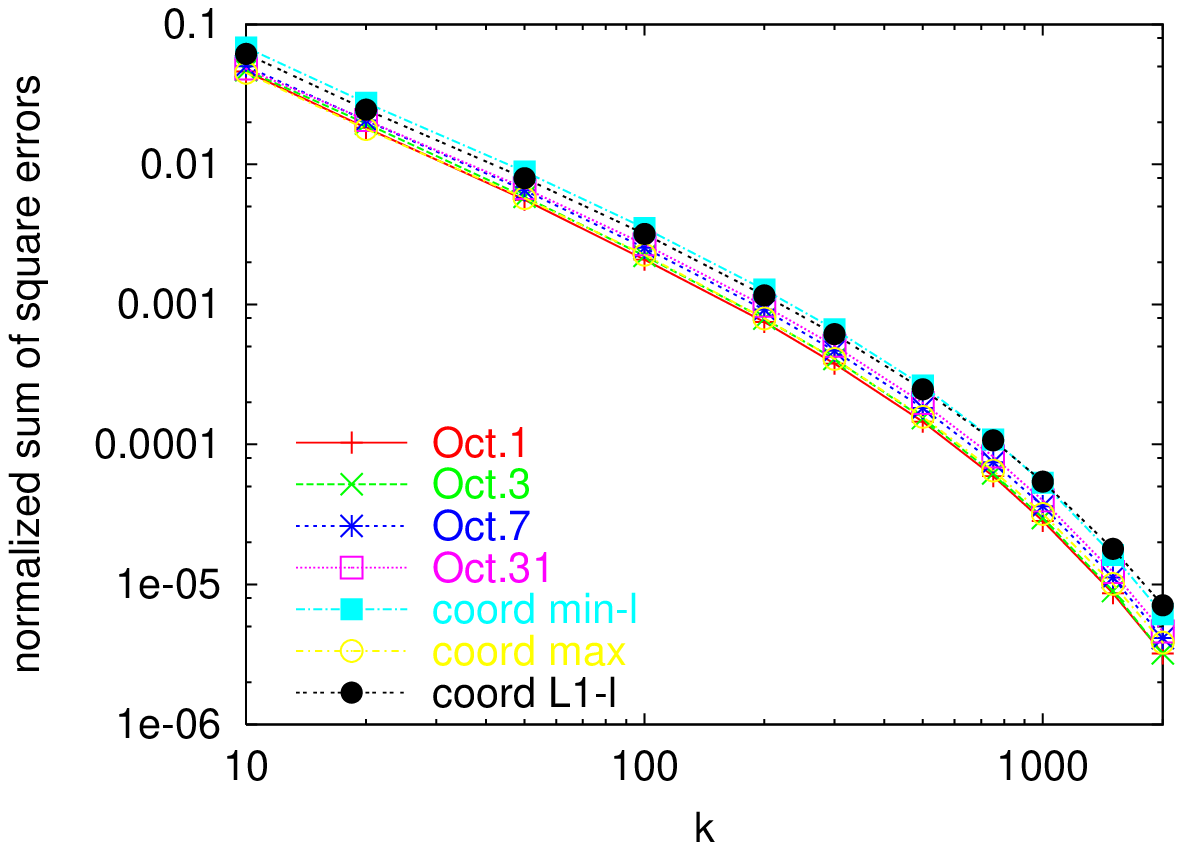,width=0.32\textwidth}
\end{tabular}
}
\caption{Stock dataset.  Left: $\cR=\{1,2\}$ (October 1-2, 2008), Middle: $\cR=\{1,\ldots,5\}$ (trading days in October 1-7, 2008), Right: $\cR=\{1,\ldots,23\}$ (all trading days in October, 2008).  Upper two rows are $\SV$ and $\nSV$ with ``high'' weights.  lower two rows are ``volume'' weights.}
\label{stock_sv:fig}
\end{figure*}
} 

\notinproc{Figures~\ref{tperiods1:fig}, \ref{tperiods2:fig}, \ref{netflix_sv:fig}
and \ref{stock_sv:fig} show}\onlyinproc{Across all datasets (Figure~\ref{selected_sv:fig} shows selected plots)},
$\SV[a_\ell^{(\min \cR)}]$, $\SV[a_\ell^{(\max \cR)}]$, and $\SV[a_\ell^{(L_1
\cR)}]$ and $\SV[a^{(b)}]$ for $b\in \cR$ are within an order of
magnitude.
On our
datasets,
$\nSV[a^{(b)}]$ and $\nSV[a_\ell^{(\max \cR)}]$
are clustered together with
$k\nSV\ll 1$ (and decreases with $k$) (theory says $(k-2)\nSV \leq 1$.)
We also observed that  $\nSV[a_l^{(L_1 \cR)}]$ and $\nSV[a_l^{(\min \cR)}]$
are typically close to $\nSV[a^{(b)}]$.
  We observe the empirical relations
{\small $\SV[a_\ell^{(\min \cR)}]  < \SV[a_\ell^{(\max \cR)}]$ (with
larger gap when the $L_1$ difference is very small), $
\SV[a_\ell^{(L_1 \cR)}] < \SV[a_\ell^{(\max \cR)}]$, and
$\SV[a_\ell^{(\min \cR)}] < \min_{b\in\cR} \SV[a^{(b)}]$.}
Empirically, the variance of
our multi-assignment estimators with respect to single-assignment weights
is significantly lower than
the worst-case analytic bounds in Section~\ref{varbounds:sec} (Lemma~\ref{disvarmin} and~\ref{disvarL1}).
 For normalized (relative) variances, we observe the ``reversed'' relations
{\small $\nSV[a_\ell^{(\min \cR)}]> \nSV[a_\ell^{(\max \cR)}]$,
$\nSV[a_\ell^{(L_1 \cR)}]> \nSV[a_\ell^{(\max \cR)}]$, and
$\nSV[a_\ell^{(\min \cR)}] > \max_{b\in\cR} \nSV[a^{(b)}]$} which
are explained by smaller normalization factors for $w^{(\min \cR)}$
and $w^{(L_1 \cR)}$.

\smallskip
\noindent
{\bf S-set versus L-set estimators.}
\notinproc{Figure~\ref{sl_ratios:fig}  quantifies
the advantage  of
the stronger l-set estimators over the s-set estimators
for coordinated sketches.  The advantage highly varies between datasets:}
\onlyinproc{To understand the advantage  of
the stronger l-set estimators over the s-set estimators, we studied the
ratios $\SV[a_s^{(\min \cR)}]/\SV[a_l^{(\min \cR)}]$ and
$\SV[a_s^{(L_1 \cR)}]/\SV[a_l^{(L_1 \cR)}]$  as a function of $k$. The advantage highly varies between datasets:}
15\%-80\% for the Netflix dataset, 0\%-9\% for IP dataset1,  0\%-20\% for IP dataset2, and 0\%-300\% on the Stocks data set.

\notinproc{
\begin{figure*}
\centerline{
\begin{tabular}{ccc}
\epsfig{figure=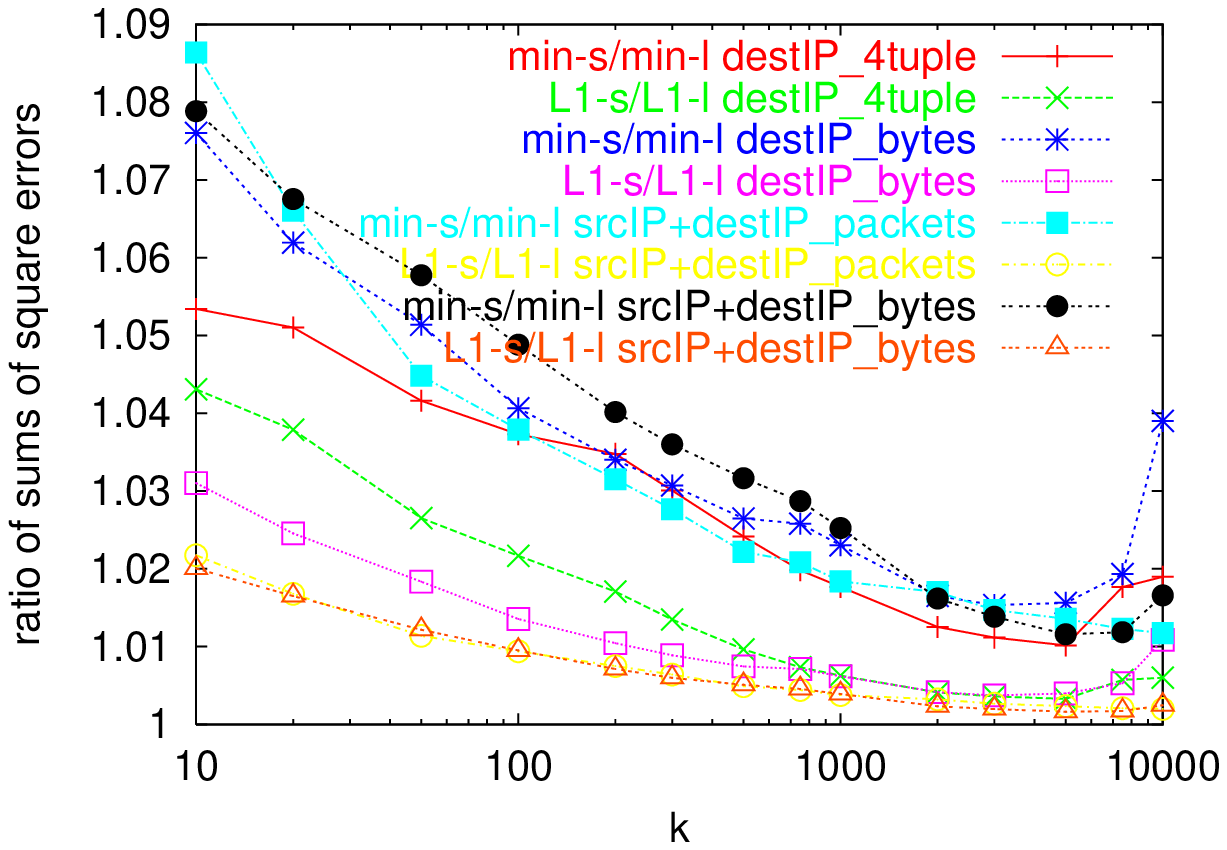,width=0.32\textwidth} &
\epsfig{figure=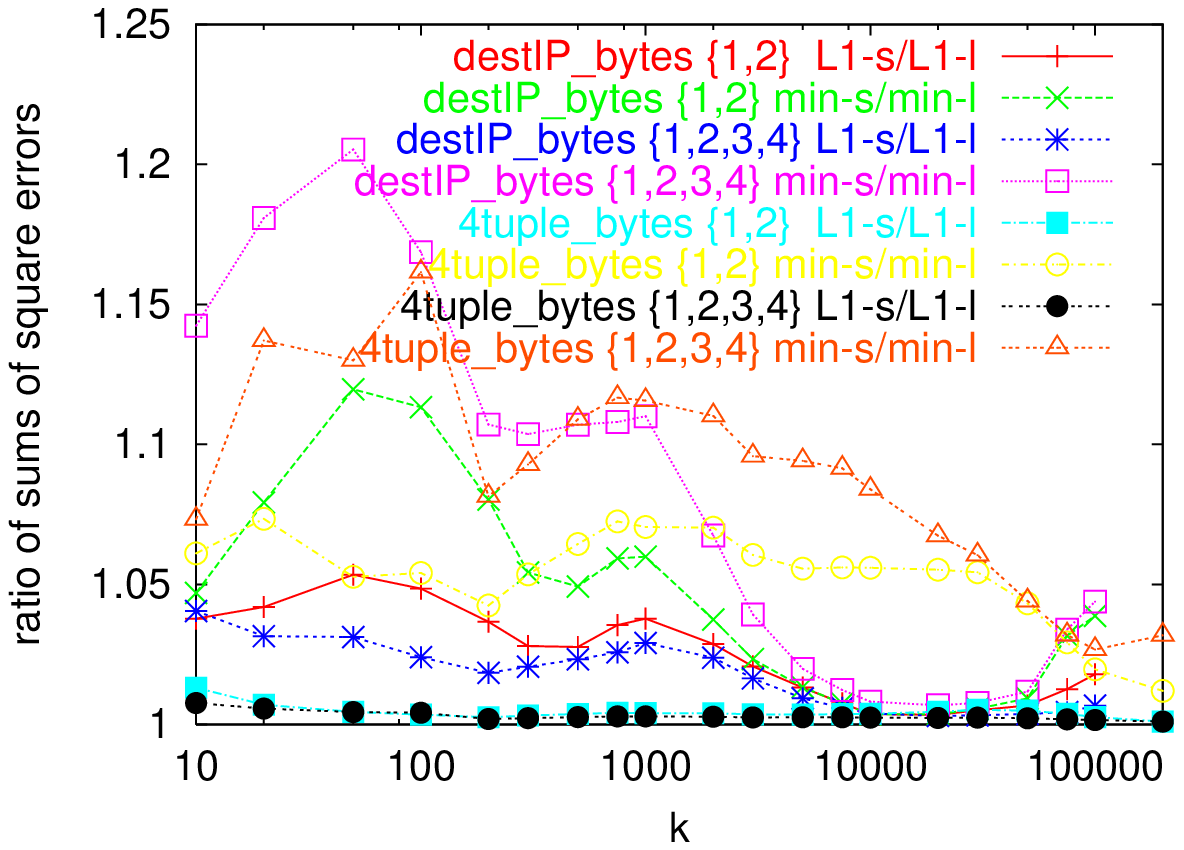,width=0.32\textwidth} &
\epsfig{figure=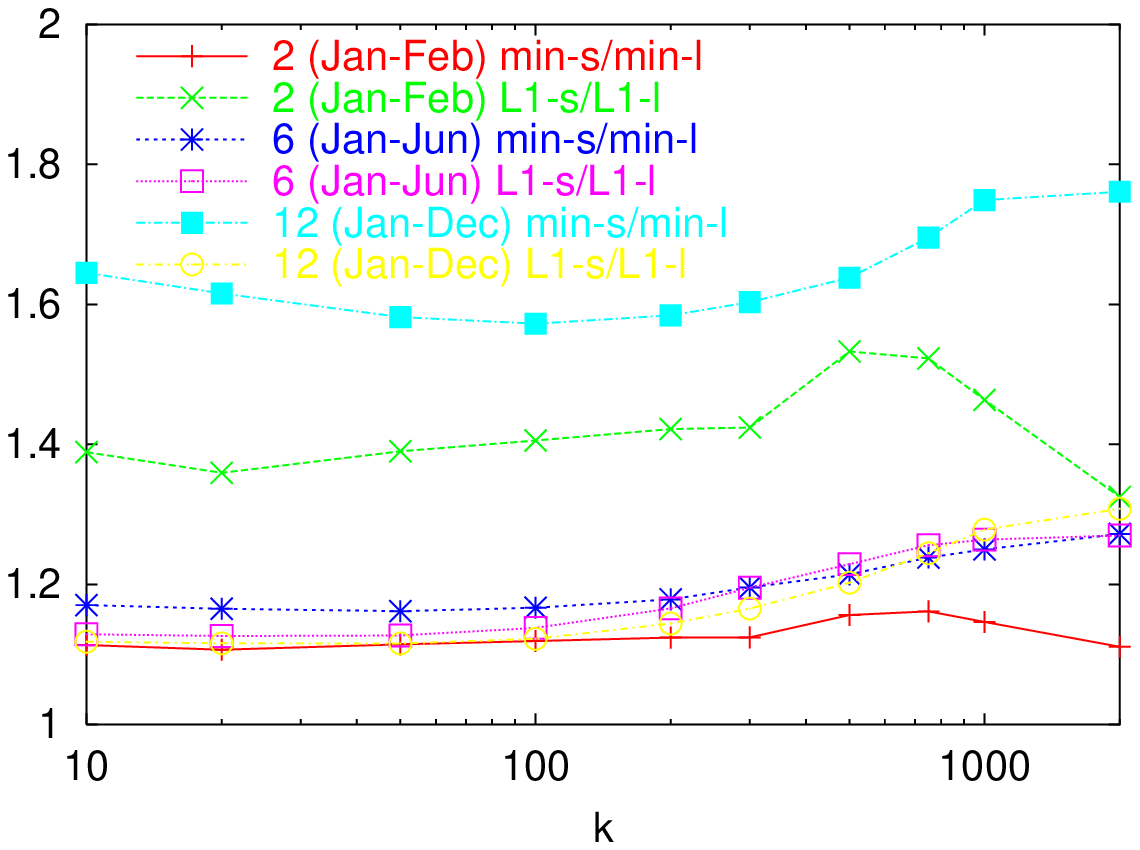,width=0.32\textwidth} \\
\epsfig{figure=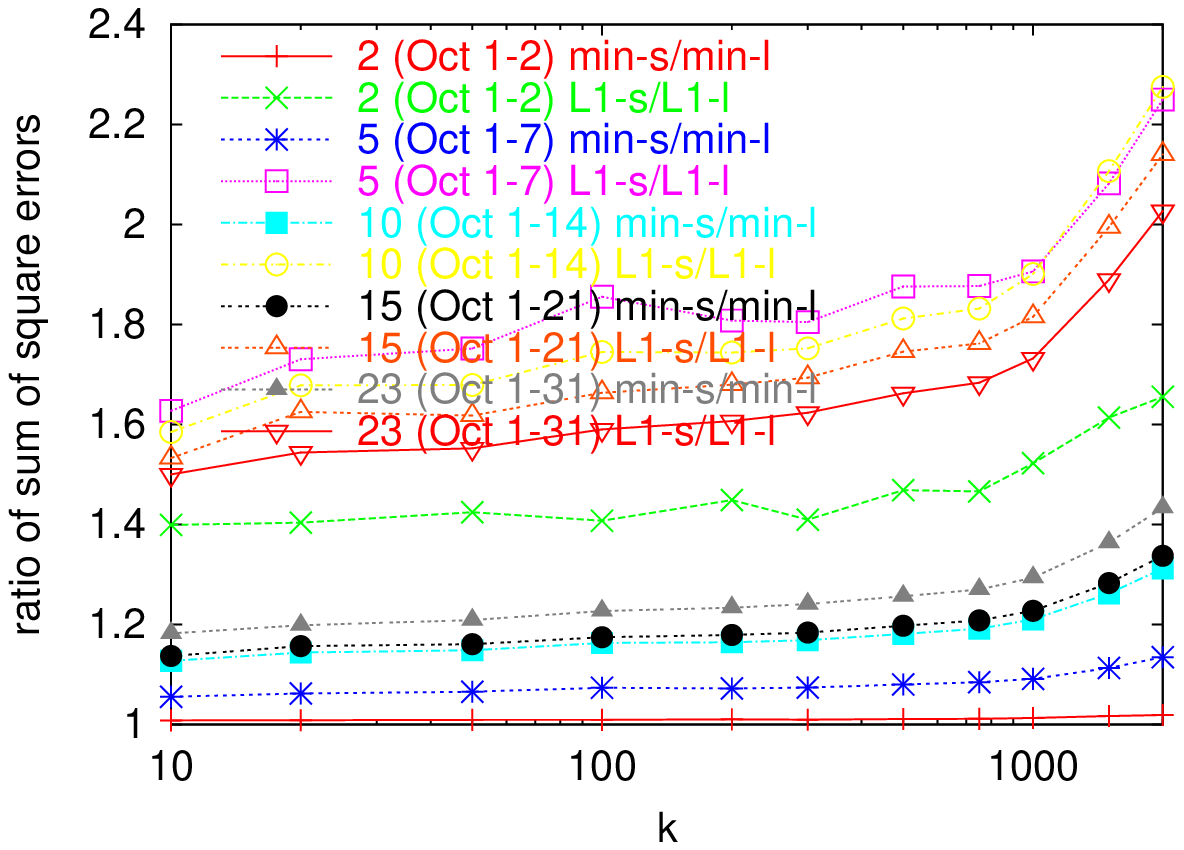,width=0.32\textwidth} &
\epsfig{figure=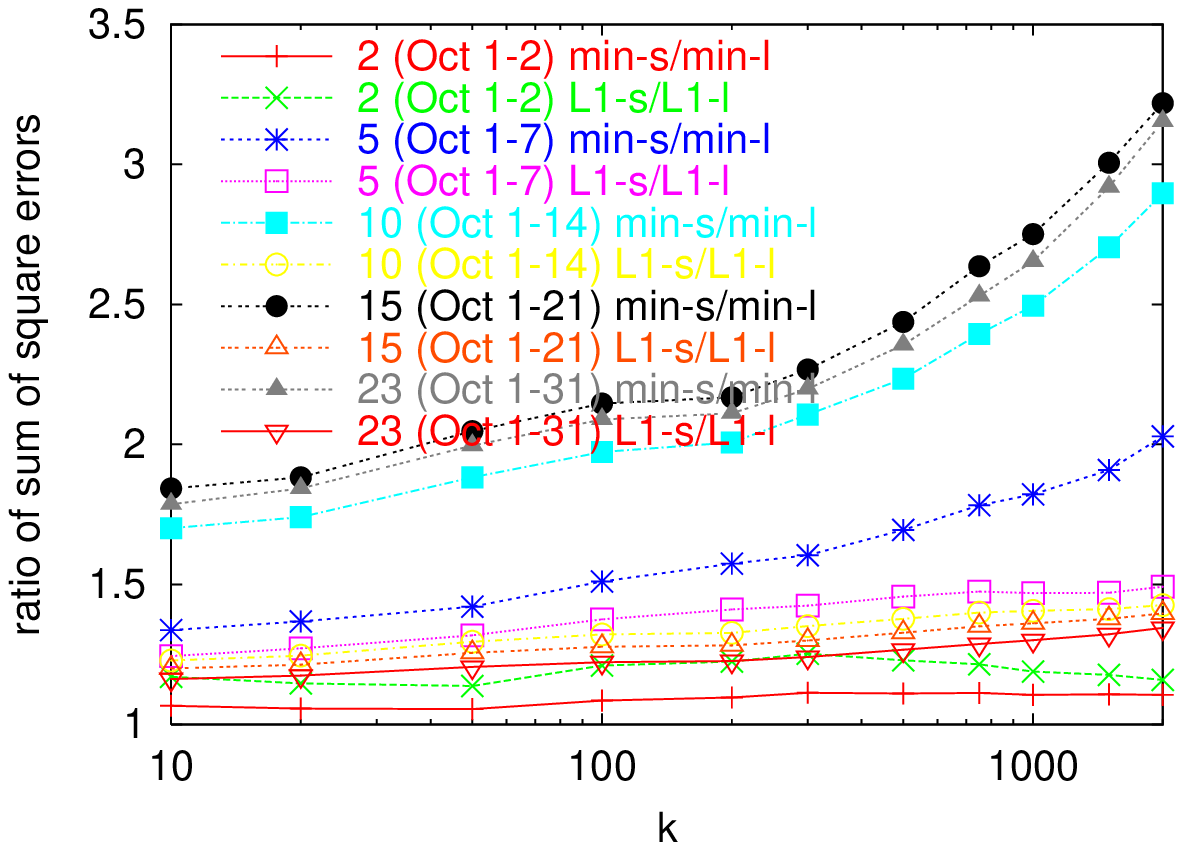,width=0.32\textwidth} &
\end{tabular}
}
\caption{Top:  IP dataset1 (left),  IP dataset2 (middle),  Netflix data set (right). Bottom: stocks dataset high-values (left),  stocks dataset volume values (middle).  $\SV$ ratio of s-set and l-set estimators for $w^{(\min \cR)}$ and $w^{(L_1 \cR)}$, $\SV[a_s^{(\min \cR)}]/\SV[a_l^{(\min \cR)}]$ and
$\SV[a_s^{(L_1 \cR)}]/\SV[a_l^{(L_1 \cR)}]$.}
\label{sl_ratios:fig}
\end{figure*}
}


\subsection{Colocated data}

We computed shared-seed coordinated and independent
sketches and show results for \IPPS\ ranks
(see Section~\ref{prelim:sec}).
Results for \EXP\ ranks were similar.

 We consider the following $w^{(b)}$-weights estimators.
 $a_c^{(b)}$: the shared-seed coordinated sketches inclusive estimator
 (Section~\ref{coloc:sec}, Eq.~(\ref{colocssc})).
  $a_i^{(b)}$: the independent sketches inclusive estimator in (Section~\ref{coloc:sec}, 
Eq.~(\ref{colocind})).  $a_p^{(b)}$: the plain bottom-$k$ sketch
 \rc\ estimator (\cite{DLT:jacm07} for \IPPS\ ranks).  Among all keys of
the combined sketch this
 estimator uses only the  keys which are part of the bottom-$k$ sketch of $b$.

We study the benefit of our inclusive estimators by comparing
them to plain estimators.
Since plain estimators can not be used effectively for multiple
assignment aggregates, we focus on single-assignment aggregates.

\noindent
{\bf Inclusive versus plain estimators.}
\looseness=-1 The plain estimators we used are
optimal for individual bottom-$k$ sketches and the benefit of
inclusive estimators comes from utilizing keys that were sampled
for ``other'' weight assignments.
We computed the ratios $$\SV[a_i^{(b)}]/\SV[a_p^{(b)}] \; \mbox{and} \;
\SV[a_c^{(b)}]/\SV[a_p^{(b)}]$$ as a function of $k$.
\notinproc{As Figures~\ref{ratiosigVdata1:fig}, \ref{ratiosigV2008080102:fig}
 and~\ref{ratiosigVstocks20081001:fig} show, the}
\onlyinproc{These} ratios vary between 0.05 to 0.9 on our datasets and shows
a significant benefit for inclusive estimators
\onlyinproc{(see Figure~\ref{ratiosigVdata1_selected:fig}).}
Our inclusive estimators are considerably more accurate
with both coordinated and independent sketches.
With independent sketches the benefit of the
inclusive estimators is larger than with coordinate sketches
since
 the independent sketches contain many more distinct keys
for a given $k$.

 \ignore{
 Combined estimators facilitate estimators
for multiple-assignment aggregates. Unbiased estimators for $\max$ can not
be obtained at all from single-assignment samples and even biased
estimators may have large variance (intuitively,
because keys with positive max but zero weight with respect to the
particular assignment have zero inclusion probability).

 We compare combined estimators for a particular
weight assignment to plain estimators on this assignment.  In
this case, the benefit of combined estimators comes from
utilizing sampled keys taken for ``other'' assignments.
}
\onlyinproc{
\begin{figure}[htbp]
\centerline{\begin{tabular}{cc}
\epsfig{figure=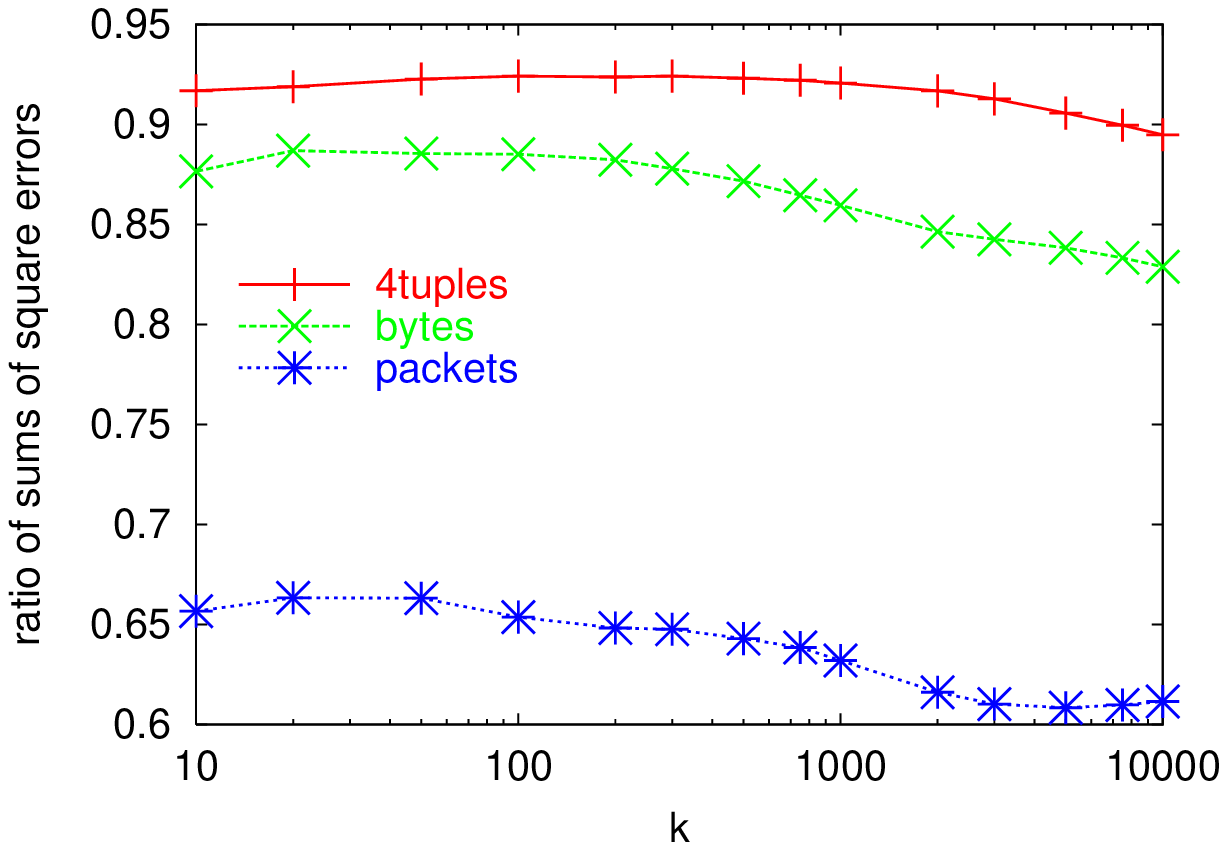,width=0.22\textwidth} &
\epsfig{figure=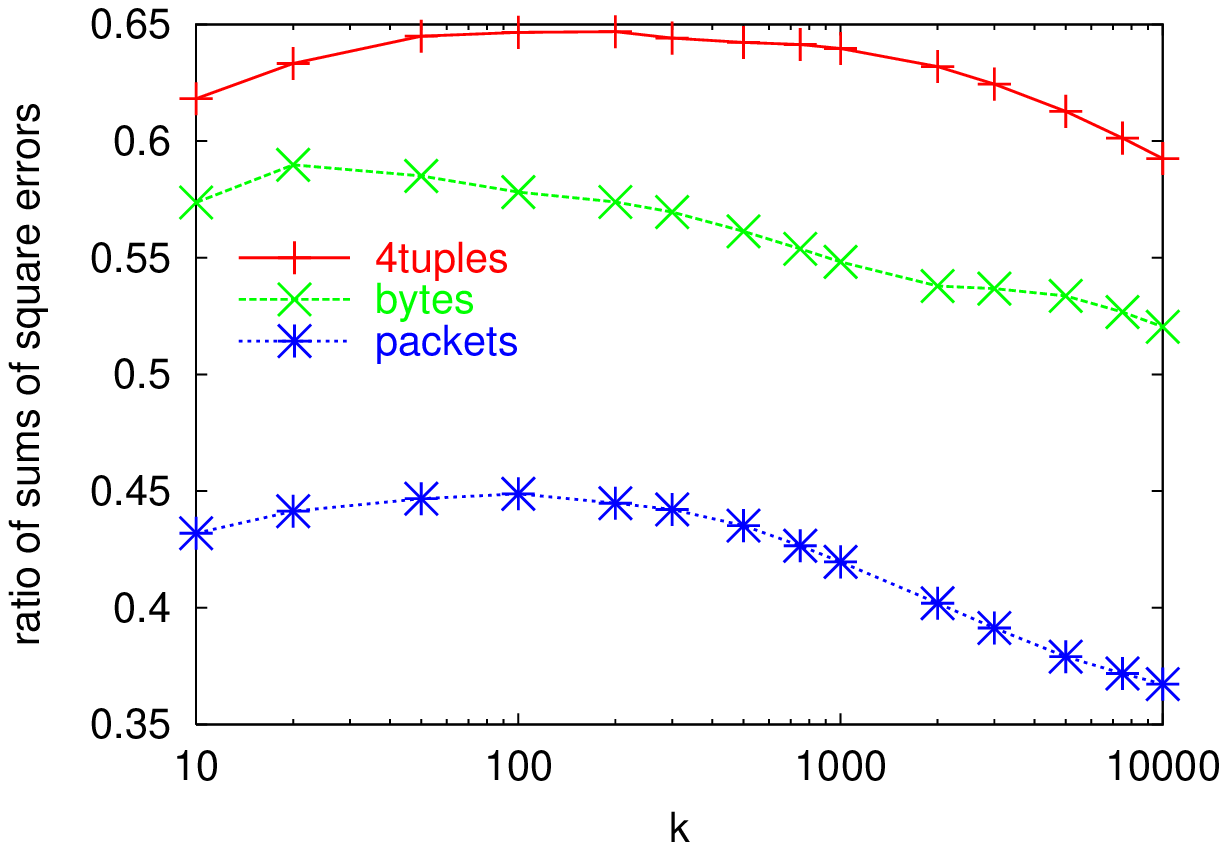,width=0.22\textwidth}
\end{tabular}
}
\caption{Inclusive versus plain estimators. IP dataset1,  key$=$4tuple.
Left:  $\SV[a_c^{(b)}]/\SV[a_p^{(b)}]$ (coordinated sketches).
Right: $\SV[a_i^{(b)}]/\SV[a_p^{(b)}]$ (independent sketches).
\vspace*{-0.3cm}
\label{ratiosigVdata1_selected:fig}
}
\end{figure}
  }

\notinproc{

\begin{figure*}[htbp]
\centerline{\begin{tabular}{cc}
\epsfig{figure=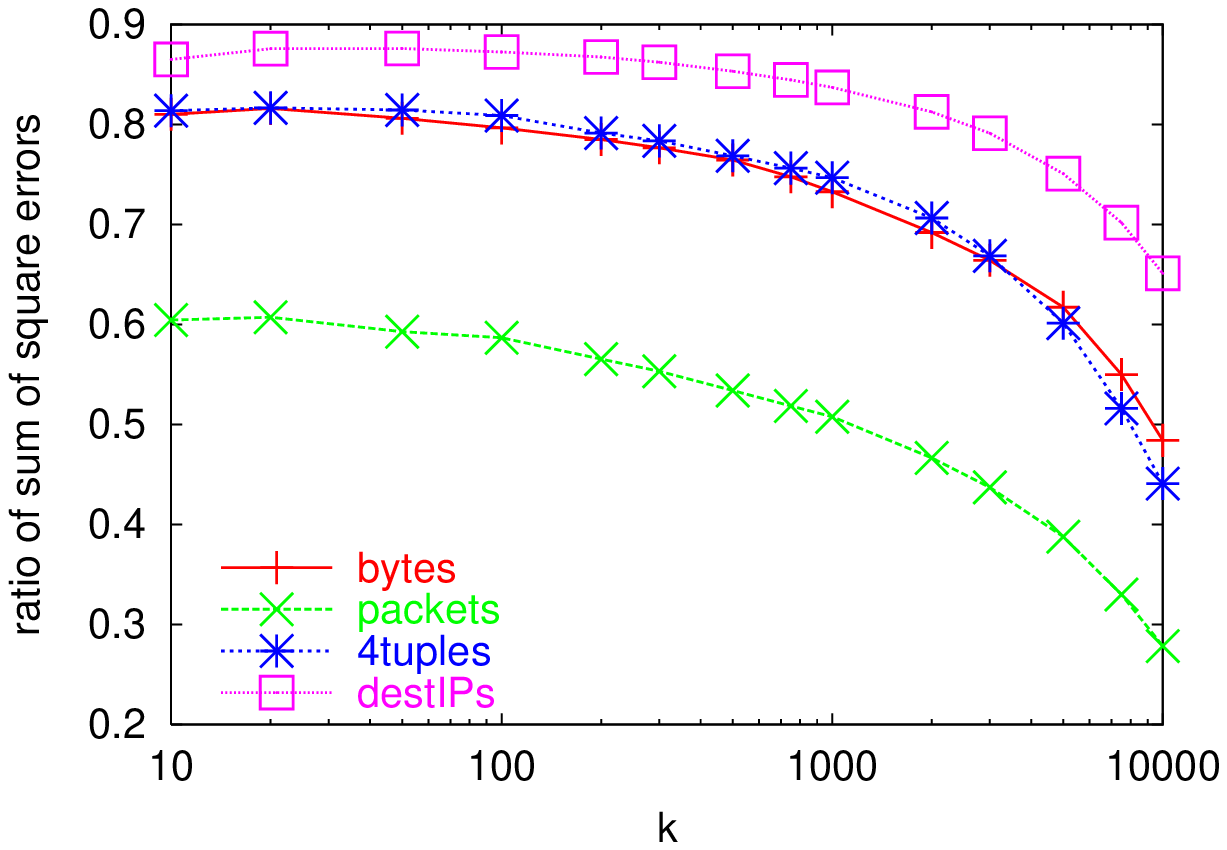,width=0.45\textwidth} &
\epsfig{figure=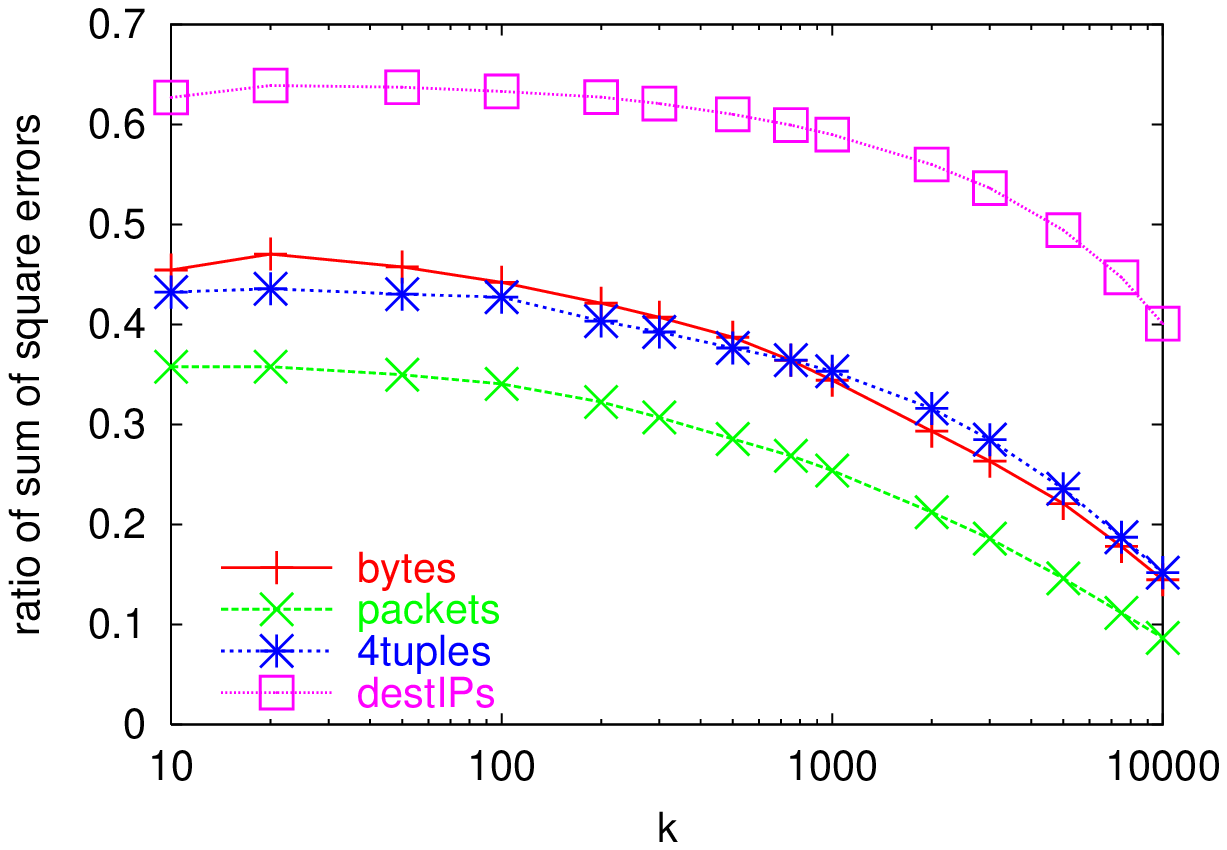,width=0.45\textwidth} \\
\epsfig{figure=multi_code/resultsMO/peering4_r50_ratioSIGV_coord.eps,width=0.45\textwidth} &
\epsfig{figure=multi_code/resultsMO/peering4_r50_ratioSIGV_ind.eps,width=0.45\textwidth}
\end{tabular}
}
\caption{IP dataset1: Top: key$=$destIP.  Bottom: key$=$4tuple.
Left: $\SV[a_c^{(b)}]/\SV[a_p^{(b)}]$.
Right: $\SV[a_i^{(b)}]/\SV[a_p^{(b)}]$.}
\label{ratiosigVdata1:fig}
\end{figure*}

\begin{figure*}[htbp]
\centerline{\begin{tabular}{cc}
\epsfig{figure=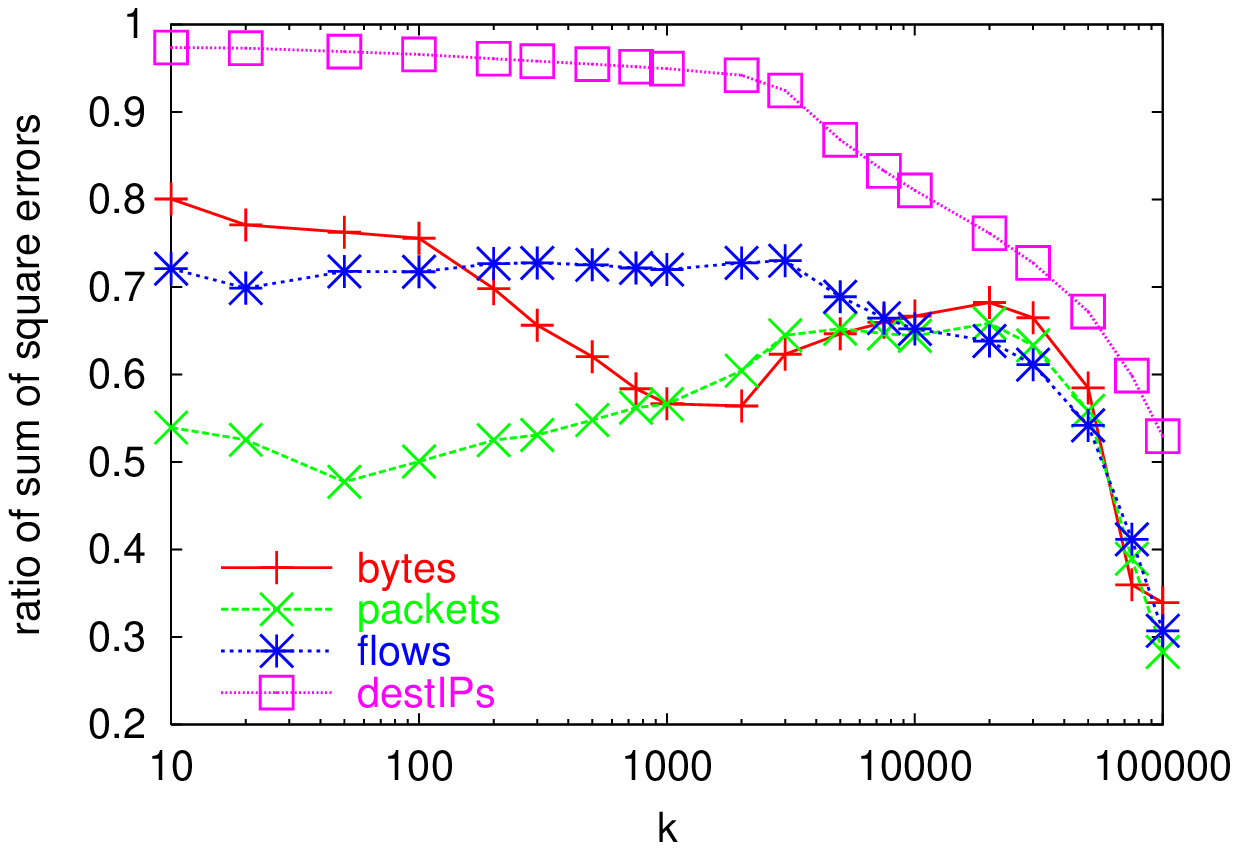,width=0.45\textwidth} &
\epsfig{figure=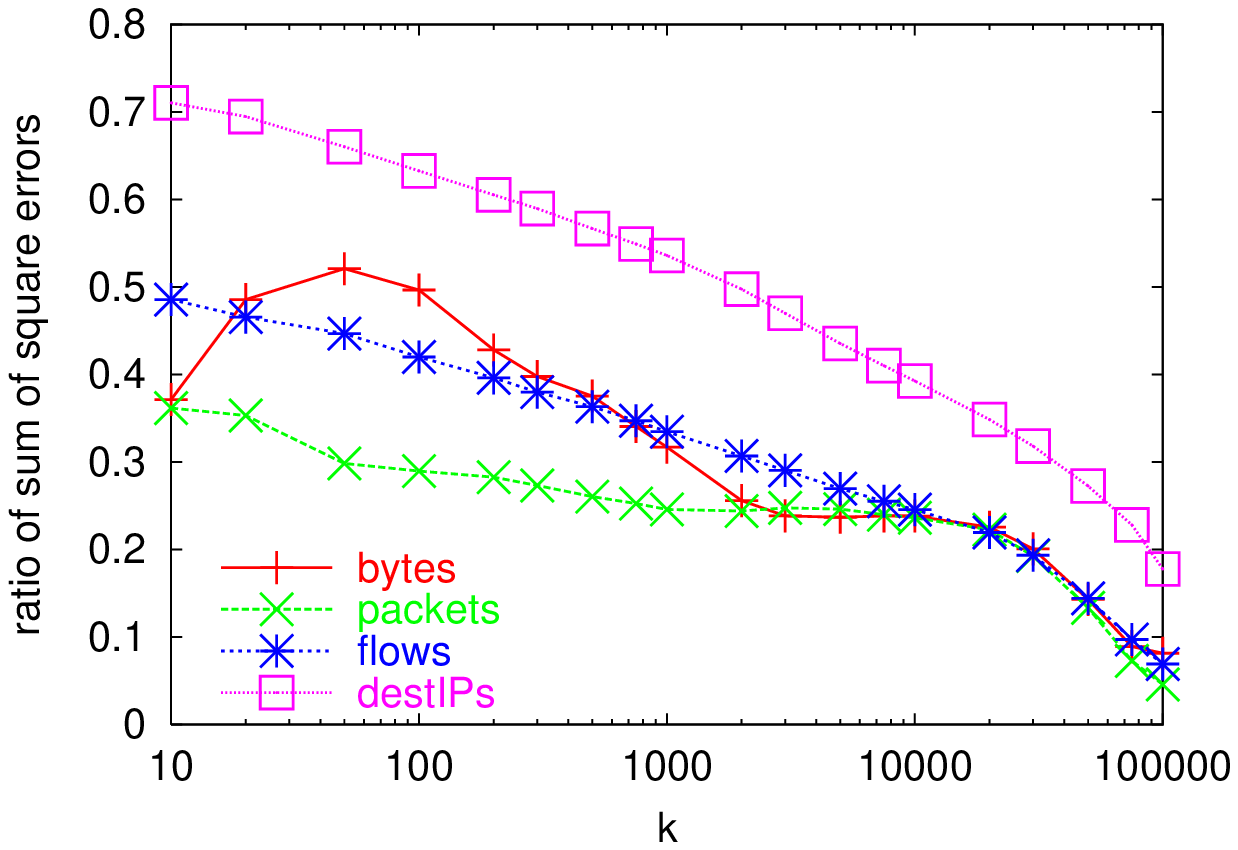,width=0.45\textwidth} \\
\epsfig{figure=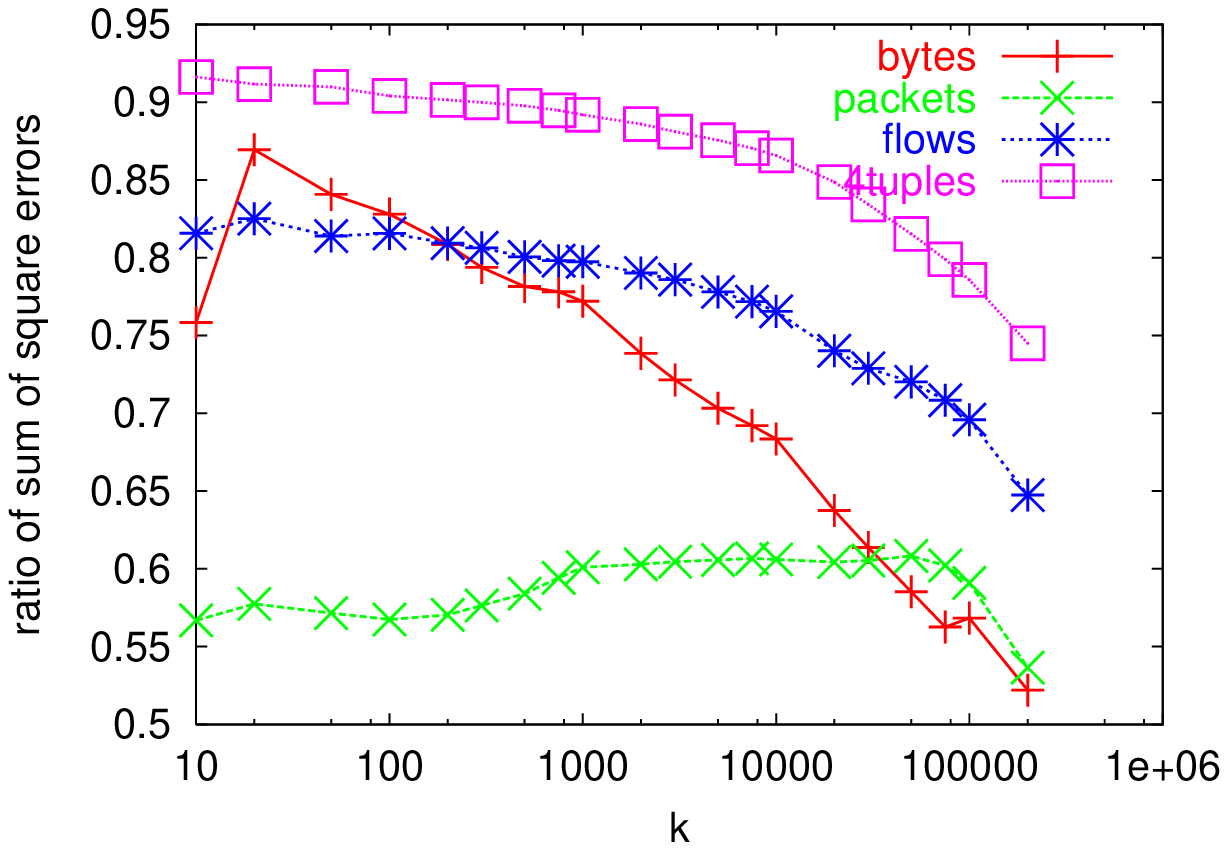,width=0.45\textwidth} &
\epsfig{figure=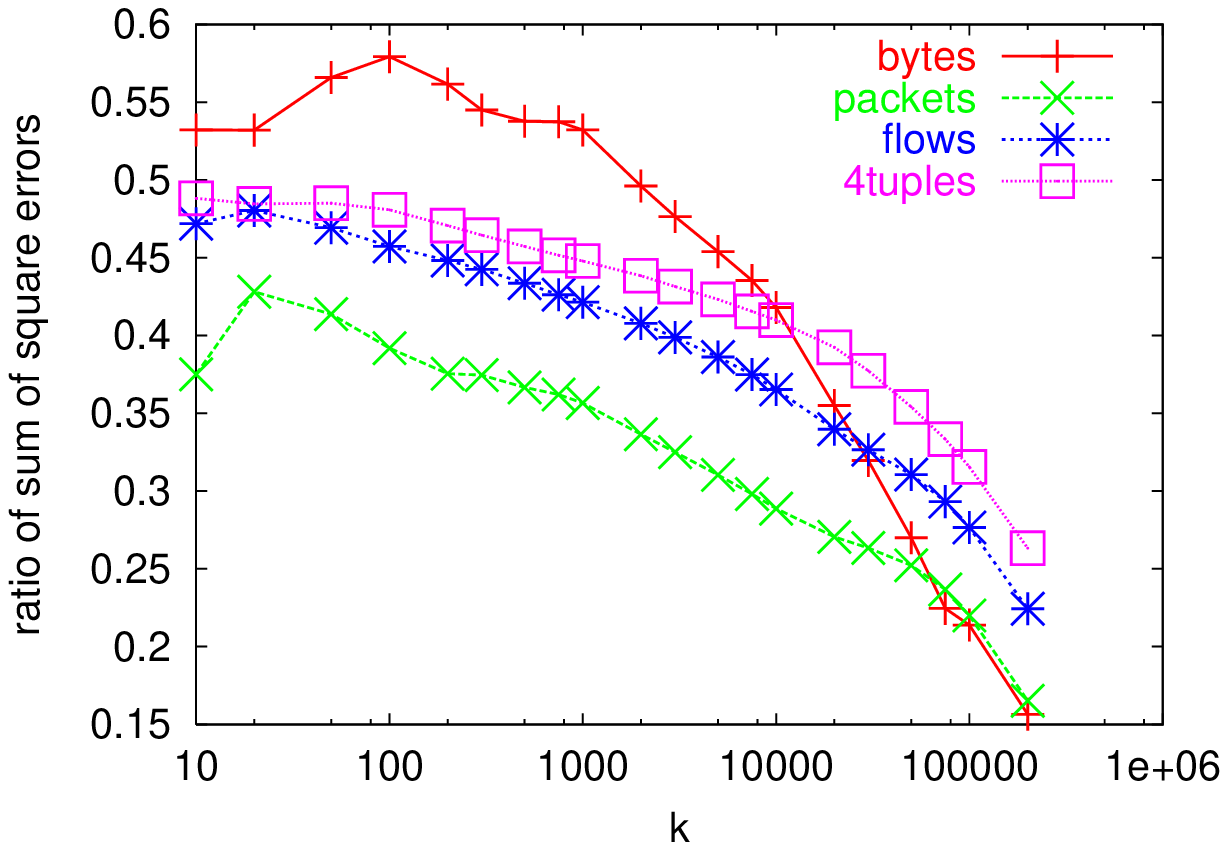,width=0.45\textwidth}
\end{tabular}
}
\caption{IP dataset2: Top: key$=$destIP.  Bottom: key$=$4tuple. Left: $\SV[a_c^{(b)}]/\SV[a_p^{(b)}]$.
Right: $\SV[a_i^{(b)}]/\SV[a_p^{(b)}]$.}
\label{ratiosigV2008080102:fig}
\end{figure*}

\begin{figure*}[htbp]
\centerline{\begin{tabular}{cc}
\epsfig{figure=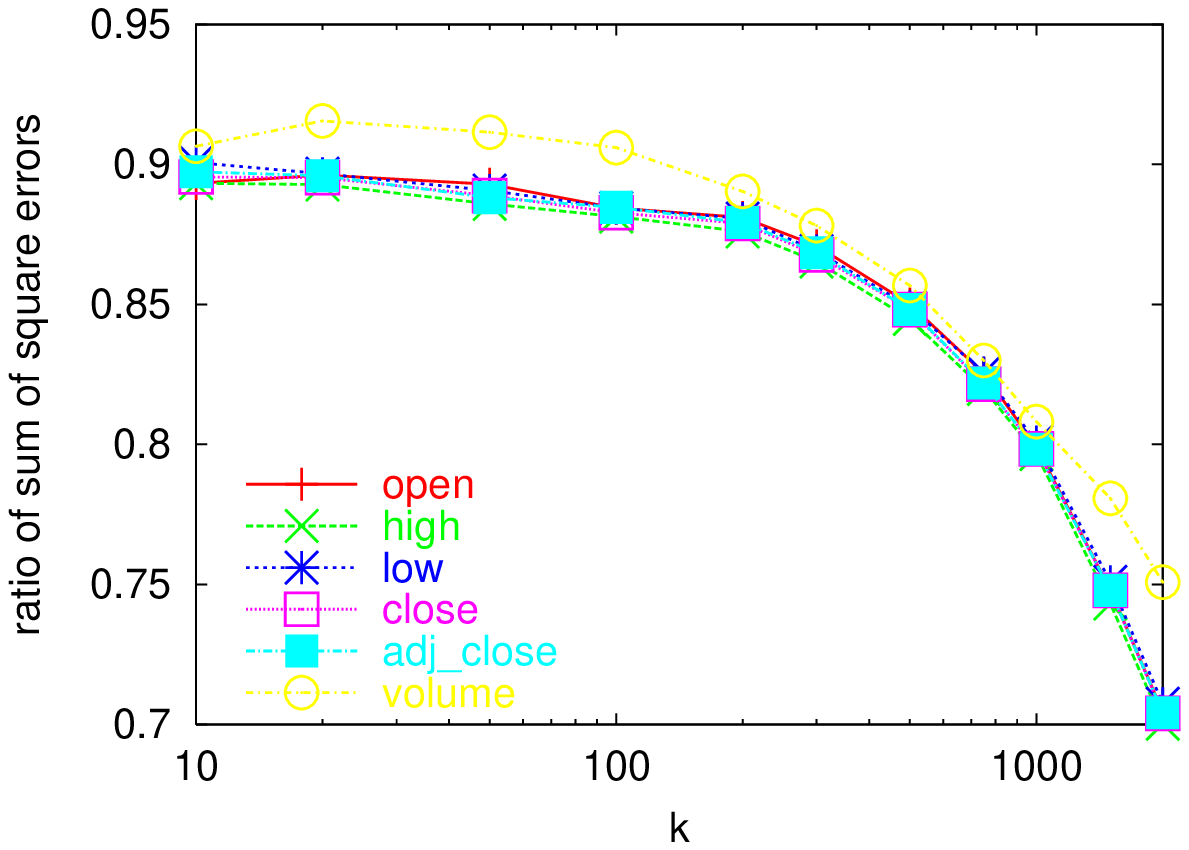,width=0.45\textwidth} &
\epsfig{figure=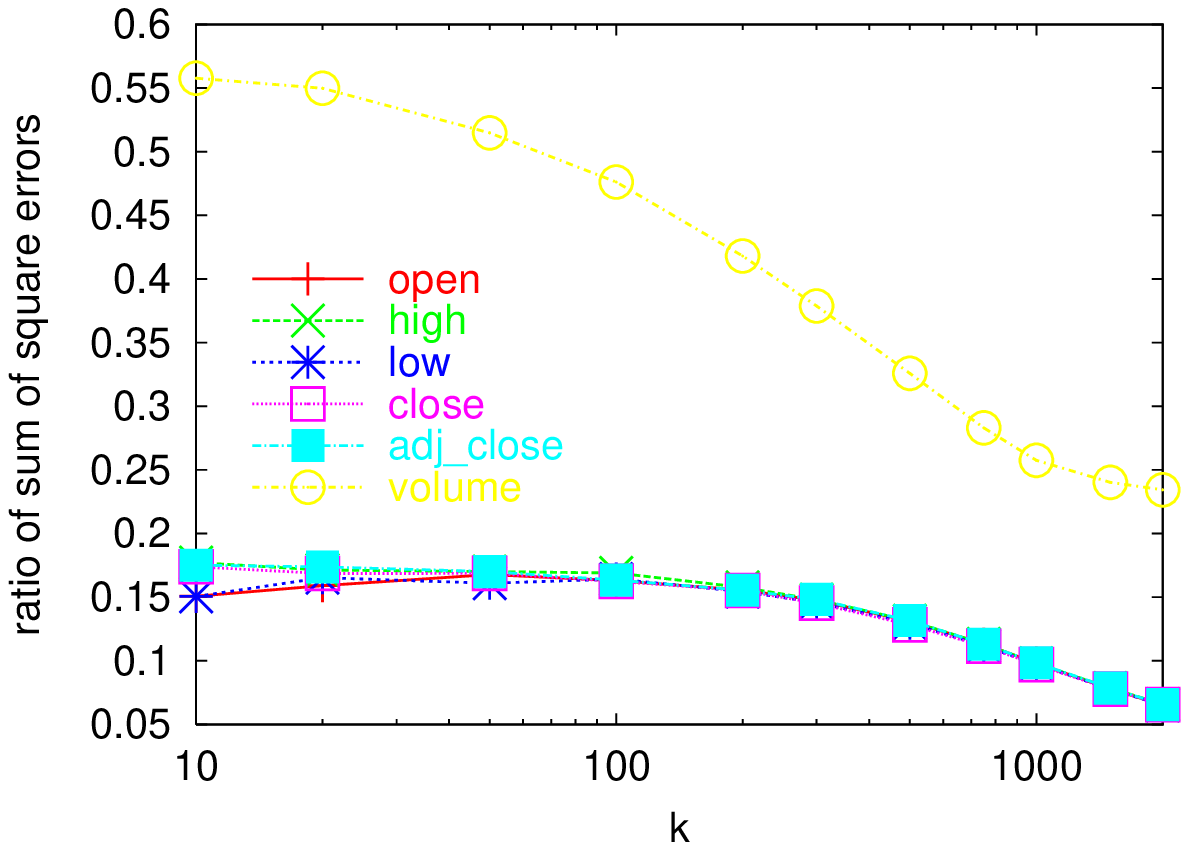,width=0.45\textwidth}
\end{tabular}
}
\caption{Stocks dataset: October 1, 2008. Left: $\SV[a_c^{(b)}]/\SV[a_p^{(b)}]$.
Right: $\SV[a_i^{(b)}]/\SV[a_p^{(b)}]$.}
\label{ratiosigVstocks20081001:fig}
\end{figure*}
}

\medskip
\noindent
{\bf Variance versus storage.}
For a fixed $k$, the plain estimator is in fact
identical for independent and coordinated bottom-$k$ sketches.
Independent bottom-$k$ sketches, however, tend to be larger than
coordinated bottom-$k$ sketches. Here we compare the performance relative to the
 {\em combined sample size}, which is
the number of distinct keys in the combined sample.
We therefore use the notation
$a_{p,i}^{(b)}$ for the plain estimator applied to independent sketches and
$a_{p,c}^{(b)}$ for the plain estimator applied to coordinated sketches.

We compare summaries (coordinated and independent) and estimators
(inclusive and plain) based on the tradeoff of variance versus summary
size (number of distinct keys).  \notinproc{
  Figures~\ref{sigVpeering1:fig}, \ref{sigVpeering4:fig},
  \ref{sigV2008080102_dests:fig}, and \ref{sigV2008080102_4tuple:fig}
  show the normalized sums of variances, for inclusive and plain
  estimators $\nSV[a_i^{(b)}]$, $\nSV[a_c^{(b)}]$,
  $\nSV_{p,c}[a^{(b)}]$, $\nSV[a_{p,i}^{(b)}]$, as a function of the
  combined sample size.}  \onlyinproc{We considered the normalized
  sums of variances, for inclusive and plain estimators
  $\nSV[a_i^{(b)}]$, $\nSV[a_c^{(b)}]$, $\nSV[a_{p,c}^{(b)}]$,
  $\nSV[a_{p,i}^{(b)}]$, as a function of the combined sample size
  (see Figure~\ref{selected_vs:fig}).}
For a fixed sketch size, plain
estimators perform worse for independent sketches than for coordinated
sketches.  This happens since an independent sketch of some fixed size
contains a smaller sketch for each weight assignment than a coordinated
sketch of the same size.
In other words the ``k'' which we use to get an independent sketch of
some fixed size is smaller than the ``k'' which we use to get
a coordinated sketch of the same size.
Inclusive
estimators for independent and coordinated sketches of the same size
had similar variance.  (Note however that for a given union size, we
get weaker confidence bounds with independent samples than with
coordinated samples, simply because we are guaranteed fewer samples
with respect to each particular assignment.)

\onlyinproc{
\begin{figure*}[htbp]
\centerline{\begin{tabular}{ccc}
\epsfig{figure=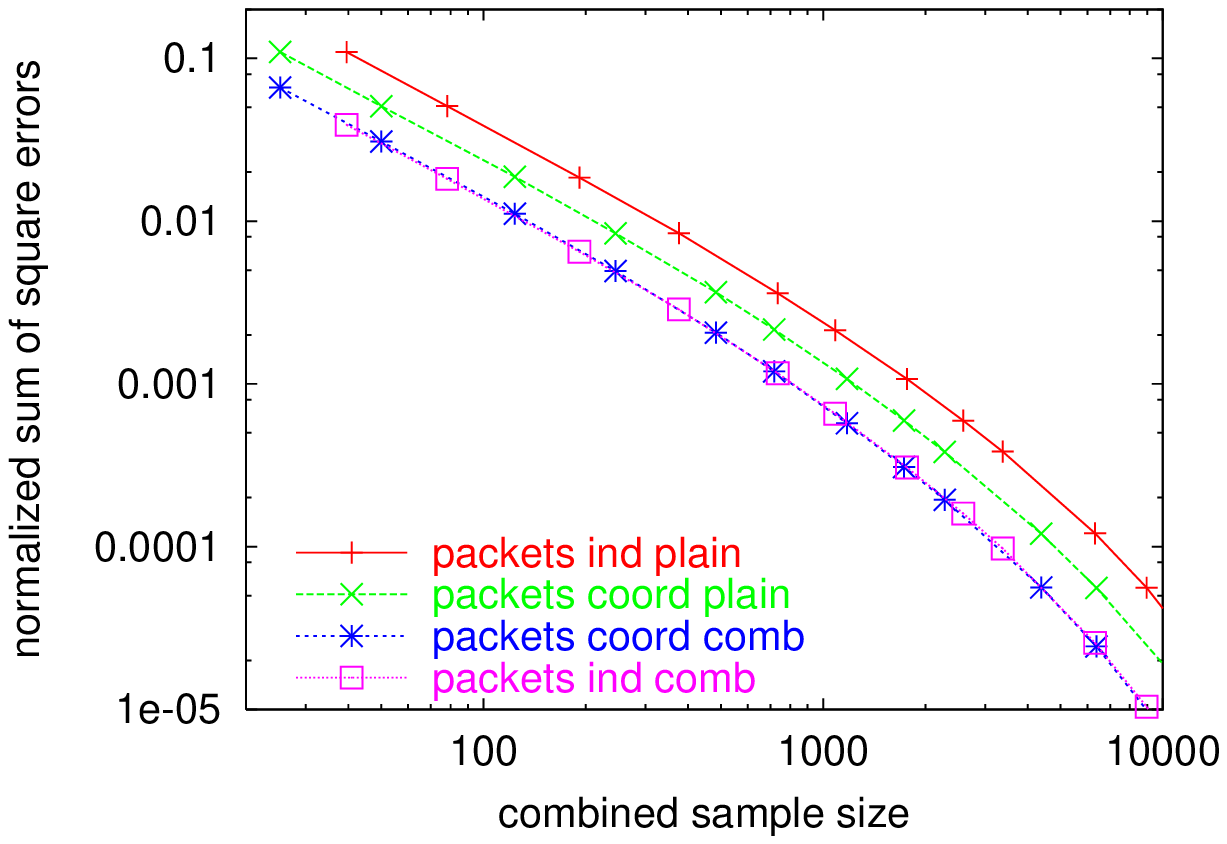,width=0.32\textwidth} &
\epsfig{figure=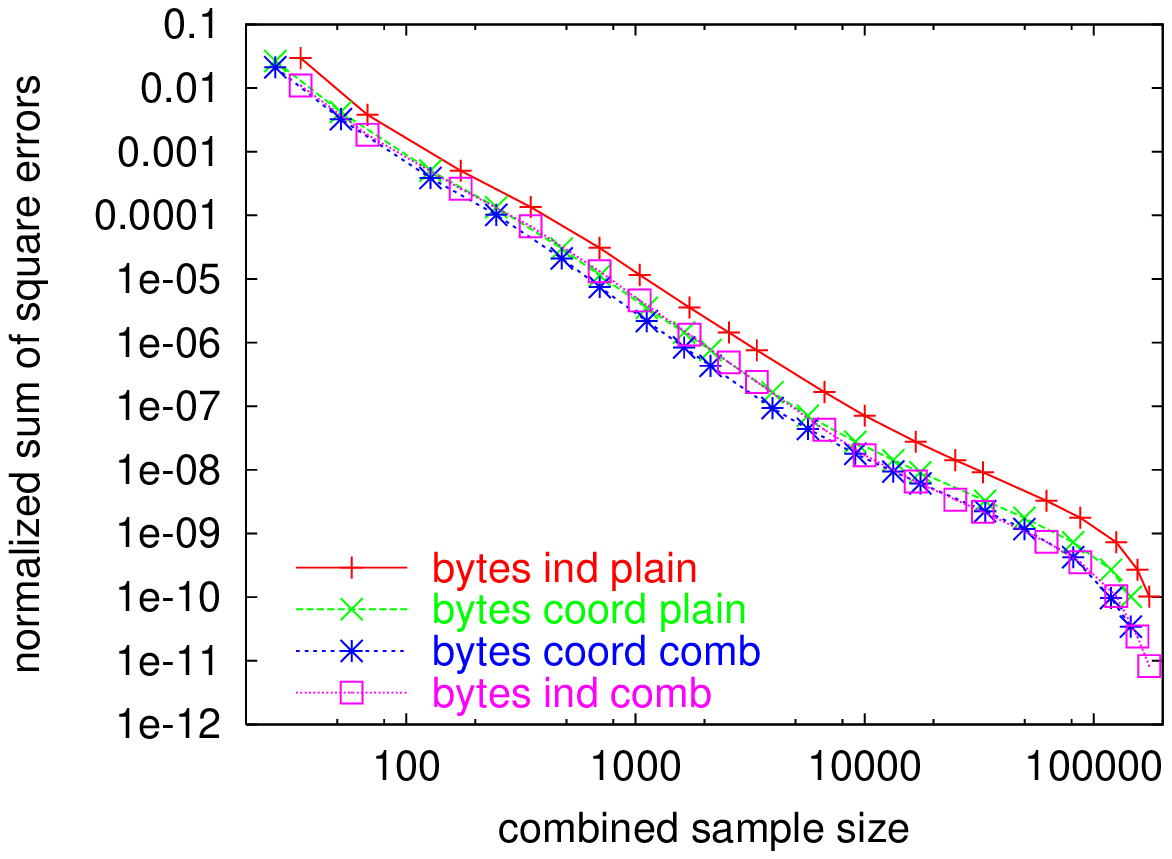,width=0.32\textwidth} &
\epsfig{figure=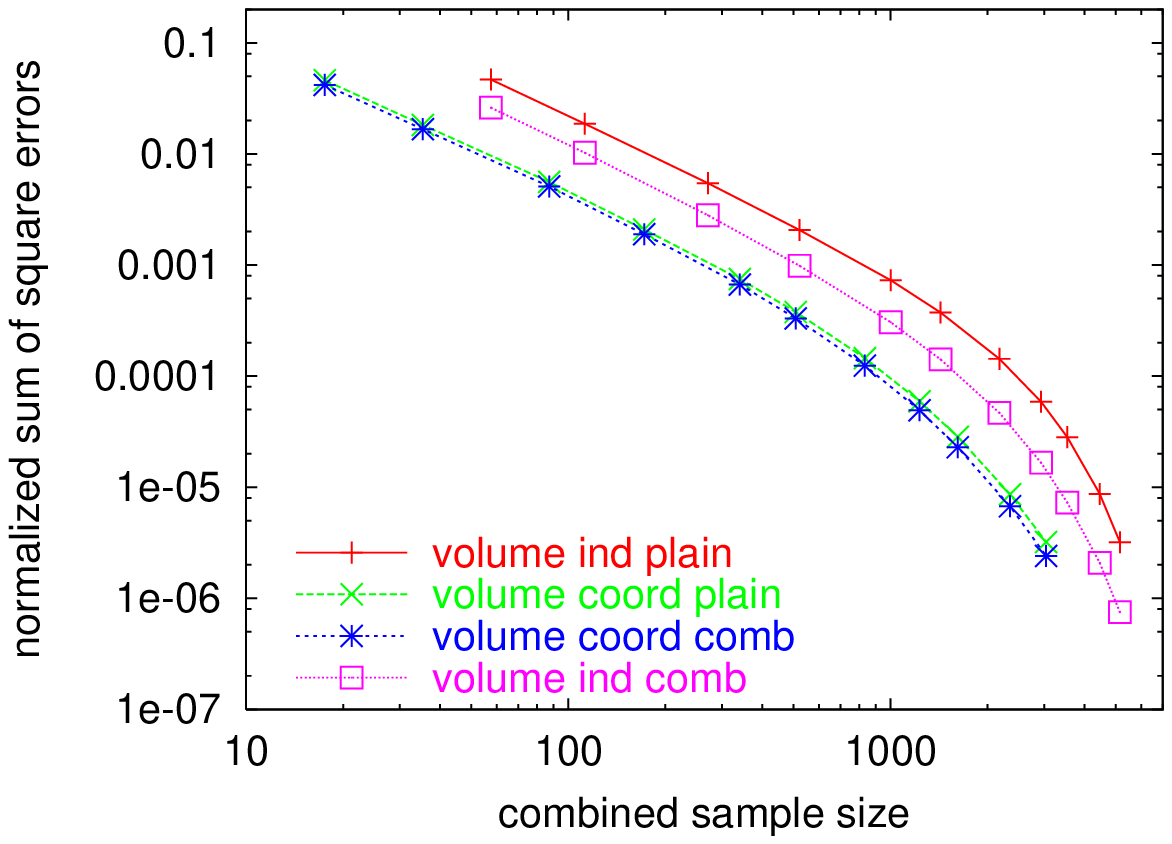,width=0.32\textwidth}
\end{tabular}
}
\caption{$\nSV[a_{c}^{(b)}]$, $\nSV[a_{p,i}^{(b)}]$,  $\nSV[a_{p,c}^{(b)}]$,
 $\nSV[a_i^{(b)}]$ as a function of combined sample size.
Left: IP dataset1 key=destIP, attribute= number of packets.
 Middle: IP dataset2 hour3: key=destIP, attribute = number of bytes; Right: Stocks dataset, volume.
\vspace{-0.6cm}
\label{selected_vs:fig}
}
\end{figure*}
}

\notinproc{
\begin{figure*}[htbp]
\centerline{\begin{tabular}{cc}
\epsfig{figure=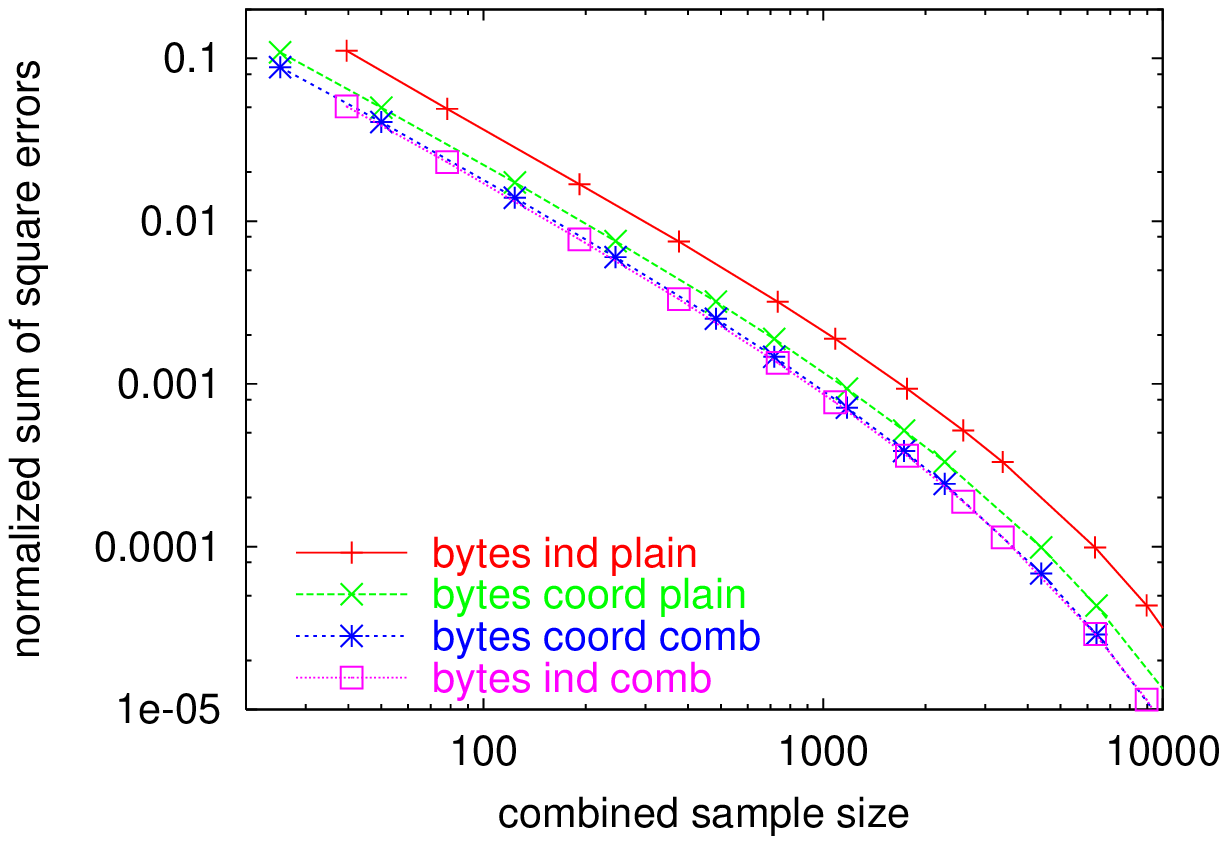,width=0.45\textwidth} &
\epsfig{figure=multi_code/resultsMO/peering1_r200_SIGV_packets.eps,width=0.45\textwidth} \\
bytes & packets \\
\epsfig{figure=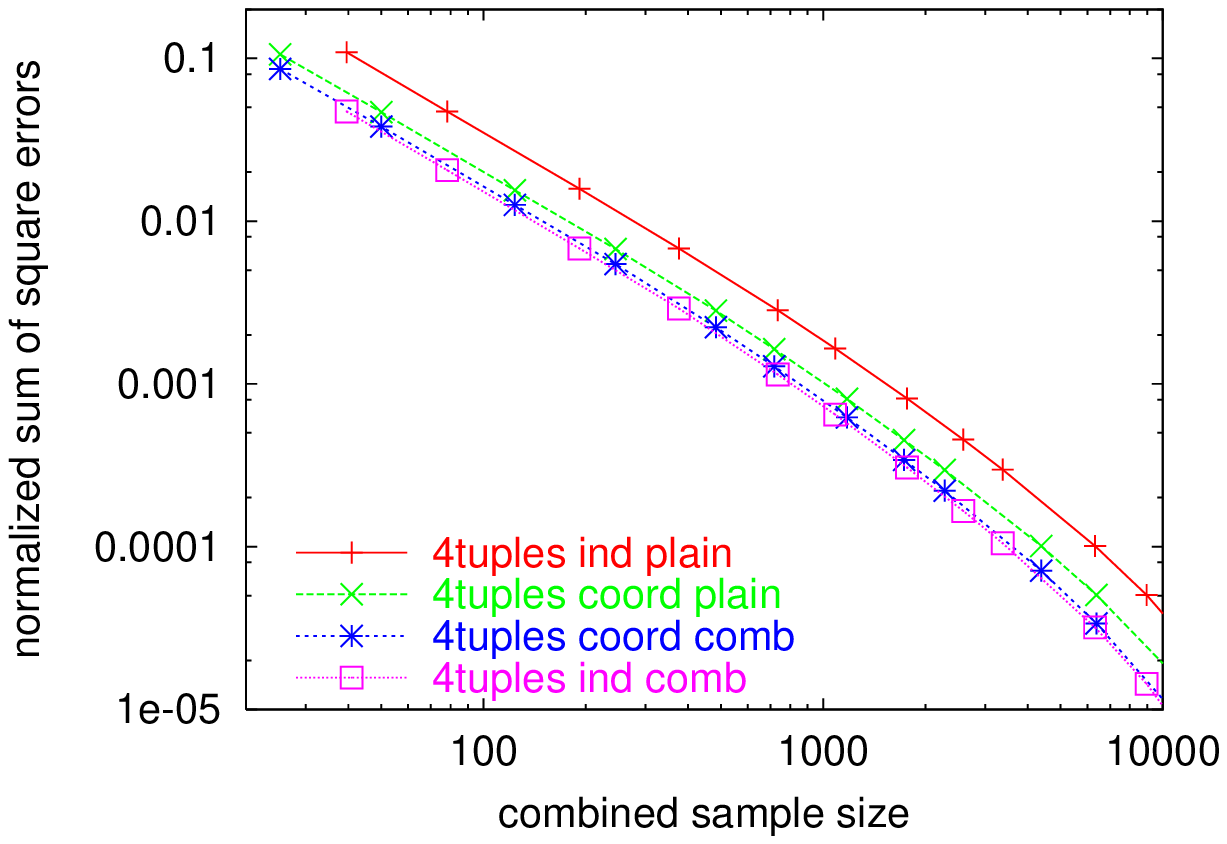,width=0.45\textwidth} &
\epsfig{figure=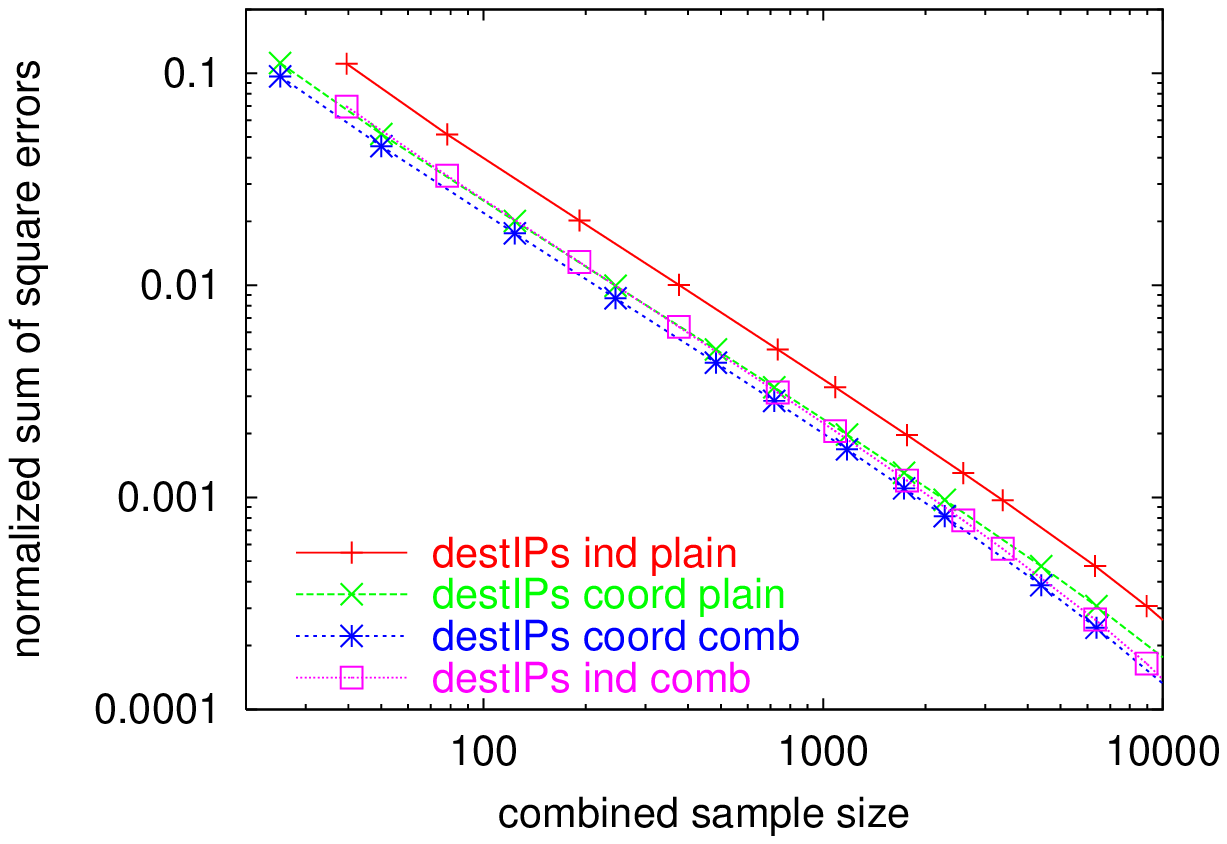,width=0.45\textwidth}  \\
4tuples & destIP
\end{tabular}
}
\caption{IP dataset1:  key$=$destIP. $\nSV[a_i^{(b)}]$,  $\nSV[a_{c}^{(b)}]$, $\nSV[a_{p,i}^{(b)}]$,  $\nSV[a_{p,c}^{(b)}]$ as a function of (combined) sample size.}
\label{sigVpeering1:fig}
\end{figure*}

\begin{figure*}[htbp]
\centerline{\begin{tabular}{ccc}
\epsfig{figure=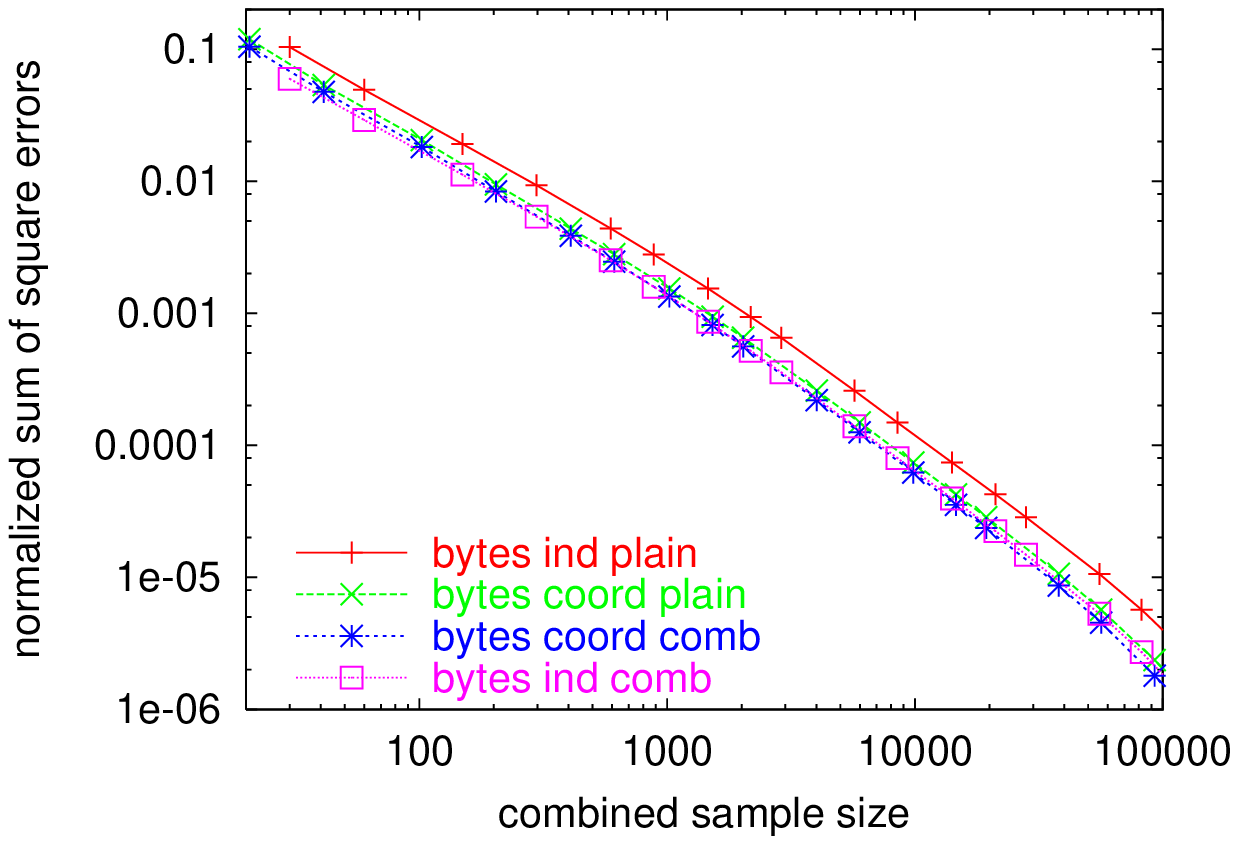,width=0.30\textwidth} &
\epsfig{figure=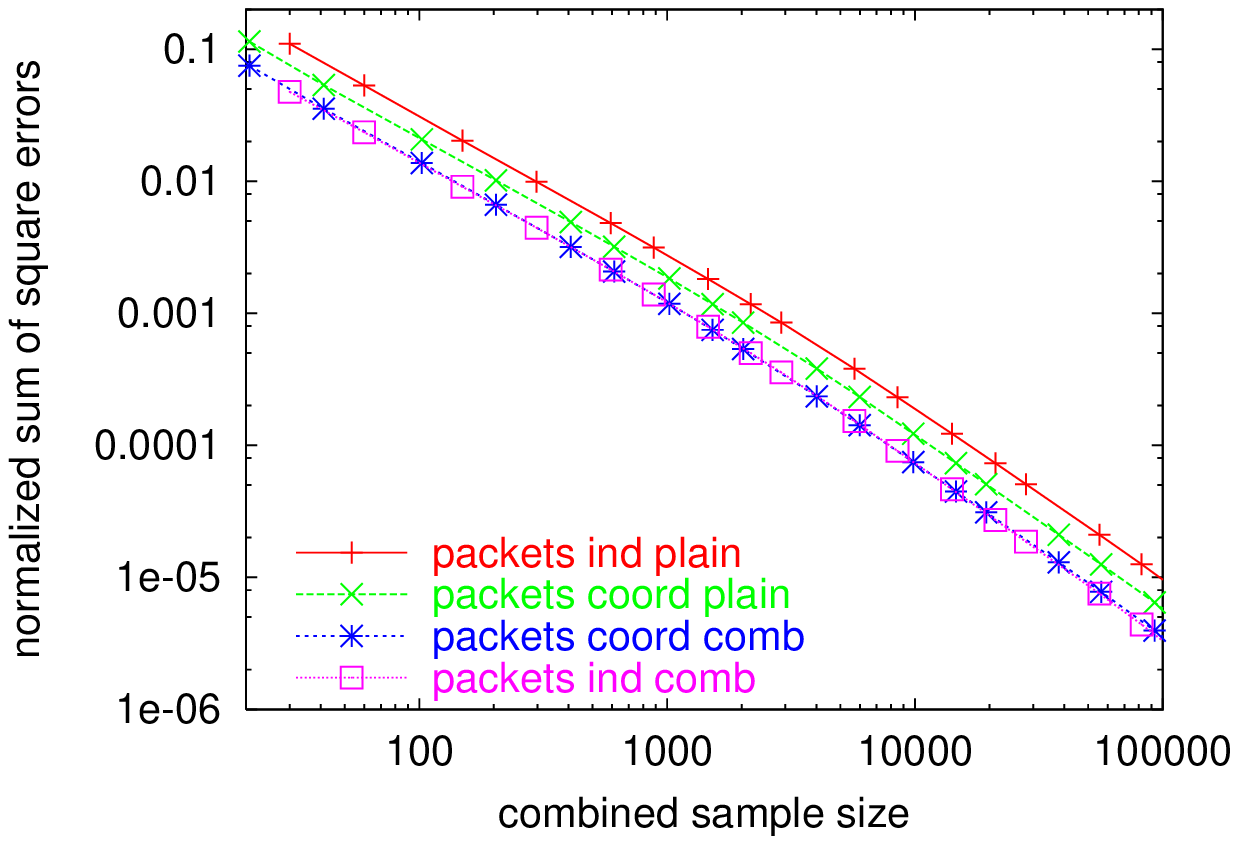,width=0.30\textwidth} &
\epsfig{figure=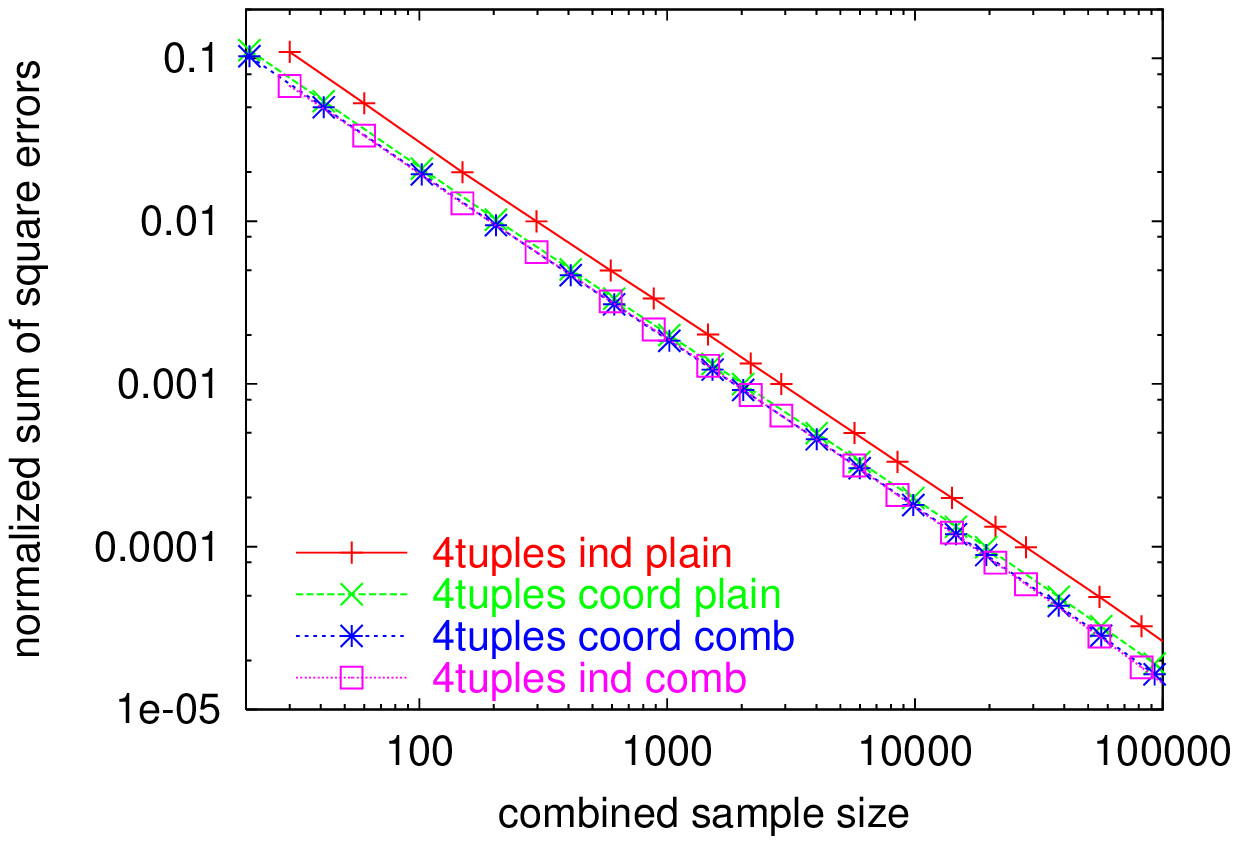,width=0.30\textwidth} \\
bytes & packets & 4tuples
\end{tabular}
}
\caption{IP dataset1:  key$=$4tuples. $\nSV[a_i^{(b)}]$,  $\nSV[a_{c}^{(b)}]$, $\nSV[a_{p,i}^{(b)}]$,  $\nSV[a_{p,c}^{(b)}]$ as a function of (combined) sample size.
 \label{sigVpeering4:fig}}
\end{figure*}

\begin{figure*}[htbp]
\centerline{\begin{tabular}{cc}
\epsfig{figure=multi_code/resultsMO/nfcapd2008080102_r100_SIGV_bytes.eps,width=0.45\textwidth} &
\epsfig{figure=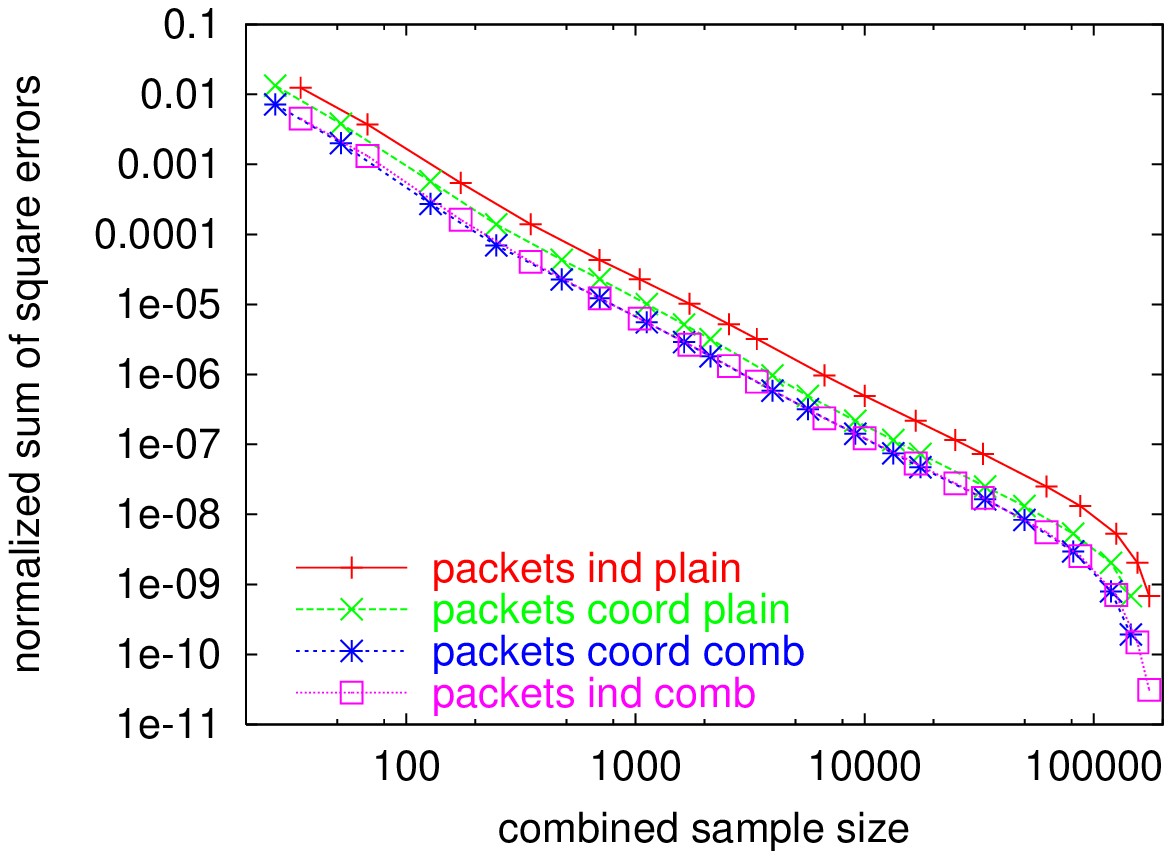,width=0.45\textwidth} \\
bytes & packets \\
\epsfig{figure=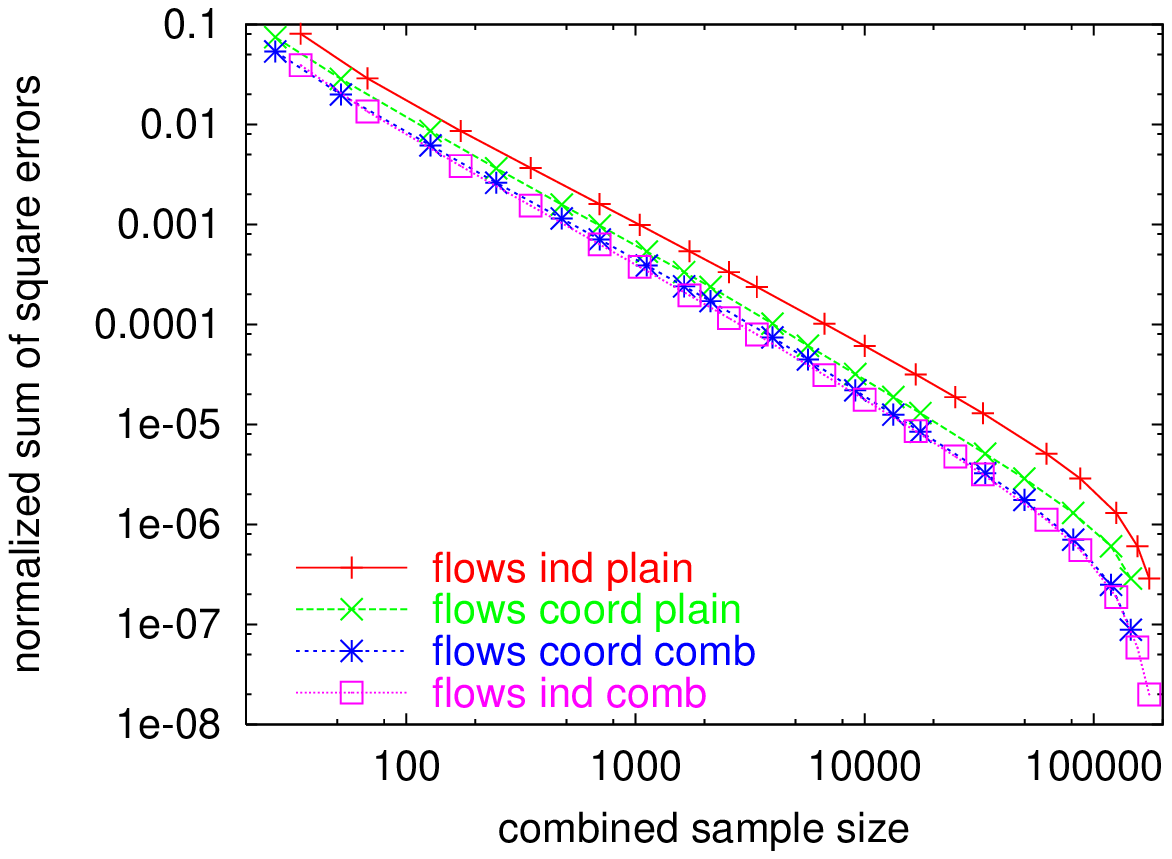,width=0.45\textwidth} &
\epsfig{figure=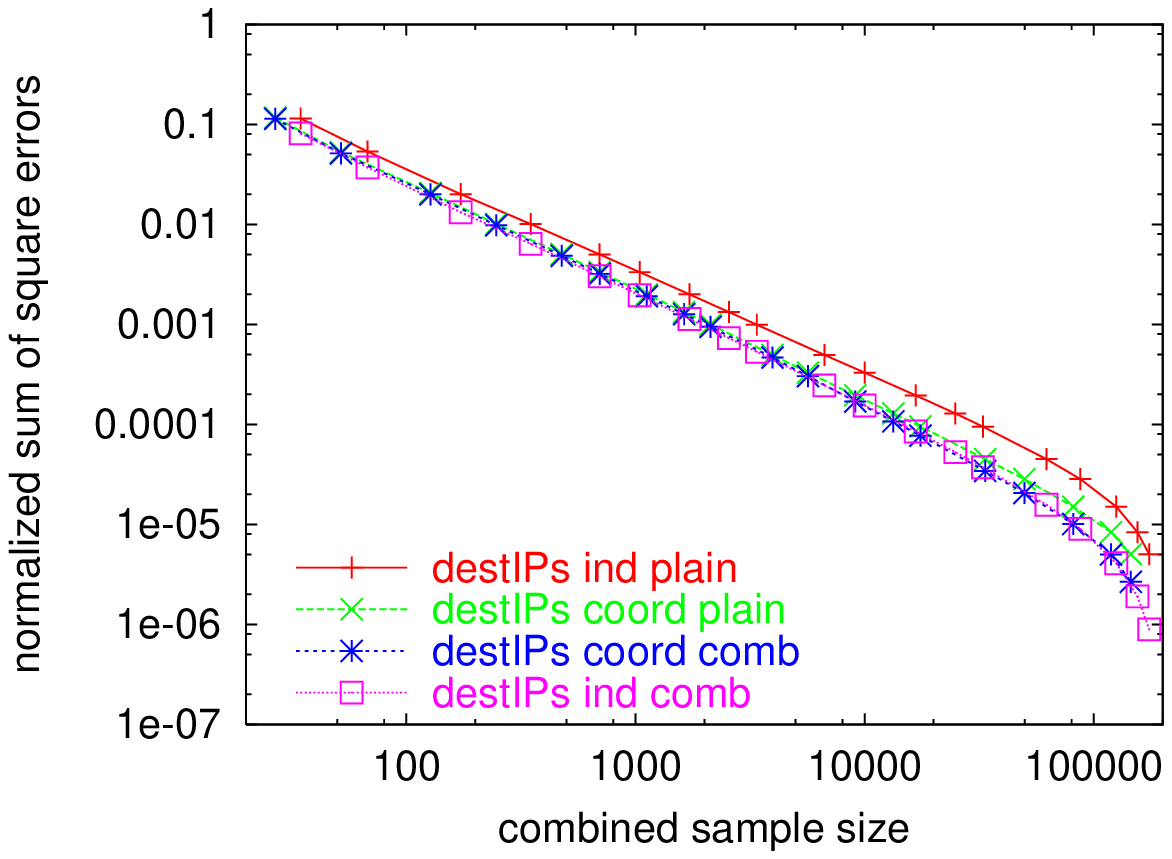,width=0.45\textwidth}  \\
4tuples & destIP
\end{tabular}
}
\caption{IP dataset2:  key$=$destIP hour3. $\nSV[a_i^{(b)}]$,  $\nSV[a_{c}^{(b)}]$, $\nSV[a_{p,i}^{(b)}]$,  $\nSV[a_{p,c}^{(b)}]$ as a function of (combined) sample size.}
\label{sigV2008080102_dests:fig}
\end{figure*}

\begin{figure*}[htbp]
\centerline{\begin{tabular}{cc}
\epsfig{figure=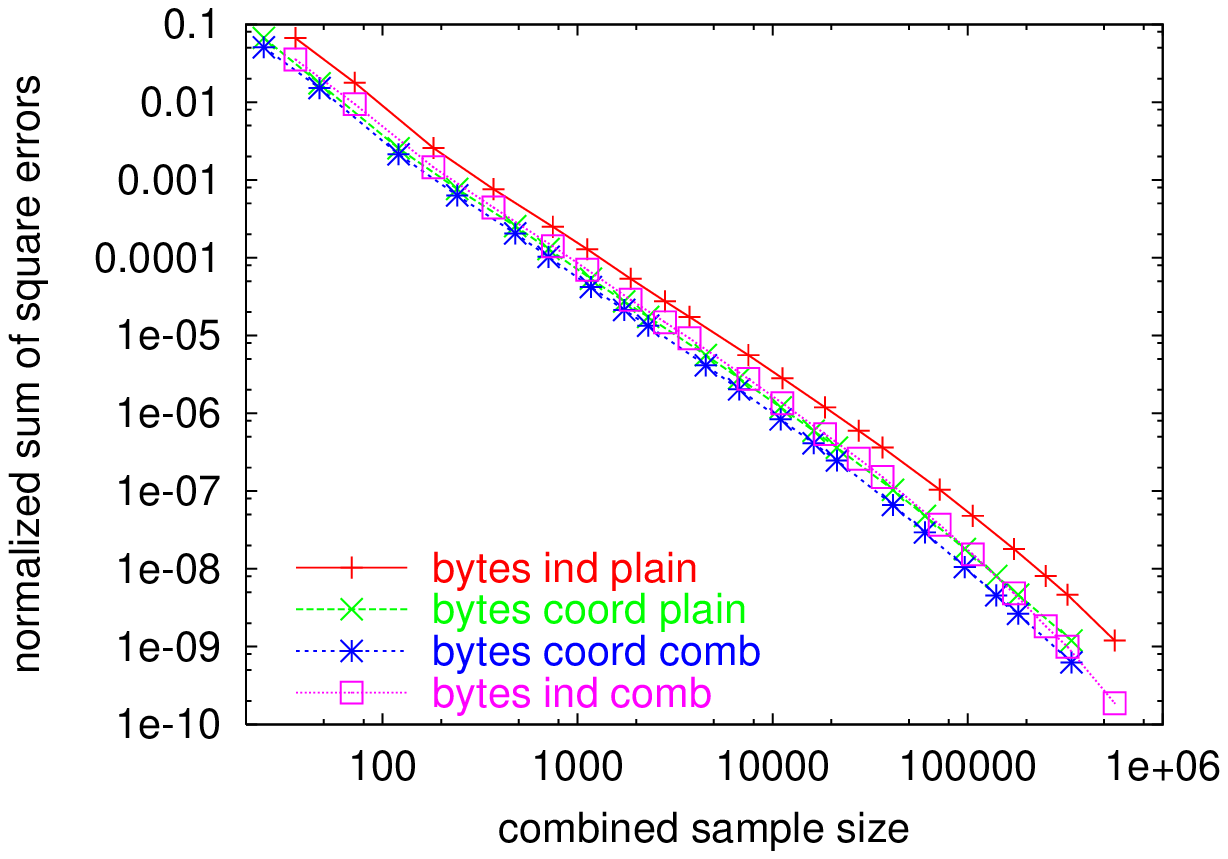,width=0.45\textwidth} &
\epsfig{figure=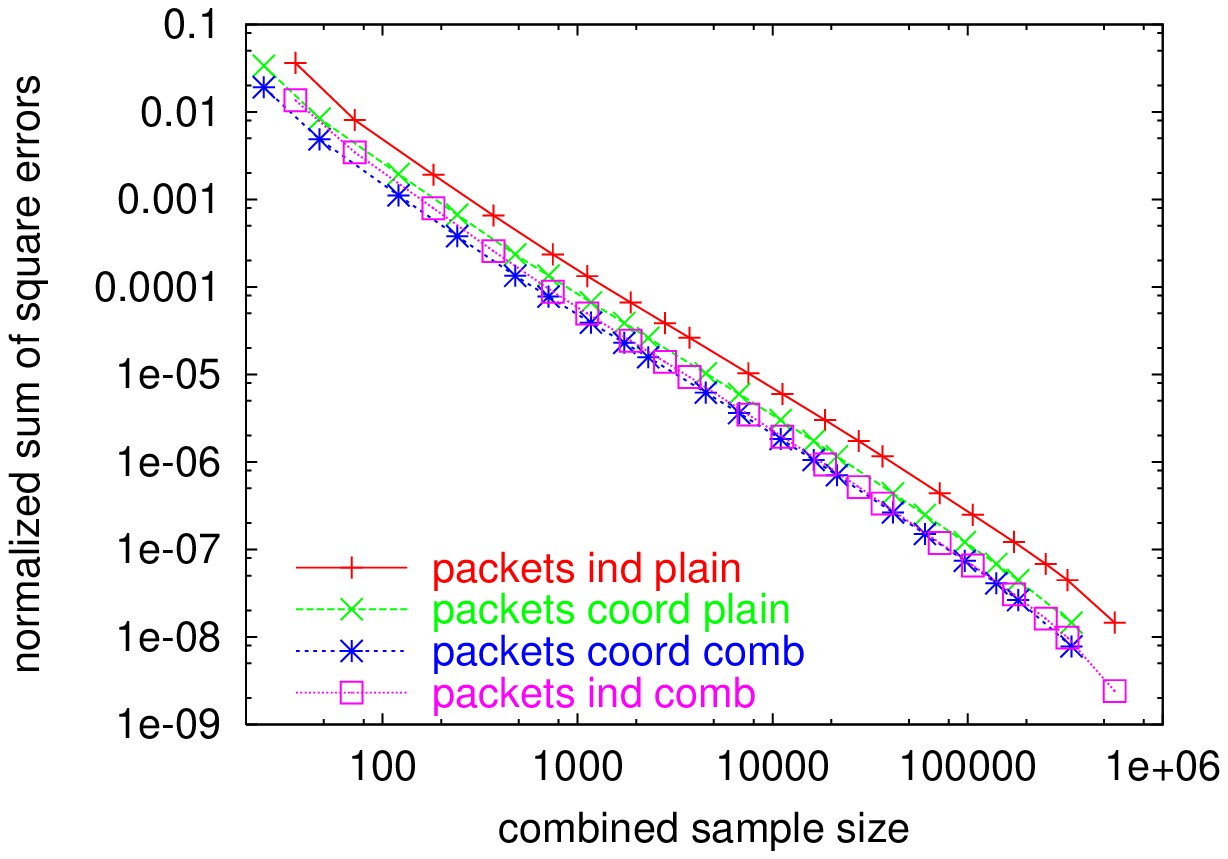,width=0.45\textwidth} \\
bytes & packets \\
\epsfig{figure=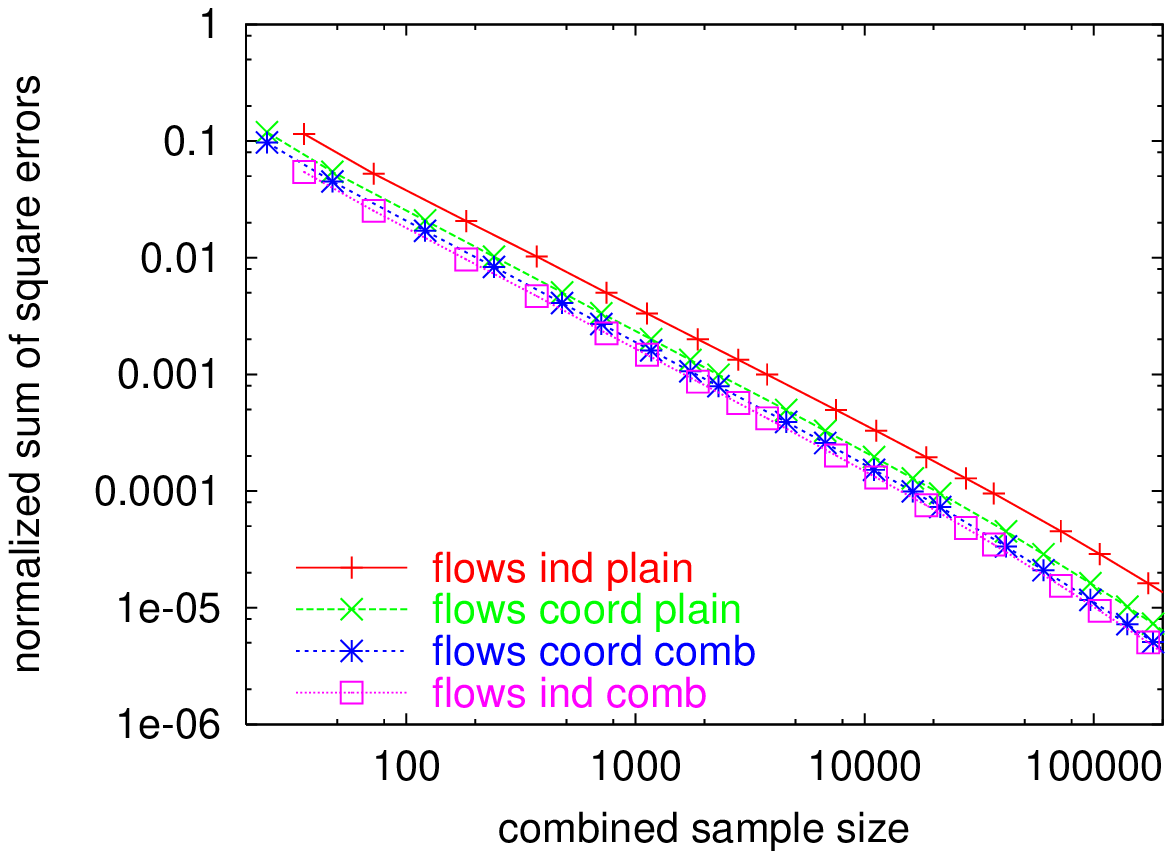,width=0.45\textwidth} &
\epsfig{figure=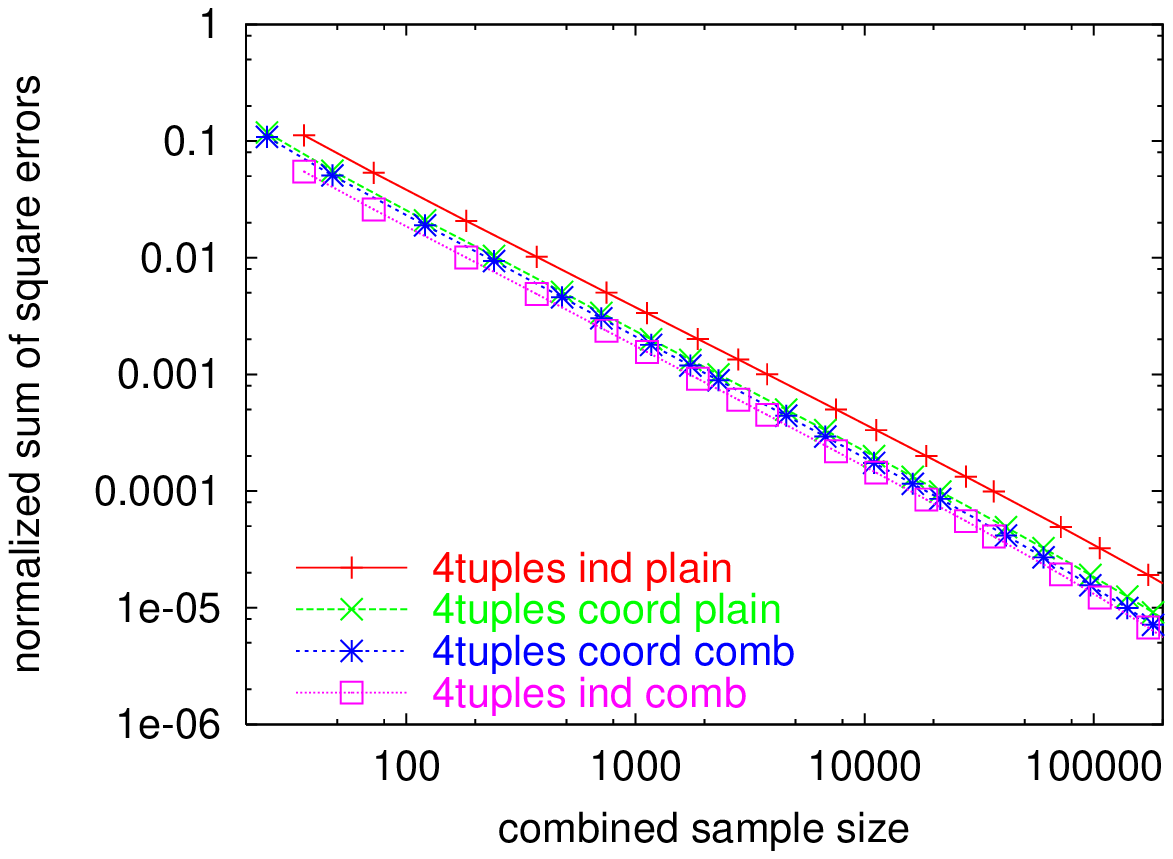,width=0.45\textwidth}  \\
flows & 4tuples
\end{tabular}
}
\caption{IP dataset2:  key$=$4tuple hour3.  $\nSV[a_i^{(b)}]$,  $\nSV[a_{c}^{(b)}]$, $\nSV[a_{p,i}^{(b)}]$,  $\nSV[a_{p,c}^{(b)}]$ as a function of (combined) sample size.
\label{sigV2008080102_4tuple:fig}}
\end{figure*}

\begin{figure*}[htbp]
\centerline{\begin{tabular}{cc}
\epsfig{figure=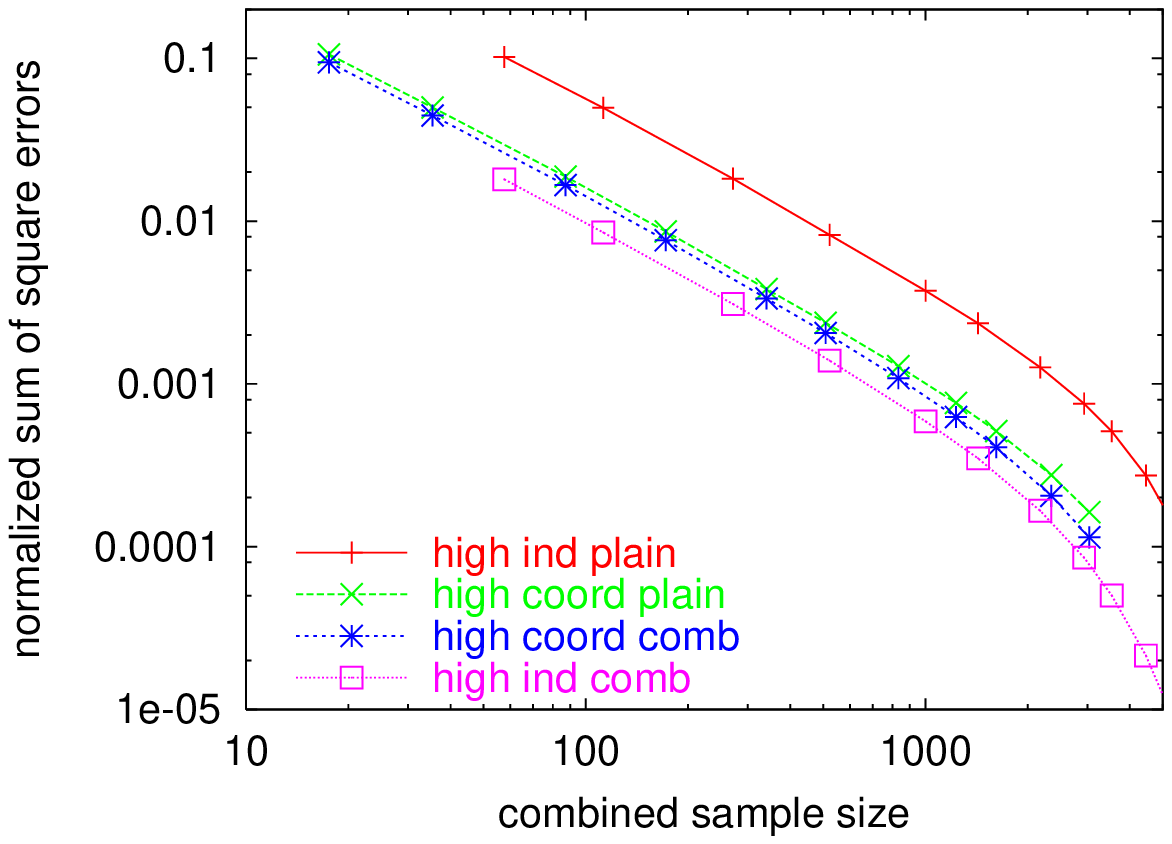,width=0.40\textwidth} &
\epsfig{figure=multi_code/resultsMO/stocks_20081001_SIGV_volume.eps,width=0.40\textwidth} \\
high & volume
\end{tabular}
}
\caption{Stocks dataset. $\nSV[a_i^{(b)}]$,  $\nSV[a_{c}^{(b)}]$, $\nSV[a_{p,i}^{(b)}]$,  $\nSV[a_{p,c}^{(b)}]$ as a function of (combined) sample size.
 \label{sigVstocks:fig}}
\end{figure*}
} 

\medskip
\noindent
{\bf Sharing index.}
 The {\em sharing index}, $|S|/(k*|\cW|)$ of a colocated summary $S$
is the ratio of the expected number of distinct keys in $S$ and the product of
$k$ and the number of weight assignments $|\cW|$.  The sharing index 
falls in the interval $[1/|\cW|,1]$ and is lower (better) when more keys are
shared between samples of different assignments.  More keys are shared
when coordination is used and when weight assignments are more similar --
when all assignments are identical and coordinated sketches are used then the 
index is exactly $1/|\cW|$.

Figure~\ref{storage:fig} plots the sharing index for coordinated and
independent bottom-$k$ sketches, as a function of $k$ for different data sets.
  Coordinated
sketches minimize the sharing index (Theorem~\ref{sharing:lemma}).
 On our
datasets, the index lies in $[0.25,0.68]$ for coordinated
sketches and in $[0.4,1]$ for independent sketches.  The sharing index
decreases when $k$ becomes a larger fraction of keys, both for
independent and coordinated sketches -- simply because it is
more likely that a key is included in a sample of another assignment.
For independent sketches, the sharing index is above $0.85$ for
smaller values of $k$ and can be considerably higher than with
coordinated sketches.

\onlyinproc{
\begin{figure}[htbp]
\centerline{\begin{tabular}{cc}
\epsfig{figure=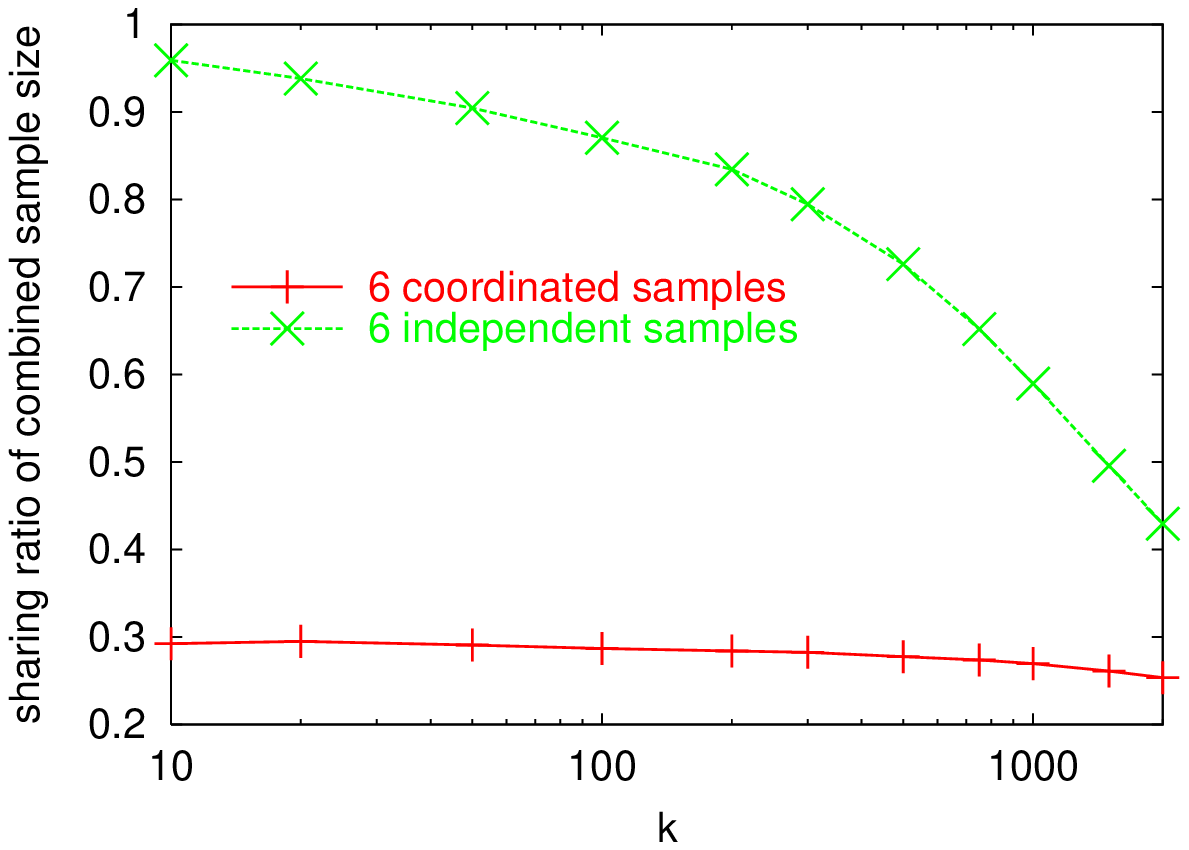,width=0.24\textwidth} &
\epsfig{figure=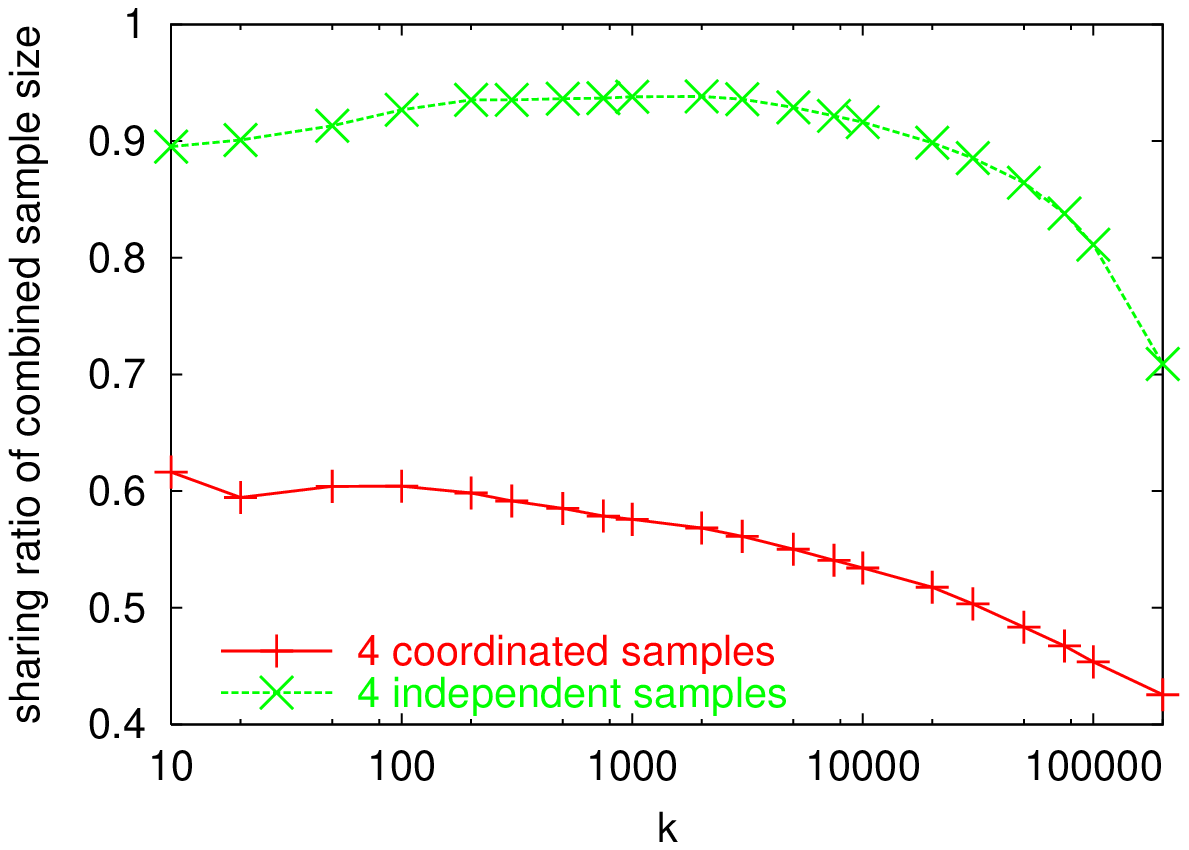,width=0.24\textwidth}
\end{tabular}
}
\caption{Sharing index of coordinated and independent sketches. Left:  Stocks dataset (6 weight assignments). Right: IP dataset2, key$=$4tuple.
\vspace*{-0.4cm}
\label{selected_storage:fig}}
\end{figure}
}

\notinproc{
\begin{figure*}[htbp]
\centerline{\begin{tabular}{ccc}
\epsfig{figure=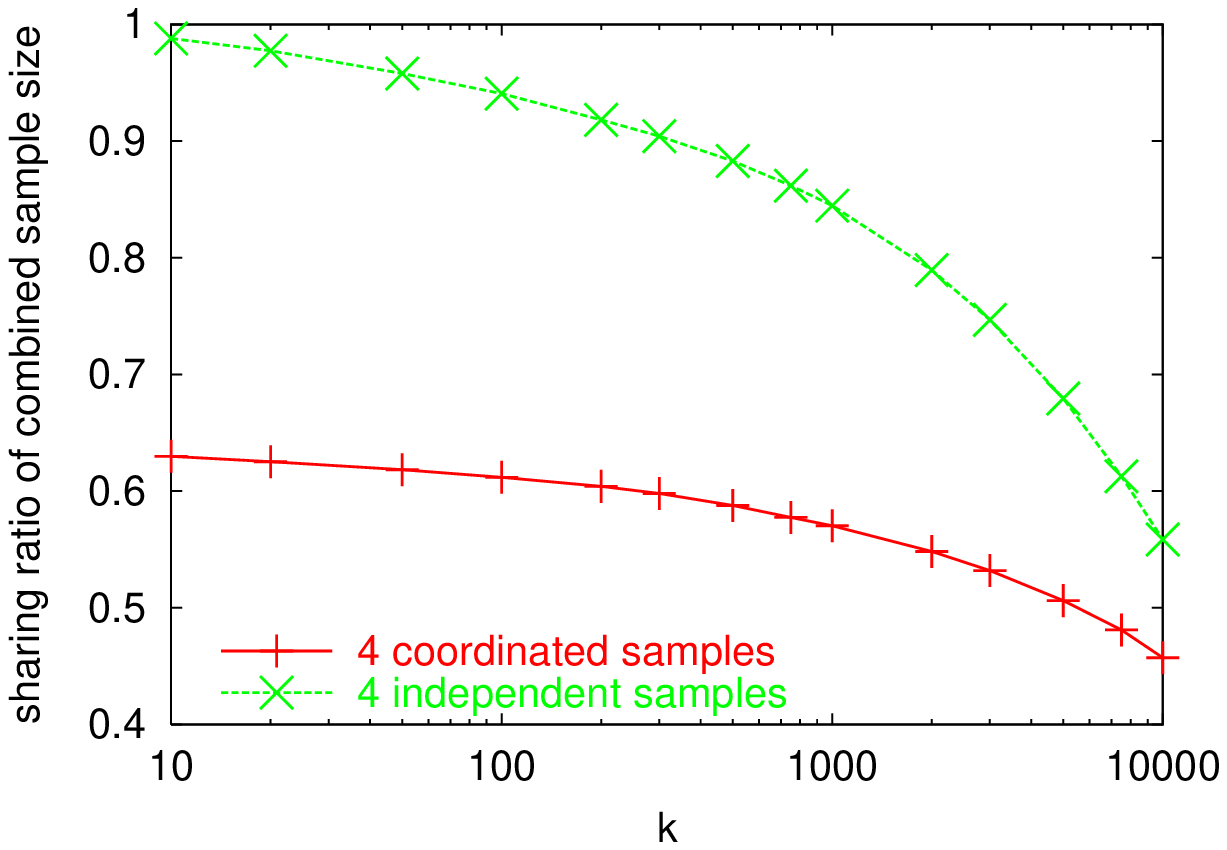,width=0.31\textwidth} &
\epsfig{figure=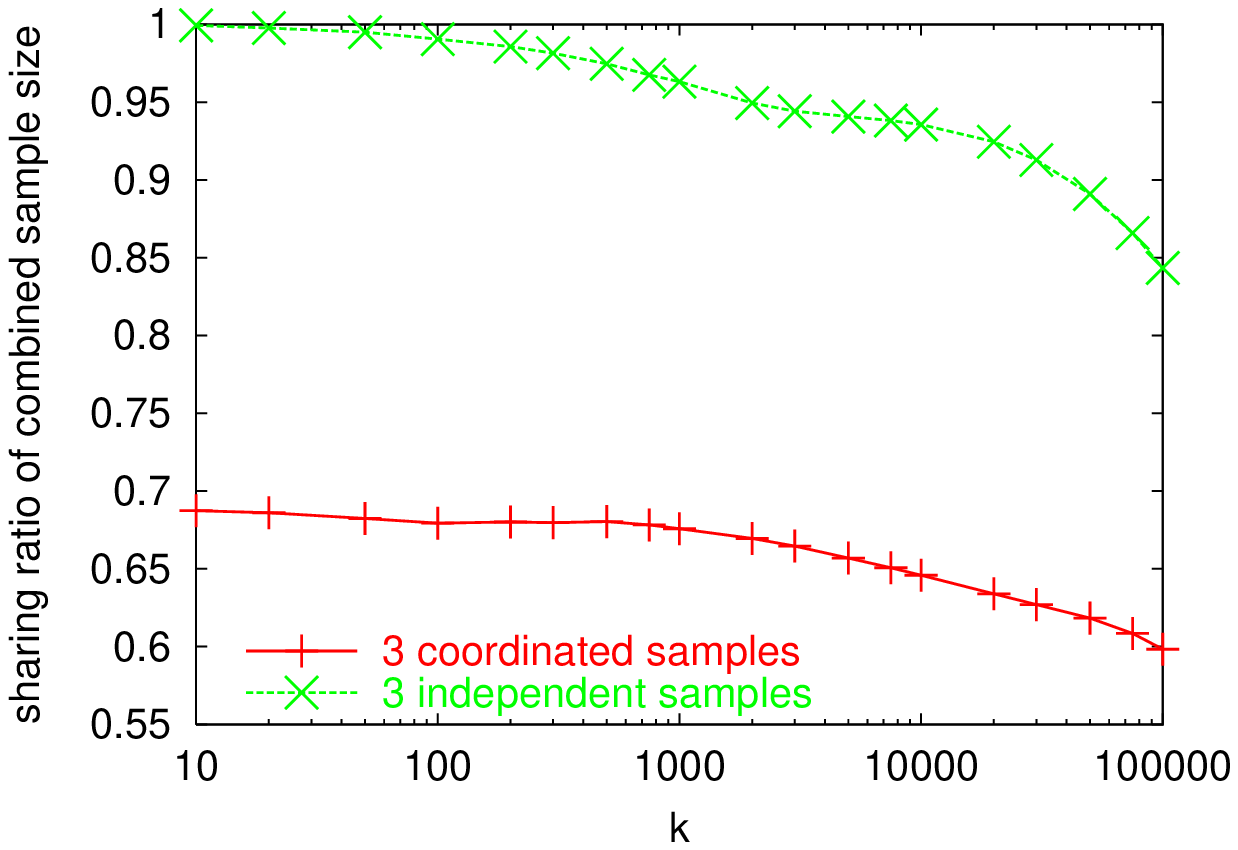,width=0.31\textwidth} &
\epsfig{figure=multi_code/resultsMO/stocks_20081001_r50_storage.eps,width=0.31\textwidth}\\
\epsfig{figure=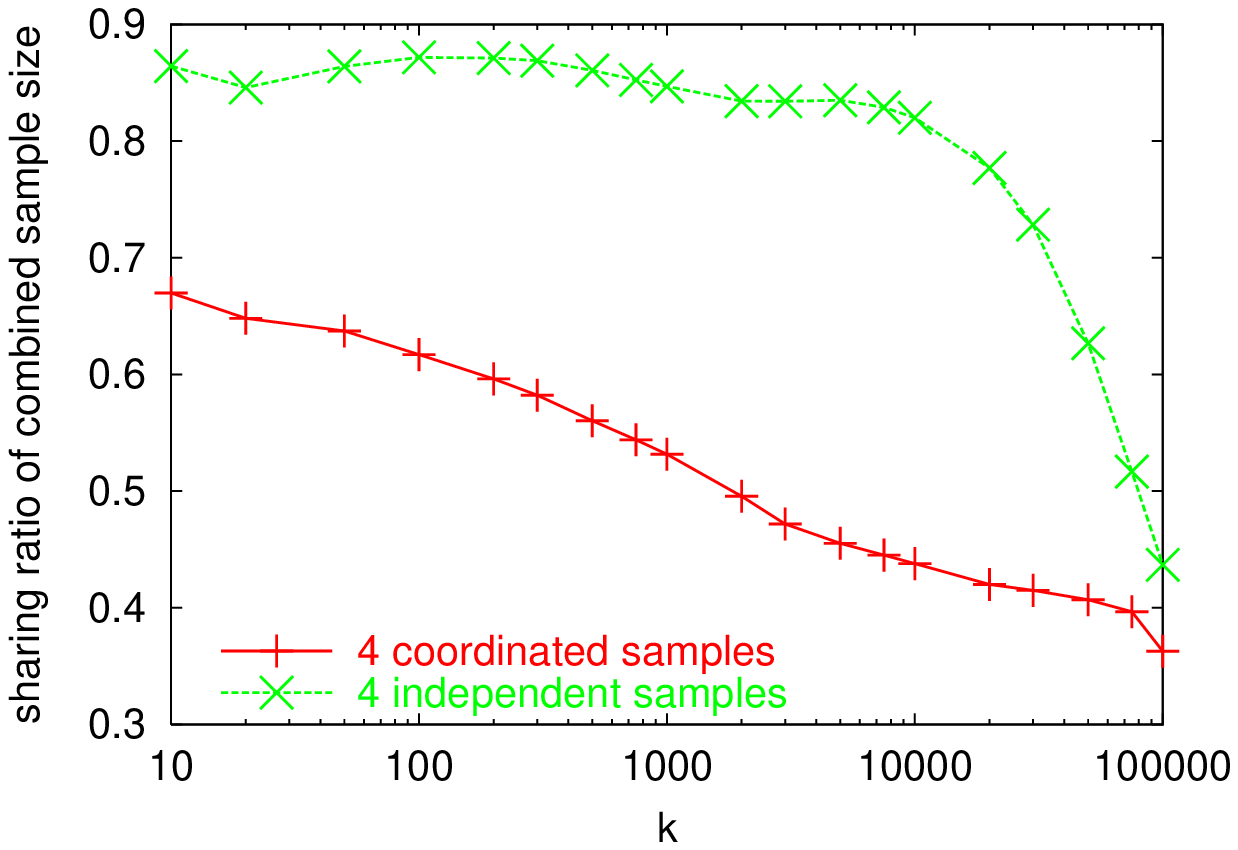,width=0.31\textwidth} &
\epsfig{figure=multi_code/resultsMO/nfcapd2008080102.4tuple_storage.eps,width=0.31\textwidth} &
\end{tabular}
}
\caption{Sharing index of coordinated and independent sketches. Left: IP dataset1, key$=$destIP (4 weight assignments: bytes, packets, 4tuples, IPdests). Middle: IP dataset1, key$=$4tuple (3 weight assignments: bytes, packets, 4tuples). Right: Stocks dataset (6 weight assignments).
Bottom: IP dataset2, key$=$destIP (left) IP dataset2, key$=$4tuple (middle)
\label{storage:fig}}
\end{figure*}
}

\section{Conclusion} \label{concl:sec}

We motivate and study the problem of summarizing data sets modeled as
keys with {\em vector} weights.  We identify two models for these data
sets, {\em dispersed} (such as measurements from different times or
locations) and {\em collocated} (records with multiple numeric
attributes), that differ in the constraints they impose on scalable
summarization.  We then develop a sampling framework and accurate
estimators for common aggregates\onlyinproc{.}\notinproc{,
including aggregations over subpopulations that are specified a
posteriori.}

Our estimators over coordinated weighted samples
for single-assignment and multiple-assignment aggregates including weighted sums and the $L_1$ difference, max, and min improve over previous methods
by orders of magnitude.
Previous methods include independent weighted samples from each
assignment, which poorly supports multiple-assignment aggregates, and
coordinated samples for uniform weights (each item has the same weight in all
assignments where its weight is nonzero), which perform poorly when, as is often
the case, weight values are skewed.  For colocated data sets, our
coordinated weighted samples achieve optimal summary
size while guaranteeing embedded weighted samples of certain sizes with respect
to each individual assignment.  We
derive estimators for single-assignment and
multiple-assignment aggregates over both independent or coordinated
samples that are significantly tighter than existing ones.


As part of ongoing work, we are applying our sampling and estimation
framework to the challenging problem of detection of network
problems. We are also exploring the system aspects of deploying our
approach within the network monitoring infrastructure in a large ISP.

 \bibliographystyle{plain}
\bibliography{cycle,replace,p2p,data_structures,varopt}

\end{document}